\documentclass[12pt,document,nofootinbib,superscriptaddress,onecolumn,preprintnumbers,balancelastpage]{article}
\pdfoutput=1

\usepackage{jheppub_1}

\usepackage{subfig}
\usepackage[countmax]{subfloat}
\usepackage{framed}
\usepackage{mdframed}

\usepackage{multirow}

\usepackage{slashed}

\usepackage{booktabs} 

\usepackage{array}

\makeatletter  
\newcommand{\thickhline}{%
    \noalign {\ifnum 0=`}\fi \hrule height 1pt
    \futurelet \reserved@a \@xhline
}

\newcolumntype{"}{@{\hskip\tabcolsep\vrule width 1pt\hskip\tabcolsep}}

\makeatother

\usepackage{latexsym}
\usepackage{amssymb}
\usepackage{epsfig,amsmath,graphics}
\usepackage{listings}
\usepackage{color}
\usepackage{dcolumn}
\usepackage{slashed}
\usepackage{cancel}
\usepackage{comment,latexsym} 
\usepackage{bm}
\usepackage{verbatim}
\usepackage{tabularx}
\usepackage{dcolumn}
\definecolor{Mahogany}{rgb}{0.62,0.24,0.15}
\definecolor{colorLink}{rgb}{0.7,0,0}
\definecolor{colorCite}{rgb}{0,.7,0}
\definecolor{colorURL}{rgb}{0,0,0.7}
\usepackage[colorlinks=true,linktocpage=true,linkcolor=black,citecolor=colorCite,urlcolor=colorURL]{hyperref}
\usepackage{setspace}
\usepackage{xspace}
\usepackage{multirow}
\usepackage{titlesec}
\usepackage[bbgreekl]{mathbbol}
\usepackage{bm}
\usepackage{indentfirst}
\usepackage{float}
\usepackage{placeins}
\usepackage{afterpage}

\usepackage{empheq}
\usepackage[most]{tcolorbox}
 \tcbset{myformula/.style={colback=white!40,colframe=black,
every box/.style={highlight math style={colback=LightBlue!50!white,colframe=Navy}}
}}

\usepackage[toc,page]{appendix}

\usepackage{etoolbox}
\appto\appendix{\addtocontents{toc}{\protect\setcounter{tocdepth}{0}}}

\usepackage{scalerel}

\usepackage{arydshln}

\allowdisplaybreaks

\expandafter\def\expandafter\normalsize\expandafter{%
    \normalsize
    \setlength\abovedisplayskip{8pt}
    \setlength\belowdisplayskip{8pt}
    \setlength\abovedisplayshortskip{8pt}
    \setlength\belowdisplayshortskip{8pt}
}

\newcommand{\NN}{\mathcal{N}}

\newcommand{\eref}[1]{Eq.~(\ref{#1})} 
\newcommand{\sref}[1]{Sec.~\ref{#1}}

\DeclareRobustCommand{\Fig}[1]{Fig.~\ref{#1}}

\DeclareRobustCommand{\Eq}[1]{Eq.~(\ref{#1})}
\DeclareRobustCommand{\Eqs}[2]{Eqs.~(\ref{#1}) and (\ref{#2})}

\newcommand{\tr}{\text{tr}}

\newcommand{\be}{\begin{equation}}
\newcommand{\ee}{\end{equation}}
\newcommand{\bea}{\begin{eqnarray}}
\newcommand{\eea}{\end{eqnarray}}

\newcommand{\nbsb}{\bar{n}\cdot \bar{\sigma}} 
\newcommand{\alc}{\mathcal{A}} 
\newcommand{\alcs}{\mathcal{A}^*} 
\newcommand{\nsb}{n\cdot \bar{\sigma}} 
\newcommand{\nbs}{\bar{n}\cdot \sigma} 
\newcommand{\ns}{n\cdot \sigma}

\newcommand{\nbp}{\bar{n}\cdot \partial} 
\newcommand{\np}{n\cdot \partial}
 
\newcommand{\bs}{\bar{\sigma}}

\preprint{MIT--CTP 4826}

\title{
Soft-Collinear Supersymmetry
}

\author[a]{Timothy Cohen,}
\author[a,b]{Gilly Elor,}
\author[c,d]{and Andrew J.~Larkoski\,}

\affiliation[a]{\footnotesize Institute of Theoretical Science, University of Oregon, Eugene, OR 97403}
\affiliation[b]{\footnotesize Center for Theoretical Physics, Massachusetts Institute of Technology, Cambridge, MA 02139}
\affiliation[c]{\footnotesize Physics Department, Reed College, Portland, OR 97202}
\affiliation[d]{\footnotesize Center for Fundamental Laws of Nature, Harvard University, Cambridge, MA 02138}

\emailAdd{tcohen@uoregon.edu}
\emailAdd{gelor@uoregon.edu}
\emailAdd{larkoski@reed.edu}

\abstract{
Soft-Collinear Effective Theory (SCET) is a framework for modeling the infrared structure of theories whose long distance behavior is dominated by soft and collinear divergences.  This paper demonstrates that SCET can be made compatible with supersymmetry (SUSY).  Explicitly, the effective Lagrangian for $\mathcal{N} = 1$ SUSY Yang-Mills is constructed and shown to be a complete description for the infrared of this model.  For contrast, we also construct the effective Lagrangian for chiral SUSY theories with Yukawa couplings, specifically the single flavor Wess-Zumino model.  Only a subset of the infrared divergences are reproduced by the Lagrangian -- to account for the complete low energy description requires the inclusion of local operators. SCET is formulated by expanding fields along a light-like direction and then subsequently integrating out degrees-of-freedom that are away from the light-cone.  Defining the theory with respect to a specific frame obfuscates Lorentz invariance --  given that SUSY is a space-time symmetry, this presents a possible obstruction.  The cleanest language with which to expose the congruence between SUSY and SCET requires exploring two novel formalisms: collinear fermions as two-component Weyl spinors, and SCET in light-cone gauge.  By expressing SUSY Yang-Mills in ``collinear superspace", a slice of superspace derived by integrating out half the fermionic coordinates, the light-cone gauge SUSY SCET theory can be written in terms of superfields.  As a byproduct, bootstrapping up to the full theory yields the first algorithmic approach for determining the SUSY Yang-Mills on-shell superspace action. This work paves the way toward discovering the effective theory for the collinear limit of $\mathcal{N} = 4$ SUSY Yang-Mills.
}

\begin{document} 
\maketitle

\setcounter{page}{2}
\begin{spacing}{1.2}

\pagebreak
\section{Introduction and Summary}
\label{sec:intro}
Soft-Collinear Effective Theory (SCET)~\cite{Bauer:2000ew, Bauer:2000yr, Bauer:2001ct, Bauer:2001yt, Bauer:2002uv, Hill:2002vw, Chay:2002vy, Beneke:2002ph} is a powerful framework whose primary purpose is to systematically isolate processes that are dominated by the soft (low momentum) and collinear divergence structure of quantum field theories; see~\cite{Becher:2014oda, StewartNotes} for reviews.  This paper provides a systematic exploration of the ideas of SCET as applied in the context of supersymmetric (SUSY) field theory.  We are all familiar with the power of SUSY to elucidate aspects of quantum field theory -- by demonstrating how SUSY can be maintained in the SCET limit, we are providing a new avenue for exploring the formal aspects of these fascinating and useful Effective Field Theories (EFTs).  Leveraging the tremendous wealth of results that exists for both SUSY field theory and even supergravity will allow us to learn more about the theoretical underpinnings of SCET while also exploring uncharted phase space of SUSY models.

There has been tremendous progress in utilizing SCET for practical applications.  In the context of QCD and electroweak theory, there are many results as applied to heavy meson decays ~\cite{Bauer:2001cu,Chay:2002vy,Beneke:2002ph,Lunghi:2002ju}, proton collisions~\cite{Bauer:2002nz,Stewart:2009yx,Mantry:2009qz,Becher:2010tm,Beneke:2010da}, electroweak showering~\cite{Chiu:2007yn, Chiu:2007dg, Bauer:2016kkv}, and WIMP dark matter annihilation~\cite{Baumgart:2014vma, Bauer:2014ula, Ovanesyan:2014fwa}.  While this list only begins to scratch the surface, it demonstrates the obvious utility of this approach.  Clearly, the structure of SCET deserves to be studied on its own merits.  There does exist a literature whose purpose is to study effective descriptions for the soft and collinear limits in their own right, specifically in the context of hard-soft-collinear factorization~\cite{Bauer:2001yt, Bauer:2002nz, Bauer:2002aj, Collins:2004nx, Manohar:2006nz, Liu:2008cc, Bauer:2010cc, Fleming:2014rea, Feige:2014wja, Rothstein:2016bsq}, the method of regions~\cite{Beneke:1997zp, Jantzen:2011nz}, higher-order terms in the soft limit expansion \cite{Laenen:2010uz,Casali:2014xpa,Broedel:2014fsa,Bern:2014vva,Larkoski:2014bxa}, symmetry constraints on the local operator structure~\cite{Marcantonini:2008qn, Manohar:2002fd}, Regge theory~\cite{Donoghue:2009cq, Donoghue:2014mpa}, the collinear anomaly/rapidity renormalization \cite{Becher:2003qh, Becher:2010tm, Chiu:2011qc, Becher:2011xn, Becher:2011pf, Chiu:2012ir}, defining the EFT in non-covariant gauges~\cite{Beneke:2002ni, Beneke:2002ph, Idilbi:2010im, GarciaEchevarria:2011md}, collinear gravity~\cite{Beneke:2012xa, Nandan:2016ohb}, and connections with the on-shell approach to amplitudes and $\mathcal{N} = 4$ supersymmetric Yang Mills~\cite{Chay:2010az, Basso:2014jfa}.    Our focus here is to lay the groundwork for understanding the interplay of SCET with SUSY theories.  

This paper provides the first steps to describing the degrees-of-freedom relevant for SCET in the language of $\NN = 1$ SUSY.  Our focus will be on two of the simplest 4-D SUSY models.  First, we will prove that SUSY Yang-Mills (SYM) can be formulated as a self-consistent SCET.  That such a SUSY theory of the collinear limit can exist could be anticipated from the SUSY relations of the collinear splitting functions \cite{Berends:1990vx,Seymour:1997kj} of adjoint quarks and gluons.  Developing this theory will require introducing the concept of ``collinear superspace"~\cite{Cohen:2016jzp}, which is the natural setting for SUSY SCET.  

For contrast, we will then demonstrate that the Lagrangian for the propagating collinear fields of the (single flavor) Wess-Zumino model only accounts for a subset of the infrared structure of the full theory.  In particular, one must include additional local operators to account for all the infrared divergences as we discuss in detail below.   This provides an example of the subtleties one can encounter when working with chiral theories.  A particularly interesting phenomenological application is Higgsstrahlung, where a collinear Higgs is emitted off a top quark.  Identifying the local operators and their organization to provide a complete low energy description of this process will be paramount for future precision studies.  For examples of inclusive calculations involving top Yukawa couplings, see~\cite{Chiu:2007dg, Chiu:2008vv, Manohar:2014vxa}

Even for these simplest cases, there are many issues that we must address: how to formulate SCET in superspace, how to match the infrared divergence structure, and how to generalize the Poincar\'e algebra.  The most obvious concern stems from the fact that SCET is formulated on the light cone.  Specifically, the collinear sector of the theory are the modes that are ``near" a light-like direction, while the fields that are ``far" from this light cone are integrated out.  This separation is controlled by a power counting parameter as described in detail below.  One of the consequences is that components of any object that carries a spacetime index will also be split apart by this power counting.  The relevance of this particular frame can be physically interpreted as characterizing the direction of the hard particle of interest.  Hence, the theory is defined on a Lorentz breaking background -- the residual space-time symmetry is encoded by requiring that the theory satisfy the so-called reparameterization invariance (RPI)~\cite{Luke:1992cs, Manohar:2002fd} (see~\cite{Heinonen:2012km} for an alternative approach).  The argument for the RPI of SYM is identical to the standard QCD case, and we will briefly review it in what follows.  However, it is a priori unclear that there should be any analogous residual SUSY invariance once the collinear frame has been chosen.  One of the new results presented below is a derivation of the RPI transformations of the supercharges that remain after taking the collinear limit of the theory.

SUSY is a transformation on fermionic coordinates.  Therefore, taking the collinear limit of superspace can be thought of in close analogy to taking the collinear limit of a (two-component) fermionic field.  Hence, we will provide the first example of two-component SCET both as a useful formalism on its own but also to pave the way to discovering the procedure for ``integrating out half of superspace."  This approach can be stated algorithmically, and we will apply it explicitly to both SYM and the Wess-Zumino models.  The resulting theories are expressed in terms of only on-shell degrees-of-freedom.  To demonstrate its equivalence to a component SCET Lagrangian, we will write the EFT in light-cone gauge, which results in removing the non-propagating modes of the gauge boson explicitly at the Lagrangian level.  There exists a large literature on light-cone SUSY~\cite{Mandelstam:1982cb}, including a superspace formalism which will be of use to us here, see \emph{e.g.}~\cite{Leibbrandt:1987qv} for a review. This on-shell formalism is entirely equivalent to the theory as expressed in more conventional Lorentz covariant gauges, a fact that will be critical to the comparison between the various ways of expressing the model.  The Lagrangian will be identical in component form to the one that lives in collinear superspace, thereby demonstrating self-consistency of the theory since it is both RPI and SUSY invariant.  

\begin{figure}[h!]
\centering
\includegraphics[width=.9 \textwidth]{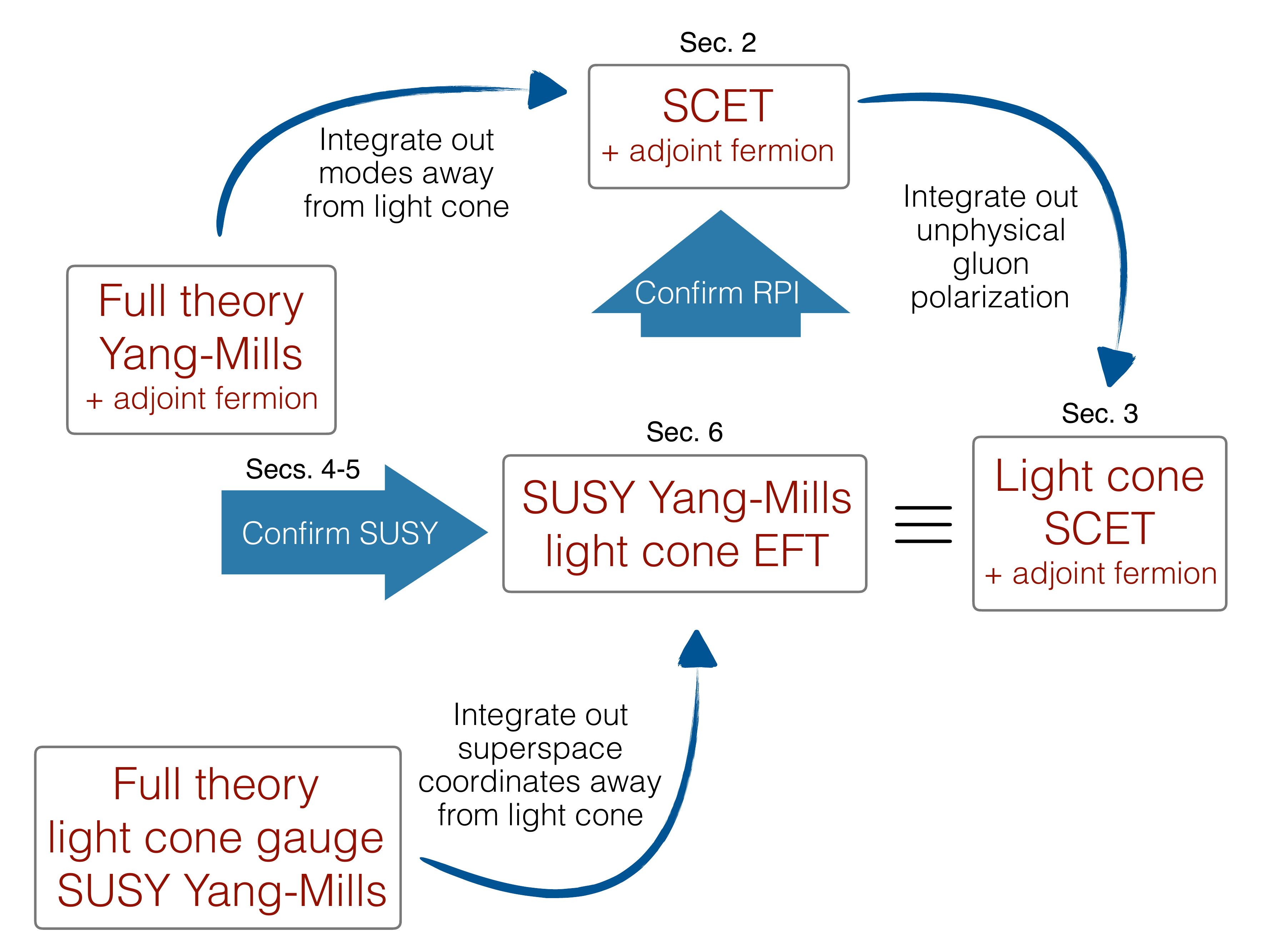}
\caption{\footnotesize{This figure provides a schematic for the main results of this paper.  Our starting point is the full Yang-Mills theory with an adjoint Weyl fermion, which we will later identify as a gaugino. In Sec.~\ref{sec:twocomp}, we apply the SCET procedure (reformulated in the language of two-component spinors) to integrate modes away from the light cone.  This yields the familiar QCD SCET Lagrangian coupled to an adjoint Weyl fermion. Section~\ref{sec:LCSCET} details the gauge fixing procedure and the corresponding light-cone SCET Lagrangian. Section~\ref{sec:susy} explores the systematics of SUSY in the collinear limit, while Sec.~\ref{sec:CollSuperspace} is devoted to the superspace formulation of the EFT.  The derivation of the SYM Lagrangian from collinear superspace is given in Sec.~\ref{sec:ymsect}.  Finally, the Wess-Zumino model is analyzed in Sec.~\ref{sec:WZModel}.}}
\label{fig:SketchOfProof}
\end{figure}

There are many issues that we do not address, some of which will be discussed in the Outlook section at the end of this manuscript.  The outline of the paper is illustrated in Fig.~\ref{fig:SketchOfProof}.  A few technical appendices are also provided.

\section{SCET for Two-Component Spinors}
\label{sec:twocomp}
Since SUSY is most naturally formulated using Weyl fermions, it is prudent to begin our exploration of SUSY SCET by deriving the effective Lagrangian for the collinear fields using two-component fermions.  Additionally, this will provide an opportunity to review some of the basics of SCET while also making our conventions and notation explicit. We will also discuss the effective theory in the presence of gauge fields, starting from the standard QCD Lagrangian with two-component fields. Finally, we will briefly review scalar SCET, since it will be relevant for our exploration of the Wess-Zumino model.

We will work in Minkowski space with signature $g^{\mu \nu} = \textrm{diag}\left(+1,-1,-1,-1\right)$.  The collinear direction is taken along the $+\hat z$ light-cone direction: $n^\mu = \left(1,0,0,1\right)$.  Then the anti-collinear direction is defined by $n^2 = 0 = \bar{n}^2$ and $n \cdot \bar{n} = 2$.  It is usually convenient to make the explicit choice $\bar{n}^\mu =\left(1,0,0,-1\right)$ although, as discussed below, RPI transformations allow shifts away from this canonical choice.  Lorentz four vectors are then expanded as
\begin{align}
\label{eq:lightconemomentum}
p^\mu =  \frac{n \cdot p}{2}  \, \bar{n}^\mu + \frac{\bar{n}\cdot p}{2}\,n^\mu + p_{\perp}^{\mu}, \quad\text{or}\quad p^\mu = \big(n\cdot p, \bar{n} \cdot p , \vec{p}_\perp\big).
\end{align}
We use the notation where $p^2 = (n\cdot p) (\bar{n}\cdot p) + p_\perp^2$, where $p_\perp^2 \equiv p_\perp \cdot p_\perp = -p_1^2 - p_2^2$.  Here, and in everything that follows, the ``$\cdot$" is a 4-vector dot-product.

The collinear limit is defined by the momentum shells which scale like $p_{n}^{\mu} \sim \Lambda (\lambda^2,1, \lambda)$, where $\Lambda$ is some dimensionful scale, and $\lambda \ll 1$ is the SCET power counting parameter.\footnote{\footnotesize{See for instance Ref.~\cite{Bauer:2000ew} for a physical example of using SCET to resum large infrared logarithms in the inclusive rate for $B\rightarrow X_s \gamma$. In this case a large separation of scales is due to $\lambda = \Lambda_\text{QCD}/m_b \ll 1$.}}  We can therefore interpret $p_n^2 \sim \lambda^2$ as the virtuality (or allowed distance from the light cone) for the collinear modes in the EFT.  Similarly, an anti-collinear momenta scales as $p_{\bar{n}}^{\mu} \sim \Lambda (1, \lambda^2, \lambda)$.  Depending on the process of interest there are also soft modes $p_{\text{s}}^{\mu} \sim \Lambda (\lambda, \lambda, \lambda)$ or ultra-soft modes $p_{\text{us}}^{\mu} \sim \Lambda (\lambda^2, \lambda^2, \lambda^2)$.  When relevant, we will use the ultrasoft scalings and will not distinguish between these further.  When discussing power counting in what follows, we will follow standard practice and work with units where $\Lambda = 1$.


\subsection{Universal Soft and Collinear Limits}
In order to motivate the general formulation of SCET for theories such as $\mathcal{N}=1$ SYM, we first review the original motivation for the development of SCET in gauge theories: universal infrared (IR) divergences.  To identify the soft and collinear limits of gauge theory amplitudes, we can employ the power counting of momenta outlined above.

Consider the simple example of photon emission from a left-handed Weyl fermion current.  The tree-level amplitude for emission is
\begin{align}\label{eq:photon_emission}
\!\!\!\!\!\!\!\!\raisebox{-0.25\height}{\includegraphics[width=3.5cm]{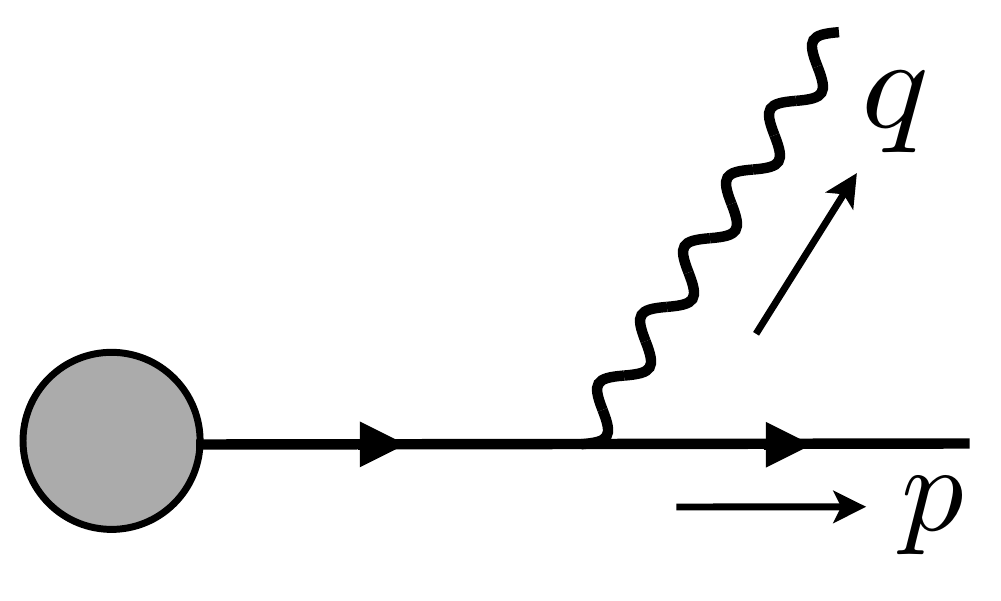}} &= x^{\dagger}_{\dot{\alpha}}(p) \big(-i\,e\, \bar{\sigma}^\mu\big)^{\dot{\alpha}\alpha} \epsilon^{*}_{\mu}(q) \frac{-i \,(p+q) \cdot \sigma_{ \alpha \dot{\beta}}}{2\, p \cdot q} \,\, \mathcal{M}^{\dot{\beta}}(p+q) \\ \nonumber
&=-e \left[ \left(
\frac{p\cdot \epsilon^*(q)}{p\cdot q} \right) \delta_{\dot \beta}^{\dot \alpha}\,x_{\dot{\alpha}}^\dagger (p)- x_{\dot{\alpha}}^\dagger (p) \frac{(q\cdot \bar \sigma)^{\dot \alpha \alpha}(\epsilon^*(q)\cdot \sigma)_{\alpha\dot\beta}}{2\,p\cdot q}
\right]\mathcal{M}^{\dot{\beta}}(p+q) ,
\end{align}
where we have used $\sigma$-matrix identities, the QED Ward identity, and the Weyl equation of motion $x^\dagger_{\dot{\alpha}} (p) (\bar{\sigma} \cdot p)^{\dot{\alpha}\alpha} = 0$.  Throughout, we follow the notation of \cite{Dreiner:2008tw} -- the momentum space wave function of a left handed Weyl spinor is denoted by $x_\alpha(p)$.  The spinor (anti-spinor) indices $\alpha\, \big(\dot{\alpha}\big)$ range from 1 to 2, and we work in the Weyl basis, see App.~\ref{app:notation} for details.   Armed with \eref{eq:photon_emission}, we will now show that only the first term is divergent in the soft limit, while both terms are divergent in the collinear limit.

To study the \emph{soft} limit, we rescale the momentum $q$ by $\lambda^2$: $q^\mu\to \lambda^2 q^\mu$ and insert it into \eref{eq:photon_emission}.  We then have
\begin{align}
\text{Eq.~\eqref{eq:photon_emission}}\to -e \left[ \left(
\frac{n\cdot \epsilon^*(q)}{\lambda^2 (n\cdot q)} \right) \delta_{\dot \beta}^{\dot \alpha}\,x_{\dot{\alpha}}^\dagger (p)- x_{\dot{\alpha}}^\dagger (p) \frac{(q\cdot \bar \sigma)^{\dot \alpha \alpha}(\epsilon^*(q)\cdot \sigma)_{\alpha\dot\beta}}{(\bar n\cdot p)(n\cdot q)}
\right]\mathcal{M}^{\dot{\beta}}(p+\lambda^2q) \,,
\end{align}
where without loss of generality, we have assumed that $p$ lies along the $n$ direction: $p^\mu =n^\mu (\bar n\cdot p) /2$.   In the limit that $\lambda\to 0$, the first term dominates, and so the leading soft limit of the amplitude is
\begin{equation}
\raisebox{-0.25\height}{\includegraphics[width=3.5cm]{Figures/M1QED.pdf}}\quad\xrightarrow[\text{soft}]{\makebox[0.8cm]{}} \quad-e  \,
\frac{n\cdot \epsilon^*(q)}{n\cdot q} \delta_{\dot \beta}^{\dot \alpha}\,x_{\dot{\alpha}}^\dagger (p)\mathcal{M}^{\dot{\beta}}(p)\,.
\end{equation}
Note that the polarization vector does not scale with $\lambda$ since the soft limit is isotropic.  By summing over all external legs that could emit the soft photon, 
we arrive at the following factor:
\begin{equation}
{\cal S}(q) = \sum_i q_i \frac{n_i\cdot \epsilon^*(q)}{n_i\cdot q}
\end{equation}
where the sum runs over all external legs $i$ with charge $q_i$.  This universal object is the so-called soft current and is gauge invariant by conservation of charge.

A similar power counting can be done for \emph{collinear} momentum $q$, and we will see that both terms in the amplitude contribute at $\mathcal{O}(1/\lambda)$.  Unlike for soft gluons, the collinear gluon polarization vector is constrained.  The Ward identity $q\cdot \epsilon^*(q)=0$ must be maintained as an expansion in $\lambda$ so that when $q$ is collinear, $n\cdot \epsilon^*(q)$ must be suppressed by a power of $\lambda$.  Additionally, the collinear completeness relation\footnote{Only the perpendicular components of the metric exist at leading power in $\lambda$ because the physical polarization is perpendicular to $q$.  Because $q$ is dominantly along the $n$ direction, the polarization vector is dominantly in the $\perp$ plane.}
\begin{equation}
\sum_{\text{spins}}\epsilon^\mu(q)\epsilon^{*\nu}(q)=-g_\perp^{\mu\nu}+{\cal O}(\lambda)
\end{equation}
implies that $\epsilon^{\nu_\perp}(q) \sim 1$.  It follows that $n\cdot \epsilon^*(q)\sim \lambda$.  We must also use the equation of motion for $x_{\dot{\alpha}}^\dagger (p)$ which, in the exactly collinear limit, reduces to
\begin{equation}
x_{\dot{\alpha}}^\dagger (p) \big(n\cdot \bar \sigma\big)^{\dot{\alpha}\alpha} = 0\,.
\end{equation}
Then, the leading collinear limit of the amplitude is
\begin{align}
&\raisebox{-0.25\height}{\includegraphics[width=3.5cm]{Figures/M1QED.pdf}} 
\quad\xrightarrow[\text{collinear}]{\makebox[0.8cm]{}} \notag\\ 
&\qquad\qquad\quad-e \left[ 
\frac{n\cdot \epsilon^*(q)}{ n\cdot q} \delta_{\dot \beta}^{\dot \alpha}\,x_{\dot{\alpha}}^\dagger (p)- x_{\dot{\alpha}}^\dagger (p) \frac{\big(q_\perp \cdot \bar \sigma_\perp\big)^{\dot \alpha \alpha}\big(\epsilon^*_\perp(q)\cdot \sigma_\perp\big)_{\alpha\dot\beta}}{(\bar n\cdot p)(n\cdot q)}
\right]\mathcal{M}^{\dot{\beta}}(p+q)\,.
\label{eq:FullThyCollinearFactor}
\end{align}
Using the collinear power counting, it is clear that terms in this expression scale like $1/\lambda$.  The term in brackets is gauge invariant in the exactly collinear limit, which will be important for comparisons made below when we express SCET in the light-cone gauge.

The existence of universal IR divergences, both soft and collinear, therefore motivates the formulation of a SCET Lagrangian.\footnote{\footnotesize{Note that there are other IR divergences in QED, for instance from fermion pair production off a photon current. These are well known and will not be discussed further here. Additionally, see~\cite{Ellis:1991qj,Peskin:1995ev,Collins:2011zzd,Schwartz:2013pla} for reviews of the IR divergences of QCD. }} An important consistency check when formulating a new SCET is that the singularity structure of the EFT must reproduce the expected IR divergences found by factorizing and power counting the full theory amplitudes, as will be further discussed in Secs.~\ref{sec:IRinLCG} and \ref{sec:IRstructureYukawa}.

\subsection{The SCET Lagrangian}
Starting with a full theory left-handed Weyl fermion $u$, our goal will be to derive projection operators that allow us to separate the collinear degrees of freedom $u_n$ from the anti-collinear degrees of freedom $u_{\bar{n}}$.\footnote{We follow the conventions of the literature and refer to this decomposition as dividing the fermion into ``collinear" and ``anti-collinear" degrees of freedom.  Note that both components are defined with respect to the same collinear direction $n$, and so there is an analogous anti-collinear fermion Lagrangian that is defined for momenta pointing along the $\bar{n}$ direction.}  Note our focus is on the left handed Weyl spinor since these two fermionic degrees of freedom arise in both the chiral and vector multiplets of $\mathcal{N}=1$ SUSY.  

The position-space Weyl equation for a left-handed two component field $u$ is
\begin{align}
\mathcal{L} = i\,u^\dag(x) \,\bar{\sigma}\cdot \partial \,u(x),
\label{eq:LFreeFermion}
\end{align}
which admits the standard plane-wave solution $u(x^\mu) =  \int \textrm{d}^3 \,p\,\, x(p)\,\text{exp}(-i\,p^\mu x_\mu)$.  In \eref{eq:LFreeFermion} and many expressions that follow, the contraction of two component spinor indices is implied.

Boosting $u(x)$ along the light-cone in the $\hat z$ direction, \emph{i.e.} $n^\mu$ direction, yields:\footnote{\footnotesize For example, see Sec.~3.3 of \cite{Peskin:1995ev}.}   
\begin{align}
x(p)\big|_n &=\left[\sqrt{E+p_z}\left(\frac{1-\sigma_z}{2}\right) + \sqrt{E-p_z}\left(\frac{1+\sigma_z}{2}\right)\right] \bar{\sigma}_0 \,\xi \notag\\
&= \Big[\sqrt{\bar{n}\cdot p}\,P_n + \sqrt{n\cdot p}\,P_{\bar{n}}\Big]\xi \sim \left[ \begin{array}{c}
\lambda^1  \\
\lambda^0 \end{array} \right] \,,
\label{eq:uCollinearLimit}
\end{align}
where $\xi$ is a two-component spinor, we have inserted the appropriate factor of $\bar{\sigma}_0$ to make the spinor index structure consistent, and for the last step we have used the collinear scalings for the momentum, $\bar{n} \cdot p \sim 1$ and $n \cdot p \sim \lambda^2$.  We see that the upper component of $u(p)|_n$ is suppressed, and the collinear fermion is given by the lower component of $u(p)|_n$. 

For concreteness, one can construct the following combinations of Pauli matrices:
\begin{align}
\label{eq:SigmaNotation}
&\left(\frac{\ns}{2}\right)_{\alpha \dot{\alpha}} =  \frac{1}{2}\big(\sigma^0 - \sigma^3\big)_{\alpha \dot{\alpha}}  = 
\left[ \begin{array}{cc}
0 & 0  \\
0 & 1  \end{array} \right]_{\alpha \dot{\alpha}} \,, &\left(\frac{\nbsb}{2} \right)^{\dot{\alpha}\alpha} =    \frac{1}{2}\big(\bs_0 - \bs_3\big)^{\dot{\alpha}\alpha}  = 
\left[ \begin{array}{cc}
0 & 0  \\
0 & 1  \end{array} \right]^{ \dot{\alpha}\alpha} \,,\notag\\[7 pt]
&\left(\frac{\nbs}{2}\right)_{\alpha \dot{\alpha}}  = \frac{1}{2}\big(\sigma^0 + \sigma^3\big)_{\alpha \dot{\alpha}} = 
\left[\begin{array}{cc}
1 & 0  \\
0 & 0  \end{array} \right]_{\alpha \dot{\alpha}}  \,, &\left(\frac{\nsb}{2}\right)^{\dot{\alpha}\alpha}  =  \frac{1}{2}\big(\bs_0 + \bs_3\big)^{\dot{\alpha}\alpha} = 
\left[\begin{array}{cc}
1 & 0  \\
0 & 0  \end{array} \right]^{ \dot{\alpha} \alpha} \,.
\end{align}
These can then be used to infer the identification of the projections operators $P_n$ and $P_{\bar{n}}$, such that
\begin{align}
u = \left(P_n + P_{\bar{n}}\right) u = u_{n} + u_{\bar{n}}.
\end{align}
Comparing to \eref{eq:uCollinearLimit} yields the explicit forms
\begin{align}
\begin{array}{ll}
P_{n} \, u_n = \frac{n\cdot \sigma}{2}\frac{\bar n\cdot \bar \sigma}{2}\,u_n = u_n\,; \quad\quad\quad\quad &P_{\bar{n}}\,u_{\bar{n}} =  \frac{\nbs}{2}\frac{\nsb}{2}\,u_{\bar{n}} = u_{\bar{n}} \,; \\
P_{\bar{n}} \, u_n = 0\,; &P_n \,u_{\bar{n}}  = 0\,.
\end{array}
\label{eq:uProjections}
\end{align}
Thus the projection operators act to project half the helicity states out, namely $u_{n,1} = 0$ and $u_{\bar{n},2} = 0$; the two component collinear/anti-collinear projection operators are equivalent to the chiral projection operators (similar expressions for a right-handed Weyl fermion $v = v_n + v_{\bar{n}}$ are given in App.~\ref{app:notation}).  

We can learn a few interesting things by analyzing the free fermion Lagrangian.  Starting with \eref{eq:LFreeFermion}, expanding out $u = u_n + u_{\bar{n}}$, and expressing it in components yields
\bea
\label{eq:fullThyLfermionComponent}
\!\!\!\!\!\!\mathcal{L} = i \,u_{n,\dot{2}}^\dagger\, n \cdot \partial\, u_{n,2} + i\, u_{\bar{n},\dot{1}}^\dagger\, \left( \bar{\sigma} \cdot \partial_\perp \right)^{\dot{1}2} \, u_{n,2} + i\, u_{n,\dot{2}}^\dagger\, \left(\bar{\sigma} \cdot \partial_\perp \right)^{\dot{2}1} \, u_{\bar{n},1} +  i\, u_{\bar{n}.\dot{1}}^\dagger\, \bar{n}\cdot \partial \, u_{\bar{n},1},
\eea
where we have used for instance $(\bar{\sigma} \cdot \partial)^{\dot{2}2} = n \cdot \partial$, as can be seen from \eref{eq:SigmaNotation}.  

The power counting of the fermion components can be derived by requiring that each term in \eref{eq:fullThyLfermionComponent} must scale as $\mathcal{L}_n^{(0)} \sim \mathcal{O}\left(\lambda^4\right)$; the collinear volume element scales as $\text{d}^4 x_n \sim \lambda^{-4}$ and the action must be unsuppressed.  This fixes\footnote{\footnotesize{The $\lambda$-scaling of the collinear fields is equivalent to the twist $\tau = d_m - s$, where $d_m$ is the mass dimension. For the fermion $\tau_{u_n} = 3/2 - 1/2 = 1$. Similarly the collinear scalar field scales as $\phi_n \sim \mathcal{O}(\lambda)$, since its mass dimension is $\big[ \phi \big] = 1$ the twist is $\tau_{\phi} = 1 - 0 = 1$. }} $u_n \sim \mathcal{O}(\lambda)$ and $u_{\bar{n}} \sim \mathcal{O}(\lambda^2)$. The scalings of the fields are summarized in Table~\ref{table:scaling}. 

Additionally, if we follow standard practice and identify $n\cdot \partial$ with the light cone time derivative, \eref{eq:fullThyLfermionComponent} implies that $u_n$ is a propagating degree of freedom, while $u_{\bar{n}}$ is not.  Since it is non-propagating, we can integrate out these anti-collinear modes by solving for the classical equation of motion:
\bea
u_{\bar{n}} =   -\frac{\nbs}{2}  \frac{1}{\bar{n} \cdot \partial}\, \big(\bar{\sigma} \cdot \partial_\perp\big) \,u_n .
\label{eq:anticollEOM}
\eea
Then the collinear and anti-collinear fermion modes can both be expressed in terms of the propagating mode 
\begin{align}
& u_n =  \left[ \begin{array}{c}
0 \\
u_2 \end{array} \right] \,,
& u_{\bar{n}} =  \left[ \begin{array}{c}
\frac{(-\partial_{\perp,1} + i\, \partial_{\perp,2})}{\nbp}u_2 \\
0 \end{array} \right] \,,
\label{eq:matrixfermions}
\end{align}
where this is for the specific choice of $\bar{n}^\nu$. Note that in \eref{eq:matrixfermions} and in all that follows we will often suppress the subscript ``$n$" and take $u_2  \equiv  u_{n,2}$, and similarly $u_{n,\dot{2}}^\dag \equiv u_2^\dag$. 

Plugging \eref{eq:anticollEOM} into the \eref{eq:fullThyLfermionComponent} yields the leading power collinear Lagrangian for a free Weyl fermion:
\begin{align}
{\cal L}_{u_n} &=  u_n^\dagger \left(i\,n\cdot \partial - \frac{\partial_\perp^2}{i\,\bar n \cdot \partial}  \right) \frac{\bar n\cdot \bar \sigma}{2}\,u_n \,,
\label{eq:Lfree}
\end{align}
where we have used $ \bar{\sigma}\cdot \partial_\perp \sigma \cdot \partial_\perp = - \partial_1^2 - \partial_2^2 \equiv  \partial_\perp^2$. 
The non-local operator $1/\nbp$ is defined in terms of its momentum space representation:
\bea 
\label{eq:fouriertrans}
\frac{1}{\nbp} \,\psi(x) = \frac{1}{\nbp} \int \text{d}^4 p\, e^{-i\,p \cdot x} \,\tilde{\psi}(p) = \int \text{d}^4 p\, e^{-i\,p \cdot x} \frac{1}{\bar{n} \cdot p}\,\tilde{\psi}(p).
\eea
Then the propagator for a collinear fermion can be extracted by inverting the free momentum space Lagrangian: 
\begin{equation}
\raisebox{-0.5\height}{\includegraphics[width=1.5cm]{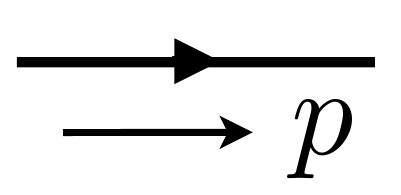}} =\,i\,\frac{n\cdot \sigma}{2}\frac{\bar n\cdot p}{(n\cdot p)(\bar n \cdot p) + p_\perp^2}\,.
\end{equation}

Next we can turn on interactions for the collinear fermions.  Of particular interest are the interactions with collinear and soft gauge bosons.  Start by expanding the full theory gauge boson fields as
\begin{align}
A^\mu = A_n^\mu  + A_{s}^\mu\,.
\end{align}
Using $A_{n}^{\mu} \sim \Lambda (\lambda^2, 1, \lambda)$ and $A_{\text{s}}^{\mu} \sim \Lambda (\lambda^2, \lambda^2, \lambda^2)$, it is straightforward to derive the leading contributions for the gauge covariant derivative $\mathcal{D}_n^\mu = (\bar{n}^\mu/2)\, n \cdot \mathcal{D}_{n,s} + (n^\mu/2)\, \bar{n} \cdot \mathcal{D}_{n} +\mathcal{D}_{\perp, n}^\mu$ acting on collinear fields:\footnote{\footnotesize{Here $\mathcal{D}$ means the usual covariant derivative. This should not be confused with the common notational practice in the SCET literature of using $\mathcal{D}$ to describe a covariant derivative after it has absorbed a collinear Wilson line.}}
\begin{align}
i \,n \cdot \mathcal{D}_{n,s} &= i \,\np + g \,n\cdot A_n + g\,n\cdot A_s  \sim \mathcal{O}\big(\lambda^2\big) \, , \nonumber \\ 
i \bar{n} \cdot \mathcal{D}_{n} &= i\, \nbp + g\, \bar{n}\cdot A_n  \sim \mathcal{O}\big(\lambda^0\big) \, , \\
i \mathcal{D}_{\perp, n}^\mu &= i\, \partial_\perp^\mu + g\, A_{n \perp}^\mu \sim \mathcal{O}(\lambda) \, , \nonumber
\end{align}
where we can treat the soft fields as background fields since they are slowly varying with respect to the collinear fields.

\begin{table}[t!]
\renewcommand{\arraystretch}{1.5}
\setlength{\arrayrulewidth}{.3mm}
\centering
\setlength{\tabcolsep}{1.1em}
\begin{tabular}{| c | c  c  c  c  c c c c c |}
    \hline
    Field &  $u_n$  &  $u_{\bar{n}}$ &  $u_s$ & $\phi_n$ & $\phi_{s}$ & $n\cdot A_n$ & $\bar{n}\cdot A_n$ & $A_{n\perp}$ & $A_s$ \\ \hline
    Scaling & $\lambda$ & $\lambda^2$ & $\lambda^3$ & $\lambda$  & $\lambda^2$ & $\lambda^2$ &$\lambda^0$ & $\lambda$ & $\lambda^2$\\ \hline
\end{tabular}
\caption{\footnotesize{The scalings of the field are chosen in such a way as to ensure the propagators are $~\mathcal{O}(1)$ \cite{Bauer:2000ew}. Alternatively the action $S = \int \text{d}^4 x_n \left( {\mathcal L}_{\phi} + \mathcal{L}_{u} + \mathcal{L}_{g}  \right)_n +   \int \text{d}^4 x_s  \left( {\mathcal L}_{\phi} + \mathcal{L}_{u} + \mathcal{L}_{g}  \right)_s $, must scale as $\mathcal{O}(1)$ which fixes the scalings of the soft and collinear fields (all $\lambda$ dependence is moved to the interactions). For instance, for soft fields $x_s \sim 1/\lambda^2$ since $p_s \cdot x_s  \sim 1$ and every component of $p_s$ scales like $ \sim \mathcal{O}(\lambda^2)$. So the soft volume element $\text{d}^4 x_s$ scales as $\ \mathcal{O}(\lambda^{-8})$ and $\mathcal{L}^{(0)}_{\textrm{s}} \sim \mathcal{O}(\lambda^8)$. } } 
\label{table:scaling}
\end{table}

The collinear field strength can be written as $g\, F_n^{\mu \nu} = i \bigl[ \mathcal{D}^\mu_n, \mathcal{D}^\nu_n \bigr]$.  This includes the full field strength for the collinear gauge boson, and also has interactions of the type $A_n - A_n - A_s$, which do not transfer enough momentum as to spoil the virtuality of the collinear gauge boson which would otherwise take it out of the regime of validity for SCET.  To leading power, each sector has its own independent gauge invariance.  For a review of the details, including a discussion of gauge fixing, the gauge boson propagator and Feynman rules, see Sec.~4.3 of~\cite{StewartNotes}. Explicitly, the gauge field strength in the collinear sector is expanded as
\bea
\nonumber  -i\, g\, F^{a \mu \nu}_n t^a = \bigl[  \mathcal{D}^\mu_n,  \mathcal{D}^\nu_n \bigr]  &=& \left( \frac{n^\mu}{2} \bigl[\nbp, \mathcal{D}_{\perp,n}^\nu \bigr] - \frac{n_\nu}{2}  \bigl[\nbp,  \mathcal{D}_{\perp,n}^\mu\bigr] \right)  \\ \nonumber
  &+& \left(  \frac{1}{4} \left( n^\mu \bar{n}^\nu - n^\nu \bar{n}^\mu \right) \bigl[\nbp,  n \cdot  \mathcal{D}_n \bigr] + \bigl[ \mathcal{D}_{\perp, n}^\mu,  \mathcal{D}_{\perp,n}^\nu \bigr] \right)  \\ 
  &+& \left( \frac{\bar{n}^\mu}{2}\bigl[n \cdot  \mathcal{D}_n,  \mathcal{D}_{\perp, n}^\nu\bigr] - \frac{\bar{n}^\nu}{2}\bigl[n \cdot  \mathcal{D}_n,  \mathcal{D}_{\perp,n}^\mu \bigr] \right) \,,
\label{eq:CollinearFieldStrength}
 \eea
 where the terms on the first, second and third line scale as  $\mathcal{O}(\lambda)$, $\mathcal{O}(\lambda^2)$ and $\mathcal{O}(\lambda^3)$ respectively. Note that upon contraction of Lorentz indices for $(F^{\mu\nu})^2$, the leading order Lagrangian density scales as $\mathcal{O}(\lambda^4)$.

Finally, one can dress the free fermion Lagrangian given above in \eref{eq:Lfree} with covariant derivatives (now being careful not to commute derivatives past gauge bosons).  This gives the two-component Lagrangian for QCD SCET:
\bea
\label{eq:qcdscetlagrangian}
\!\!\!\!\!\!\!\!\mathcal{L} = u_n^{\dagger}\left( i\, n \cdot \mathcal{D}_{n,s} + i \,\bar{\sigma} \cdot \mathcal{D}_{\perp,n} \frac{1}{i \,\bar{n}\cdot \mathcal{D}_n} i\, \sigma \cdot \mathcal{D}_{\perp,n} \right) \frac{\bar{n}\cdot \bar{\sigma}}{2}\,  u_n- \frac{1}{4}\big(F_n^{\mu \nu} \big)^2  \equiv \mathcal{L}_u + \mathcal{L}_g\,.
\label{eq:LSCET}
\eea
This expression will be used below to describe the SUSY vector multiplet, with $u_n$ identified as a collinear gaugino.  For completeness, App.~\ref{sec:CovariantGaugeSingularities} shows that this Lagrangian reproduces the correct collinear factor computed in \eref{eq:FullThyCollinearFactor}.  This result can also be obtained by starting with the four-component QCD SCET Lagrangian and projecting out two-components of a four-component fermion, $\psi_D \rightarrow \psi_{D,L} = P_L \psi_D$.

Interactions between soft and collinear fields can be removed from the Lagrangian at leading power by making the BPS field redefinition \cite{Bauer:2001yt}.  For any collinear field $X_n$, we can multiply $X_n$ by a Wilson line for soft gluons to define a new field $\widetilde{X}_n$ as
\begin{equation}
\widetilde{X}_n = Y_n \,X_n\,.
\end{equation}
The soft Wilson line $Y_n$ is defined by 
\begin{align}
 &Y_n^\dagger \,Y_n=1\,, &\mathcal{D}_s \,Y_n = 0\,,
\end{align}
and $\widetilde{X}_n$ has no couplings to soft gluons present in the Lagrangian.  Therefore, we will typically ignore the soft/ultrasoft gauge boson fields in what follows.

Finally for scalars we can write (for a review of scalar SCET see \cite{Becher:2014oda}) 
\bea
\phi = \phi_n + \phi_{\bar{n}} + \phi_s\,.
\eea
The free scalar Lagrangian is trivial in the sense that we do not have to integrate out any high virtuality modes, and to leading order the collinear, anti-collinear, and soft sectors do not mix. Thus for each sector we simply expand the derivative operators in the free Lagrangian $\partial^\mu \phi \partial_\mu \phi$ in powers of $\lambda$, yielding (for simplicity we omit possible gauge interactions):\footnote{Note that if one performs a field redefinition, $\phi_n \rightarrow \phi_n/\sqrt{\bar{n}\cdot \partial}$, see \emph{e.g.}~\cite{Chiu:2007dg}, then this kinetic term exactly mirrors that of the fermions given in \eref{eq:Lfree}, which makes the SUSY invariance of the free theory manifest.}
\bea
\mathcal{L}_{n} = - \phi_n^* \Box \phi_n = - \phi_n^* \big( \nbp\, \np + \partial^2_\perp \big)\phi_n\,.
\label{eq:scalarLfree}
\eea
Note that $\Box \sim \mathcal{O}(\lambda^2)$ so the collinear scalar (as well as the soft and anti-collinear scalars) scales as $\mathcal{O}(\lambda)$. This ensures that \eref{eq:scalarLfree} scales as $\mathcal{O}(\lambda^4)$ and similar arguments can be made for the anti-collinear and soft scalar.  See Table~\ref{table:scaling} for a summary. 

\subsection{RPI in Two-Components}
\label{sec:RPItrans}
As previously discussed, the light cone expansion breaks Lorentz invariance, and therefore \eref{eq:LSCET} does not have manifest Lorentz symmetry. However, \eref{eq:LSCET} is still invariant under the subset of unbroken Lorentz generators. Additionally, the Lagrangian must be invariant under a residual symmetry of the broken Lorentz generators, reparameterization invariance (RPI).  Note that if all orders in $\lambda$ were included, SCET must be equivalent to the full theory where full Lorentz symmetry would be restored. Order-by-order, the theory must track the broken Lorentz generators in terms of RPI.  Therefore, RPI can be thought of as a consistency condition that should be verified when constructing new SCETs from the top down.  Alternatively working from the bottom up, one should only write down operators that are consistent with RPI.  

Practically, RPI can be characterized by noting that any choice of $n^\mu$ and $\bar{n}^\mu$ which satisfy the conditions $n \cdot \bar{n} = 2$ and $n^2 = 0=\bar{n}^2$ must yield the same EFT, namely \eref{eq:LSCET}. These conditions are invariant under three different kinds of reparameterization:\\
\vspace{-15pt}
\begin{center}
\renewcommand{\arraystretch}{1.2}
\setlength{\arrayrulewidth}{.3mm}
\setlength{\tabcolsep}{1.2em}
\begin{tabular}{c|c|c}
RPI-I & RPI-II & RPI-III \\
\hline
$n_\mu \rightarrow n_\mu + \Delta_\mu^{\perp}$ & $n_\mu \rightarrow n_\mu$ & $n_\mu \rightarrow e^\alpha n_\mu$ \\
$\bar{n}_\mu \rightarrow \bar{n}_\mu$ & $\bar{n}_\mu \rightarrow \bar{n}_\mu + \epsilon_\mu^\perp$ & $\bar{n}_\mu \rightarrow e^{-\alpha} \bar{n}_\mu$
\end{tabular} 
\end{center}
where $\bar{n}\cdot \epsilon^\perp = n \cdot \epsilon^\perp = \bar{n} \cdot \Delta^\perp = n \cdot \Delta^\perp = 0$.  

In this section, we address RPI by expanding the full Lorentz invariance of the theory along the light cone defined by $n$ and $\bar{n}$.  While these arguments have previously appeared in the literature for four-component spinors~\cite{Manohar:2002fd}, this is the first derivation that specializes to two-component notation.  We will provide many details so that this section is self contained.  

\subsubsection*{The RPI Generators and Algebra}
The Poincar\'e group, relevant to the full Lorentz invariant theory, is defined by
\bea
\label{eq:poincare1}
&&\Big[P_\mu, P_\nu \Big] = 0\,;\\ 
\label{eq:poincare2}
&&\Big[M^{\mu\nu},P^\rho \Big] = i\,g^{\mu\rho}\,P^\nu-i\,g^{\nu\rho}\,P^\mu \,; \\
\label{eq:poincare3}
&&\Big[M^{\mu \nu}, M^{\kappa \rho}\Big] = - g^{\mu \kappa} M^{\nu \rho} - g^{\nu \rho} M^{\mu \kappa} + g^{\mu \rho} M^{\nu \kappa} + g^{\nu \kappa}M^{\mu \rho}\, ,
\eea
Rotations in the perpendicular plane, generated by $J_3$, are unbroken.  The vectors $n^\mu$ and $\bar{n}^\mu$ break five of the Lorentz generators, corresponding to the following:
\begin{align}
\label{eq:RPI1}
R_1^\nu &= \bar{n}_\mu\, M^{\mu\nu_\perp}\,, \\
\label{eq:RPI2}
R_2^\nu &= n_\mu\, M^{\mu\nu_\perp}\,,  \\
\label{eq:RPI3}
R_3 &= n_\mu \,\bar{n}_\nu \,M^{\mu\nu}\,.
\end{align}
These are the RPI-I, RPI-II, and RPI-III transformations, respectively.  The symbol $\nu_\perp$ denotes that this index only takes values for directions orthogonal to $n$ and $\bar n$.   Note that the Lorentz algebra given in \eref{eq:poincare3} can be used to verify that the combined algebra of RPI and the unbroken Lorentz subgroup closes~\cite{Manohar:2002fd}.

\newpage

Now that we have the algebra, it is possible to infer the action of the RPI transformations along with their power counting properties.  Starting with \eref{eq:poincare2} the algebra of RPI with the various components of $P^\mu$ is:
\begin{alignat}{3}
&\big[
R_1^\nu,n\cdot P
\big] = 2\,i\,P_\perp^\nu  \,,\quad
&&\big[
R_2^\nu,n\cdot P
\big] = 0  \,,\quad
&&\big[
R_3,n\cdot P
\big] = -2\,i\,n\cdot P  \,,\notag\\
&\big[
R_1^\nu,\bar{n}\cdot P
\big] = 0  \,,\quad
&&\big[
R_2^\nu,\bar n\cdot P
\big] = 2\,i \,P_\perp^\nu  \,,\quad
&&\big[
R_3,\bar{n}\cdot P
\big] = 2\,i\,\bar{n} \cdot P  \,,
\label{eq:RPIMomentumAlgebra}
\\
&\big[
R_1^\nu,P^{\rho}_\perp
\big] = -i\,g_\perp^{\nu\rho}\,\bar{n}\cdot P  \,,\quad
&&\big[
R_2^\nu,P^{\rho}_\perp
\big] = -i \,g_\perp^{\nu\rho}\,n\cdot P \,,\quad
&&\big[
R_3,P^{\rho}_\perp
\big] = 0  \,.\notag
\end{alignat}
Assuming $P$ is collinear, it also possible to use these results to infer how the generators of RPI scale:
\begin{align}
& R_1^\nu \sim \lambda^{-1}\,,\quad && R_2^\nu \sim \lambda\,, \quad && R_3 \sim 1\,.
\end{align}
We see that RPI-I and RPI-II scale non-trivially with $\lambda$.  This can be understood intuitively by realizing that RPI-I and -II are changes to the directions orthogonal to the collinear particle's momentum direction (the $+\hat z$ direction in the conventions taken here), thereby imbuing them with $\lambda$ dependence.  RPI-III is essentially a boost in the $\hat z$ direction, and it therefore is insensitive to the notion of distance from the light-cone.  Practically, this scaling behavior implies that their action will lead to mixing of terms that power-count differently.  This must be the case since, as discussed previously, the EFT must be invariant under the three RPI transformations at every order in $\lambda$ to ensure full Lorentz invariance when all-orders are included.  

\subsubsection*{The RPI Transformations}
\label{sec:twocompRPI}

Next, we motivate the RPI transformations of the EFT fields for a general theory with a fermion and gauge boson, and will conclude this section with a summary of these transformations in Table~\ref{table:RPI}.  For the gauge boson, the RPI transformations of the $n\cdot A_n$ and $\bar n\cdot A_n$ components follow from the RPI transformations of $n^\mu$ and $\bar n^\mu$.  The $A_{n \perp}^{\mu}$ component transformations can be determined by demanding that the full vector $A_n^{\mu}$ is invariant under RPI.  From Table~\ref{table:RPI}, we see that RPI-I and RPI-II mix the different components of the gauge field.  

Deriving the fermion transformations requires a bit more care.  For example, there is a subtlety regarding what is actually meant by $u_n$, since it is defined by a projection $P_n u_n = u_n$ and the $n$/$\bar{n}$ directions are modified by the action of RPI.  We can determine the RPI transformations by requiring that the full spinor, $u=u_n+u_{\bar n}$, is RPI-invariant.  Using the expression for $u_{\bar n}$ in \eref{eq:anticollEOM}, its RPI-1 transformation is
\begin{align}
u=&\left[1-\frac{\nbs}{2}  \frac{1}{\bar{n} \cdot \partial}\bar{\sigma} \cdot \partial_\perp\right]u_n \nonumber\\
&\xrightarrow[\text{RPI-I}]{\makebox[1.0cm]{}} \bigg[
1-\frac{\nbs}{2}  \frac{1}{\bar{n} \cdot \partial}\bar{\sigma} \cdot \partial_\perp\notag+\frac{\nbs}{2}  \frac{1}{\bar{n} \cdot \partial}\bigg(\frac{\Delta_\perp\cdot \bar \sigma_\perp}{2}\bar n \cdot \partial +\frac{\bar n\cdot \bar \sigma}{2}\Delta_\perp\cdot \partial_\perp\bigg)\bigg]\nonumber \\
&\qquad\qquad\times\big(1+M_{u_n}(\Delta_\perp)\big)u_n\,,  
\end{align}
where we have assumed that the RPI transformation of $u_n$ is linear and generated by the matrix $M_{u_n}(\Delta_\perp)$.  Demanding that all terms proportional to $\Delta_\perp$ sum to zero and eliminating those terms that are identically zero, we find
\begin{equation}
M_{u_n}(\Delta_\perp)=\frac{\Delta_\perp\cdot \sigma}{2}\frac{\bar n\cdot \bar \sigma}{2}\,.
\end{equation}
Using this result, it is clear that RPI-I acts to rotate in the helicity component that has been integrated out:
\bea
u_{n} = \left[ \begin{array}{c}
0 \\
u_2  \end{array} \right] \qquad  \xrightarrow[\text{RPI-I}]{\makebox[1.0cm]{}}    \qquad
\left[ \begin{array}{c}
\frac{1}{2}(  \Delta_\perp \cdot \sigma)_{1 \dot{2}} u_2  \\
u_2 \end{array} \right] \,.
\eea

Applying the same argument for RPI-II, we see that it effectively acts as a rescaling of the fermion fields $u_n$
\bea
u_{n} = \left[ \begin{array}{c}
0 \\
u_2  \end{array} \right]   \qquad \xrightarrow[\text{RPI-II}]{\makebox[1.0cm]{}}    \qquad
\Bigg[ \begin{array}{c}
0  \\
\big( 1 +  \frac{1}{2}(\epsilon_\perp \cdot \sigma  )_{2\dot{1}} \frac{1}{ \bar{n}\cdot \mathcal{D}_n} (\mathcal{D}_{\perp,n} \cdot \bar{\sigma} )^{\dot{1}2}   \big) u_2 \end{array} \Bigg] \,.
\eea

Finally, consider the RPI-III transformations for the fermion.  The fermion Lagrangian must be invariant under RPI-III, and in particular the term
\bea
\mathcal{L}_u = i \,u^{\dagger}_n\, \frac{\Box}{\nbp} \frac{\bar{n} \cdot \bar{\sigma} }{2} u_n \qquad  \xrightarrow[\text{RPI-III}]{\makebox[1.0cm]{}}   \qquad u^{\dagger}_n\, \frac{\Box}{e^\alpha \,\nbp} \frac{e^\alpha \,\bar{n} \cdot \bar{\sigma} }{2} u_n\,,
\label{eq:RPI-IIIFermion}
\eea
where we note that $\Box$ is an RPI invariant.  Therefore, the full two-component collinear spinor $u_n$ does not transform under RPI-III.\footnote{Note, however, a confusion arises when we specify explicit vectors $n^\mu = (1,0,0,1)$ and $\bar{n}^\mu = (1,0,0,-1)$, such that the term in \eref{eq:RPI-IIIFermion} can be written as
\bea
\mathcal{L}_u = u^{\dagger}_{\dot{2}} \frac{\Box}{\nbp} u_{2} \rightarrow u^{\dagger}_{\dot{2}} \frac{\Box}{e^\alpha \,\nbp} u_{2}\,.
\eea
This would lead one to the \textit{incorrect} inference that the collinear degree of freedom $u_{n,2}$ inherits a transformation $u_{2}\rightarrow e^{-\alpha/2} u_{2}$.}

\begin{table}[h!]
\renewcommand{\arraystretch}{1.5}
\setlength{\arrayrulewidth}{.3mm}
\centering
\small
\setlength{\tabcolsep}{0.95em}
\begin{tabular}{ |c | c | c | c|}
    \hline
  &  RPI-1 & RPI-2 &   RPI-3   \\ \hline \hline
 $n \rightarrow$  & $ n + \Delta_\perp$  &  $n $ &    $ e^\alpha\, n $ \\ \hline
$\bar{n} \rightarrow $ &     $\bar{n}$   & $ \bar{n} + \epsilon_\perp$ &    $ e^{-\alpha}\, \bar{n}$  \\ \hline
 $ u_{n} \rightarrow$ &   $ \left(1 + \frac{1}{4}\sigma \cdot \Delta_\perp \nbsb \right) u_{n}$  &$ \left(1+\frac{\epsilon^\perp \cdot \sigma}{2} \frac{1}{i\,\bar n\cdot \partial}\,i\,\partial_\perp \cdot \bar \sigma\right)u_n$ &   $   u_{n} $ \\ \hline
  $n\cdot A_n \rightarrow$ &  $ n\cdot A_n + \Delta_\perp \cdot A_n$  &$ n\cdot A_n$ &  $ e^\alpha \,n\cdot A_n$ \\ \hline
  $\bar{n}\cdot A_n \rightarrow $ &  $\bar{n}\cdot A_n$ &    $ \bar{n}\cdot A_n + \epsilon_\perp \cdot A_n$  & $e^{-\alpha}\,\bar{n} \cdot A_n$ \\ \hline
 $A_{n\perp}^{\mu} \rightarrow$ &   $A_{n\perp}^{\mu}  - \frac{1}{2} \Delta_{\perp}^{\mu} \bar{n}\cdot A_n - \frac{\bar{n}^\mu}{2} \Delta_\perp \cdot A_{n\perp}$  & $ A_{n\perp}^{\mu}  - \frac{\epsilon^{\mu}_{\perp}}{2} n \cdot A_n - \frac{n^\mu}{2} \epsilon_\perp \cdot A_n$ & $ A^{\mu}_{n\perp} $ \\ \hline
$\phi_n \rightarrow $ &    $\phi_n$  &$ \phi_n$ &   $ \phi_n$  \\ \hline
\end{tabular}
\caption{The RPI transformations for the vectors $n$ and $\bar{n}$, fermions, gauge fields, and scalars. Here $\Delta_\perp \sim \mathcal{O}(\lambda)$ parametrizes deviations away from the collinear $n^\mu$ direction, and $\epsilon_\perp \sim \mathcal{O}(1)$ parametrizes deviations away from the anti-collinear $\bar{n}^\mu$ direction while maintaining the constraint $n \cdot \bar{n} = 2$; unlike RPI-I, the deviation away from the anti-collinear direction may be large. The components of the derivative $( n\cdot \partial,\bar{n}\cdot \partial, \partial_{\perp\mu})$, gauge field, and gauge covariant derivative $\mathcal{D}_n^\mu = \partial^\mu - i \,g \,A_n^\mu$ all have the same transformations.  Note that there is some subtlety by what specifically is denoted by $u_n$; the details given in the text should resolve any ambiguities.  }
\label{table:RPI}
\end{table}

Given these explicit transformations, it is clear that there is a non-trivial interplay between RPI and gauge invariance, which manifests as a shift of the polarization vector.  The rules in Table~\ref{table:RPI} can be used to show that the SCET Lagrangian for a collinear gauge boson coupled to a collinear fermion, \eref{eq:LSCET}, is an RPI invariant. This is a standard result in four component notation, see \emph{e.g.}~\cite{StewartNotes}.  The procedure in two-component spinors is trivially generalized from the four component case, and as such we do not review this calculation here.  


\FloatBarrier

\section{SCET in Light-Cone Gauge}
\label{sec:LCSCET}
We will find it useful to work with an EFT Lagrangian for $\mathcal{N} = 1$ pure SYM that makes SUSY manifest; to this end, it will be convenient to work with only the physical gauge boson polarizations.  This can be done by employing a Lorentz non-covariant gauge.  The gauge choice $\bar{n}\cdot A_n = 0$ defines the so-called Light-Cone Gauge (LCG)~\cite{Kogut:1969xa, Bjorken:1970ah, Tomboulis:1973jn, Mandelstam:1982cb, Leibbrandt:1983pj}, see \cite{Leibbrandt:1987qv} for a review. 
Additionally, once LCG has been fixed, the gauge field mode $n\cdot A_n$ is non-propagating, and can be integrated out.  The resulting Lagrangian can be expressed using only transverse polarizations.  In this gauge, collinear Wilson lines equal the identity, see Appendix~\ref{app:wilson}. Note that for the remainder of the paper we will often drop the subscript  ``$n$" on the components of the collinear gauge boson when it is clear by context.

As described in \sref{sec:twocomp}, deriving a Lagrangian for collinear fermions requires integrating out half the fermionic degrees of freedom.   Hence, the SCET model with collinear fermions and gauge bosons expressed in LCG contains two bosonic and two fermionic modes.  This equal number of degrees of freedom hints at the possibility of a SUSY relation between these fields.  However, note that gauge and Poincar\'e invariance (and therefore RPI) are not manifest in non-covariant gauges such as LCG.\footnote{\footnotesize{This issue also arises in the standard approach to SYM theories since SUSY is not manifest in the Wess-Zumino gauge.}}  This is why we discussed RPI in Sec.~\ref{sec:twocomp} using an (equivalent) off-shell description of the model.  We leave exploring the action of RPI on the LCG fields to future work.

The rest of this section provides a detailed discussion of the LCG Lagrangian derivation -- some readers may wish to skip ahead to the final result given in Eqs.~(\ref{eq:LuSummary}) and (\ref{eq:LGsummary}); for previous work, see \emph{e.g.}~\cite{Beneke:2002ph, Beneke:2002ni}.  To begin, we simply set $\bar{n} \cdot A = 0$.  Then we will solve for the equation of motion of the non-propagating gauge mode $n \cdot A$ in order to integrate it out.  It is convenient to re-organize the transverse components of the gauge field $A_{\perp \mu} = \left( 0, A_{1}, A_{2}, 0 \right)$ into a complex scalar $\alc$:
\bea
\partial_\perp \cdot A_{\perp} = - \partial^* \alc - \partial \alc^*\,,
\eea
where we have introduce the notation
\begin{align}
\label{eq:partial}
& \frac{\sigma \cdot \partial_\perp}{\sqrt{2}}=
\frac{1}{\sqrt{2}} \left[ \begin{array}{cc}
0 & \partial_1 - i\, \partial_2 \\
\partial_1 + i\, \partial_2  & 0  \end{array} \right]_{\alpha \dot{\alpha}}
\equiv 
\left[ \begin{array}{cc}
0 & \partial^* \\
\partial  & 0  \end{array} \right]_{\alpha \dot{\alpha}}\,.
\end{align}
Similarly, we can write the light-cone scalar as
\begin{align}
& \frac{\sigma \cdot A_\perp}{\sqrt{2}} 
\equiv 
\left[ \begin{array}{cc}
0 &\alcs \\
\alc  & 0  \end{array} \right]_{\alpha \dot{\alpha}}   \,; &
\frac{\bar{\sigma}\cdot A_\perp}{\sqrt{2}} \equiv \left[ \begin{array}{cc}
0 &-\alcs \\
-\alc  & 0  \end{array} \right]^{ \dot{\alpha} \alpha}\,.
\end{align}
Likewise for the transverse gauge covariant derivative:\footnote{\footnotesize{In the notation of Leibbrandt \cite{Leibbrandt:1987qv} $A_T \equiv \alcs$ and $A_{\bar{T}} \equiv \alc$. }}
\begin{align}
& \frac{\sigma \cdot \mathcal{D}_\perp}{\sqrt{2}}  = \frac{1}{2}
\left[ \begin{array}{cc}
0 & \mathcal{D}_1 - i \,\mathcal{D}_2 \\
\mathcal{D}_1 + i \,\mathcal{D}_2 & 0  \end{array} \right]_{\alpha \dot{\alpha}}  = 
\left[ \begin{array}{cc}
0 & \partial^* - i \,g \,\alc^* \\
\partial - i\,g\,\alc & 0  \end{array} \right]_{\alpha \dot{\alpha}} 
 \equiv 
\left[ \begin{array}{cc}
0 & \nabla^* \\
\nabla & 0  \end{array} \right]_{\alpha \dot{\alpha}} \,.
\label{eq:nablaDef}
\end{align}
We note that while we will stick to this standard notation, it can be confusing since $\nabla^* \neq \text{complex conjugate}\big(\nabla\big)$ as is clear from \eref{eq:nablaDef}; the $^*$ is equivalent to complex conjugation for $\alc$ and $\partial$.  The $\sigma$-matrix contracted with derivatives, gauge fields, and therefore covariant derivatives can be written in the following matrix form:
\begin{align}
& \sigma^\mu \partial_\mu = 
\left[ \begin{array}{cc}
\np & \sqrt{2}\,\partial^*  \\
\sqrt{2} \,\partial & \nbp  \end{array} \right]_{\alpha \dot{\alpha}},  \,
& \sigma^\mu A_\mu = 
\left[ \begin{array}{cc}
n \cdot A & \sqrt{2}\,\alc^*  \\
\sqrt{2}\, \alc & \bar{n}\cdot A  \end{array} \right]_{\alpha \dot{\alpha}}.
\end{align}
Now that we have defined all the relevant notation, we can proceed to derive the Lagrangian.

\subsection{The Abelian Theory in LCG}
\label{sec:AbelianTheoryLCG}
As a warm up, we begin with the model of a free $U(1)$ collinear gauge boson:
\begin{align}
\mathcal{L}_{U(1)} &= -\frac{1}{4} F_{\mu \nu}F^{\mu \nu} = -\frac{1}{2}\big( \partial_\mu A_\nu \,\partial^\mu A^\nu - \partial_\mu A_\nu \,\partial^\nu A^\mu \big)\,.
\label{eq:LAbelian}
\end{align}
Using the following identities (assuming LCG such that $\bar{n} \cdot A = 0$)
\begin{align}
\partial_\mu A_\nu \,\partial^\mu A^{\nu} &=  (\partial_{\perp \mu} A_{\perp \nu})( \partial_\perp^\mu A_\perp^\nu) + \frac{1}{2}(\nbp A_{\perp \mu}) (\np A_\perp^\mu) + \frac{1}{2} (\np A_{\perp \mu})(\nbp A_\perp^\mu) \, , \nonumber \\ 
\partial_\mu A_\nu \,\partial^\nu A^{\mu} &= \frac{1}{4}\left(  \nbp\, n\cdot A \right)^2 + (\partial_{\perp \mu} A_{\perp \nu})( \partial_\perp^\nu A_\perp^\mu)\\
&\quad + \frac{1}{2} (\nbp A_{\perp \mu})(\partial_\perp^\mu n\cdot A) + \frac{1}{2}(\partial_{\perp \mu} n\cdot A)( \nbp A_\perp^\mu)\,, \nonumber
\end{align}
it is straightforward to expand \eref{eq:LAbelian} in light-cone coordinates: 
\begin{align}
2\mathcal{L}_{U(1)} = \,&A_{\perp \mu} (\partial_\perp^2 A_\perp^{\mu}) + \big(\partial_\perp \cdot A_\perp\big)^2 + \frac{1}{4}\left(  \nbp \,n\cdot A \right)^2\notag\\
&+ (\nbp\, n\cdot A)  (\partial_\perp \cdot A_\perp) + A_{\perp \mu} (\nbp \,\np A_{\perp}^{\mu} )\,,
\label{eq:LAbelianLCcoords}
\end{align}
where we have used $\partial_{\perp \mu} A_{\perp \nu}\, \partial_\perp^\mu A_\perp^\nu - \partial_{\perp \mu} A_{\perp \nu} \,\partial_\perp^\nu A_\perp^\mu=- A_{\perp \mu} (\partial_\perp^2 A_\perp^\mu) - (\partial_\perp \cdot A_\perp)^2$.  This form is convenient for converting to the $\mathcal{A}$  scalars.   We have the following identities: 
\begin{align}
\partial_\perp^2  &= - \partial_1^2 - \partial_2^2 =-2\partial \partial^*\,; \label{eq:LCscalarIdentity1}\\ 
\partial_\perp \cdot A_\perp &= - \partial^* \alc - \partial \alc^*\,; \label{eq:LCscalarIdentity2}\\
 A_{\perp \mu} (\partial_\perp^2 A_\perp^{ \mu}) + (\partial_\perp \cdot A_\perp)^2 &=  \Big[ \alc^* (\partial \partial^* \alc) + \alc (\partial \partial^* \alc^*) \Big] + \Big[ (\partial \alc^*)^2+(\partial^* \alc)^2 \Big]\,;  \label{eq:LCscalarIdentity3}\\
A_{\perp \mu} (\nbp\, \np A_\perp^\mu) &= - \alc(\nbp \,\np \alc^*) + \textrm{h.c.}\,, \label{eq:LCscalarIdentity4}
\end{align}
where we have integrated by parts to combine factors.  The Lagrangian becomes
\begin{align}
2 \mathcal{L}_{U(1)} =&\Big[-\alc( \nbp \,\np - \partial \partial^*) \alc^{*} + \textrm{h.c.} \Big]   + \Big[ (\partial \alc^{*})^2+(\partial^* \alc)^2 \Big] \notag\\ 
&-( \partial^* \alc + \partial \alc^{*})(\nbp \,n\cdot A)+ \frac{1}{4}\left(  \nbp \,n\cdot A \right)^2\,. 
\label{eq:LU1} 
\end{align}
Identifying $n\cdot\partial$ as the lightcone time derivative, we see that the only propagating degrees of freedom are $\mathcal{A}$ and $\mathcal{A}^*$.  Therefore, it is prudent to integrate out $n\cdot A$:
\begin{align}
\frac{\delta  \mathcal{L}_{U(1)}}{\delta n\cdot A} = 0 \qquad \Longrightarrow \qquad n\cdot A  & =  \frac{2}{\nbp} \left( \partial^* \alc + \partial \alc^{*}  \right) \,.
\label{eq:eomfree}
\end{align}

If $\mathcal{A}$ can really be interpreted as a complex scalar, then $ \mathcal{L}_{U(1)}$ should reduce to the free Klein-Gordon equation once we integrate out $n\cdot A$ and fix the gauge $\bar{n}\cdot A = 0$.  We can see this explicitly by plugging \eref{eq:eomfree} into \eref{eq:LU1}, which yields
\bea
\mathcal{L}_{U(1)} =  - \alc^* \Box \alc\,,
\eea
the kinetic term for a complex scalar.

\subsection{Interactions and IR Structure in LCG}
\label{sec:IRinLCG}
The LCG EFT must reproduce the expected collinear and soft limits of the full theory which were discussed in Sec.~\ref{sec:twocomp}. Since the soft and collinear sectors do not mix at leading order in the $\lambda$ expansion, the soft limits are identical to the full theory and are thus trivially consistent. To check agreement in the collinear limit, we can consider the example of collinear photon emission in the EFT and show that this reproduces the expected leading order divergence structure of the full theory.  Since the collinear splitting factor is gauge invariant on its own in the collinear limit, we expect to find the exact same expression in LCG.  For simplicity, we will show this agreement for SCET QED in LCG.

First, we must construct the Feynman rules for the $\mathcal{O}(e)$ interaction vertex coupling $A_\perp$ to a charged fermion.  It is convenient to work with the perpendicular components of the Lorentz vector gauge field $A_\perp^\mu$ rather then LC scalars so that the explicit dependence on the outgoing polarization vector is manifest.  The coupling to a charged fermion comes from
\begin{align}
\mathcal{L}_u \supset e\, u^{\dagger}_n  (n\cdot A ) \left(\frac{\nbsb}{2}\right)  u_n - i\, u^\dagger_n \Big[ (\bar{\sigma} \cdot \mathcal{D}_{\perp,n})  \frac{1}{\nbp} (\sigma \cdot \mathcal{D}_{\perp,n})  \left(\frac{\nbsb}{2}\right) u_n\Big] \,,
\end{align}
where $\mathcal{D}^\mu_\perp = \partial^\mu_\perp - i e A^\mu_\perp$ is the covariant derivative.  Expanding the term in brackets yields
\begin{align}
&- i\, u^\dagger_n \bigg[ (\bar{\sigma} \cdot \mathcal{D}_{\perp,n})  \frac{1}{\nbp} (\sigma \cdot \mathcal{D}_{\perp,n})  \left(\frac{\nbsb}{2}\right) u_n\bigg] \nonumber\\ 
& \quad \quad \quad  \supset  \,\,  \bigg[ \frac{\partial_\perp^\nu}{\nbp} u^\dagger_{n,p} \bigg]  \bar{\sigma}_{ \nu}   \sigma_{\mu} \,  A^\mu_\perp  \frac{\nbsb}{2} u_{n,p'}+ u^\dagger_{n,p}\, \bar{\sigma}_{ \mu} \sigma_{ \nu}  \,A_\perp^\mu \bigg[ \frac{\partial_\perp^\nu}{\nbp}   \frac{\nbsb}{2} u_{n,p'}   \bigg]   \,.
\end{align}
where we have included label momentum indices $p$ and $p'$ on the fermions, and have integrated by parts to get the second line.  This gives a contribution to the $u$-$u^\dag$-$A^\mu$ vertex which goes as $\left(    \frac{ \bar{\sigma} \cdot  p_{\perp_{}}^{}}{\bar{n}\cdot p} \sigma^\mu   +  \bar{\sigma}^\mu  \frac{\sigma \cdot p'_{\perp_{}}}{\bar{n}\cdot p'}   \right) \frac{\bar{n}\cdot\bar{\sigma}}{2}$;  these are the same spin-dependent terms that appear in the standard QCD-SCET Feynman rule, see App.~\ref{sec:CovariantGaugeSingularities}.

\begin{figure}[t!]
	\begin{center}
		\includegraphics[width=11cm]{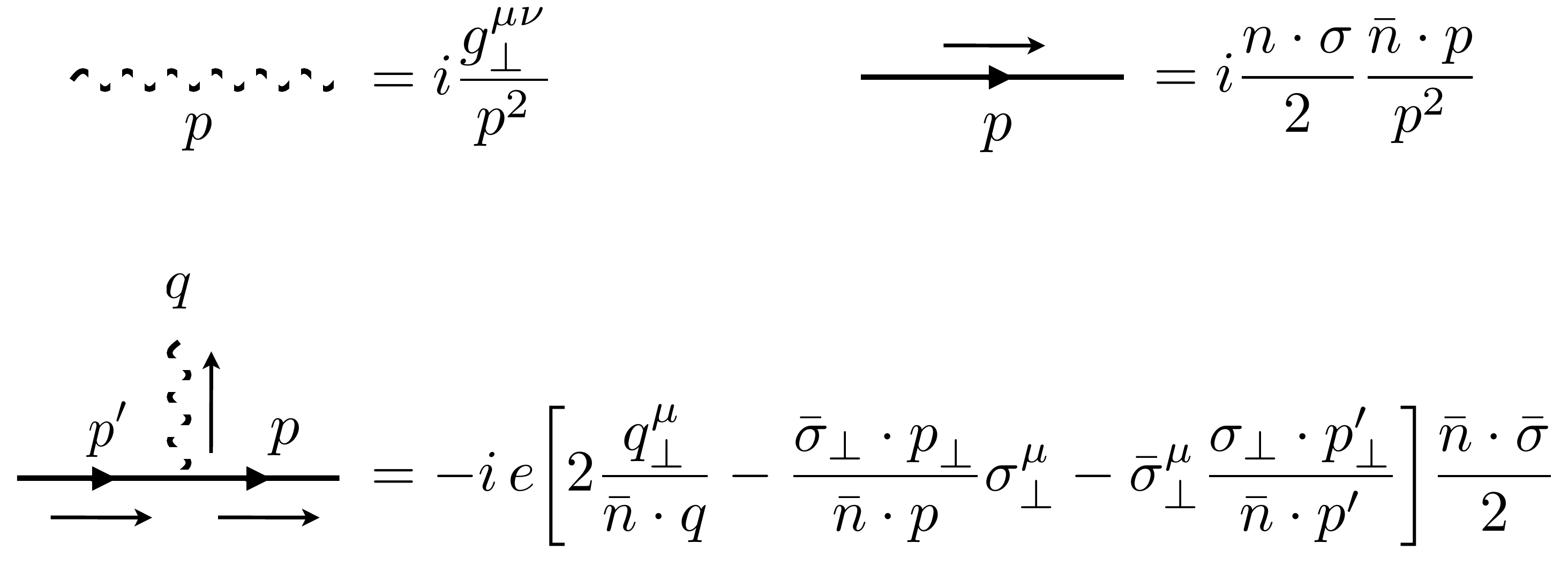}
	\end{center}
 	\caption{
Subset of the Feynman rules for the Abelian collinear LCG model with a charged fermion.  The dashed wavy line is a light cone gauge boson.} 
 	\label{fig:LCGFeynRules}
 \end{figure}

Next we integrate out the non-propagating mode $n\cdot A$:
\bea
\frac{\delta \mathcal{L}_{U(1)}}{\delta n\cdot A} = 0 \quad \Longrightarrow \quad n\cdot A =  - \frac{2}{\nbp}\, \partial_\perp \cdot A_\perp -\frac{2\,e}{(\nbp)^2} \Big[ u^\dagger_n   \left(\frac{\nbsb}{2}\right)  u_n \Big] \, ,
\eea
which will lead to some new interactions. Plugging this into the pure gauge Lagrangian \eref{eq:LAbelianLCcoords} yields no $\mathcal{O}(e)$ contribution.  Then the only term we need to include to derive the $\mathcal{O}(e)$ vertex comes from the fermion Lagrangian:
\begin{align}
\mathcal{L}_u\supset  e \,u^{\dagger}_n  (n\cdot A)  \left(\frac{\nbsb}{2}\right)  u_n  
&= e \, u^{\dagger}_n  \left( - \frac{2}{\nbp} \partial_\perp \cdot A_\perp -\frac{2\,e}{(\nbp)^2} \Big[ u^\dagger_n   \left(\frac{\nbsb}{2}\right)  u_n \Big] \right)  \left(\frac{\nbsb}{2}\right)  u_n \notag \\ 
&\supset   - 2 \, e  \,  \frac{1}{\nbp}  (\partial_\perp \cdot A_\perp )    \Big[ u^{\dagger}_n  \left(\frac{\nbsb}{2}\right)  u_n \Big] \,,
\end{align}
resulting in a contribution to the interaction vertex proportional to $q^{ \mu}_\perp/\bar{n}\cdot  q$.  The full Feynman rule is given in Fig.~\ref{fig:LCGFeynRules}.

To check that we reproduce the expected collinear factor, we compute the following diagram for emission of a collinear Abelian gauge boson off a collinear charged fermion:
\begin{align}
\raisebox{-0.25\height}{\includegraphics[width=3.8cm]{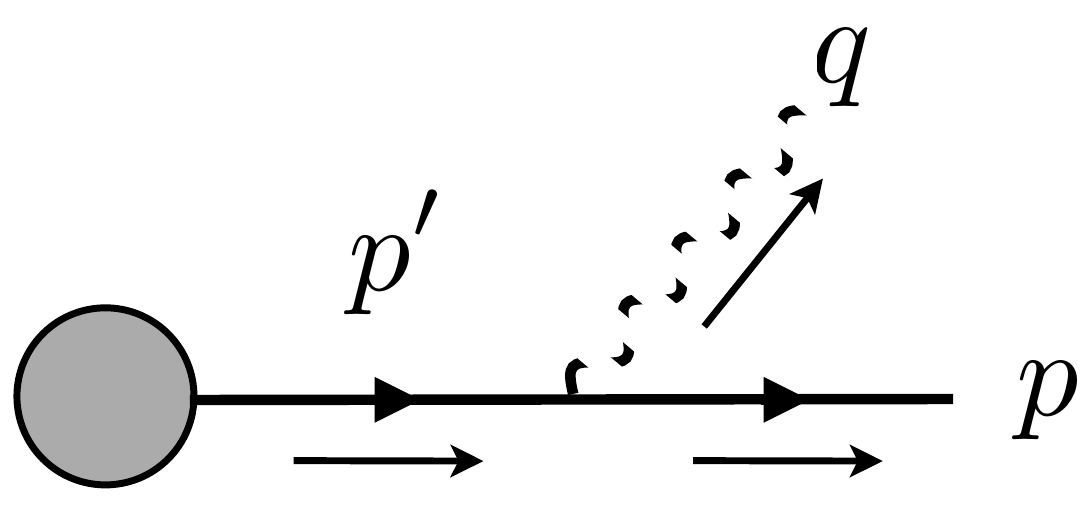}}
  &= -i\, e\, x^\dagger_n (p) \left[  2 \,\frac{q^{ \mu}_\perp}{\bar{n}\cdot  q }-  \bar{\sigma}_{\perp}^\mu  \frac{\sigma_{\perp} \cdot q_\perp}{\bar{n}\cdot p'}  \right] \epsilon^*_{ \mu} (q)  \frac{i \, \bar{n}\cdot p'}{p'^2} \mathcal{M}(p')\, ,
\label{eq:LCGphotonemission}
\end{align}
where we have aligned $p$ with the collinear direction $p^{\mu} = \frac{n^\mu}{2} \bar{n}\cdot p$, such that $p_\perp = 0$, and we have absorbed the projection operator $( \nbsb/2)(\ns/2)$ into the definition of $\mathcal{M}(p')$.  

We can simplify the expression using the QED Ward identity:
\bea
q \cdot \epsilon^* = 0 \,\, \Rightarrow \,\, q_\perp \cdot \epsilon^* = - \frac{1}{2} (\bar{n}\cdot \epsilon^*) (n \cdot q) - \frac{1}{2} (n\cdot \epsilon^*) (\bar{n}\cdot q) =  - \frac{1}{2} (n\cdot \epsilon^*) (\bar{n}\cdot q) \, ,
\eea
where we have used the LCG condition $\bar{n} \cdot \epsilon = 0$.  Conservation of momentum implies $p' = p + q$.  Then using the sigma matrix identity $ \bar{\sigma}^\mu \sigma^\nu = -  \bar{\sigma}^\nu \sigma^\mu+ 2\, g^{\mu \nu}$ and the Weyl equation of motion $x_n^\dag \,\bar{\sigma}\cdot n = 0$, the amplitude simplifies to
\begin{equation}
\raisebox{-0.25\height}{\includegraphics[width=3.8cm]{Figures/M1EFT.pdf}} = -\, e\, x^\dagger_n (p) \left[ \frac{(n\cdot \epsilon^*)}{(n \cdot q) } -    \frac{  (\bar{\sigma}_\perp \cdot q_\perp) (\sigma_\perp \cdot \epsilon^*)}{(\bar{n} \cdot p)(n \cdot q)}   \right]   \mathcal{M}(p+q),\,
\label{eq:LCGcollinearFactor}
\end{equation}
which exactly agrees with \eref{eq:FullThyCollinearFactor} and with App.~\ref{sec:CovariantGaugeSingularities}.  Therefore, the LCG Lagrangian has the same collinear structure as the full theory.  Note that this agreement is non-trivial since it required including terms that result from integrating out $n\cdot A$.

\subsection{The Non-Abelian Theory in LCG}
Now we move on to the derivation of the Lagrangian for the non-Abelian EFT.  The starting point is the adjoint QCD SCET Lagrangian given above in \eref{eq:qcdscetlagrangian}.  For convenience, let $\mathcal{L}_u$ contain the interactions between the gauge bosons and fermions, and $\mathcal{L}_g$ be the pure gauge part of the Lagrangian, which is the same as YM theory:
\bea
 \mathcal{L}_g = -\frac{1}{4} F_{\mu \nu}^a F^{\mu \nu a} &=& -\frac{1}{4} \left(  \partial_\mu A_{\nu}^a - \partial_\nu A_{ \mu}^a \right) \left( \partial^\mu A^{\nu a} - \partial^\nu A^{\mu a} \right) \nonumber\\ 
&& -\frac{1}{2} g f^{abc} A_{ \mu}^b A_{ \nu}^c \left( \partial^\mu A^{\nu a} - \partial^\nu A^{\mu a} \right)-\frac{1}{4} g^2 f^{abc} f^{ade} A_{ \mu}^b A_{ \nu}^c A^{\mu d} A^{\nu e} \nonumber \\ 
&\equiv&  \mathcal{L}_{g,0} +\mathcal{L}_{g,1} +\mathcal{L}_{g,2} 
\label{eq:CollinearKineticLagrangian}
\eea 
Note that the subscript labels on each $\mathcal{L}_{g,i}$ refers to the number of factors of $g$ that are present \emph{before} integrating out $n\cdot A^a$.  Plugging in the equation of motion for $n\cdot A^a$ leads to terms involving different powers of $g$.

The rest of this subsection follows the same procedure as for the Abelian model.  For completeness, we will derive the gauge-boson fermion interactions as well, starting from
\begin{align}
\mathcal{L}_u = i\, u^\dag \bar{\sigma}\cdot \mathcal{D}\, u\,.
\end{align}
The first step is to derive the equations of motion for $n\cdot A^a$.
\subsection*{Integrating out $n\cdot A^a$}
Working with a non-Abelian theory will not change the argument given in the previous section that $n\cdot A^a$ is non-propagating with respect to LC time.  Hence, we need to derive the equation of motion for $n \cdot A $.  Setting $\delta \mathcal{L} / \delta n\cdot A^a = 0$, gives
\begin{align}
 \frac{\delta \mathcal{L}_u}{\delta n \cdot A^b} &= g \,u_2^{\dag a} \big(t^b\big)^{ac} u_2^c = i\, g\, f^{abc} u_2^{\dag a}u_2^c\,; \\
\frac{\delta \mathcal{L}_g}{\delta n\cdot A^d} &= -\frac{1}{2} \left(\frac{\delta F_{\mu \nu}^a}{\delta n\cdot A^d} \right) F^{a \mu \nu}\,.
\end{align}
Making use of \eref{eq:CollinearFieldStrength} we find:
\bea
\frac{\delta (\mathcal{L}_u + \mathcal{L}_g)}{\delta n \cdot A^a} 
&=& g \,u_2^\dag \,u_2 t^a- \frac{i}{4}t^a \Big\{  (\nbp )^2 n \cdot A +2\, \mathcal{D}_{\perp \mu} (\nbp A_\perp^\mu ) \Big\} \,,
\eea
where we have used $\bigl[ \nbp, n\cdot A^a t^a \bigr] \, u^b_2 = \nbp (n\cdot A^a \, t^a \, u^b_2) - n\cdot A^a \, t^a \, (\nbp \, u^b_2) = \nbp \, (n\cdot A^a \, t^a) \, u^b_2$. In order to convert to LC fields and derivatives, note that
\bea
 -  \mathcal{D}_{\perp \mu}( \nbp A_\perp^\mu) &=&  \nabla^* (\nbp \alc) + \nabla (\nbp \alcs) \,.
\eea 
Plugging this in and solving the above for $n \cdot A^a$ yields:

\label{eq:eomgaugeadj}
\begin{empheq}{align}
n \cdot A^b &=  -\frac{4\,i\,g}{(\nbp)^2} u_2^{\dag a} u_2^c \big(t^b\big)^{ac} + \frac{2}{(\nbp)^2}\Big \{ \nabla^{*bc} (\nbp \alc^c) + \nabla^{bc} (\nbp \alc^{*c})\Big \} \nonumber\\
& = \frac{2}{\nbp} \left( \partial^* \alc^b + \partial \alc^{*b}  \right)+4\, g\, f^{abc}\left( \frac{1}{(\nbp)^2} (u^{\dag a}_2 u^c_2) \right)  \label{eq:ncdotA} \\ 
 &\quad -  2 \,g\, f^{abc}\left( \frac{1}{(\nbp)^2} \Big[  \alc^{*a} \,\nbp \alc^c + \alc^a\, \nbp \alc^{*c} \Bigr]\right)\,,\nonumber
\end{empheq}
\noindent where in the second line we assumed the fermions are in the adjoint representation with generators $(t^e)^{bc} = - i \,f^{ebc}$ and we expanded $\nabla^{bc} =  \delta^{bc} \,  \partial  - g\,  f^{ebc}\,  \alc^e$. Note that the $\mathcal{O}(g^0)$ terms exactly match \eref{eq:eomfree} above.  The next step is to substitute \eref{eq:ncdotA} into the Lagrangian \eref{eq:qcdscetlagrangian}.

\subsubsection*{Deriving $\mathcal{L}_{g,0}$ in LCG}
To begin, we will derive the terms which come from the part of the gauge Lagrangian without explicit $g$-dependence.  First note that the derivation of the kinetic term follows exactly the same steps as in Sec.~\ref{sec:AbelianTheoryLCG}, so we will not repeat them here.  The same manipulations to get the Lagrangian written in terms of light-cone scalars and $n\cdot A^b$ are useful here.  Starting with the Lagrangian given in \eref{eq:LU1}, and plugging in \eref{eq:ncdotA} for $n\cdot A^b$ yields the total $\mathcal{L}_{g,0}$ LCG Lagrangian: 
\begin{align}\label{eq:Lg0}
\mathcal{L}_{g,0}=  &- \alc^{*b}\Box \alc^b \\ \nonumber
&-2 \,g^2  f^{abc} f^{ebh}   \left( \frac{1}{\nbp} \left( u^{\dag a}_2 u^c_2  \right) \right)  \left(  \frac{1}{\nbp} \Big[  \alc^{*e}\,\nbp \alc^h + \alc^e\, \nbp \alc^{*h}   \Bigr]\right)    \\ \nonumber
&+  2\, g^2  f^{abc} f^{ebh}  \left( \frac{1}{\nbp} \left( u^{\dag a}_2 u^c_2  \right) \right)\left( \frac{1}{\nbp} \left( u^{\dag e}_2 u^h_2  \right) \right) \\ \nonumber
&+  \frac{ g^2 }{2}  f^{abc} f^{ebh} \left(  \frac{1}{\nbp} \Big[  \alc^{*a}\nbp \alc^c + \alc^a \nbp \alc^{*c}   \Bigr]\right)  \left(  \frac{1}{\nbp} \Big[  \alc^{*e}\nbp \alc^h + \alc^e \nbp \alc^{*h}   \Bigr]\right) \,.
\end{align}
Note that the tri-linear light-cone scalar/fermion interaction cancels when combining terms. This interaction is not generated by the kinetic Lagrangian. This is in analogy to what we will find for the Wess-Zumino model in Sec.~\ref{sec:WZModel}. 

\subsubsection*{Deriving $\mathcal{L}_{g,1}$ in LCG}
Starting with the $\mathcal{L}_{g,1}$ Lagrangian:
\begin{align}
 \mathcal{L}_{g,1}  = -\frac{1}{2}\,g \,f^{abc} A_{ \mu}^b A_{  \nu}^c \left( \partial^\mu A^{\nu a} - \partial^\nu A^{\mu a} \right)\,,
\end{align}
  we can expand these interactions in light-cone coordinates:
\begin{align}
 \mathcal{L}_{g,1}= -\frac{1}{2}\,g\, f^{abc} \Big[ A_{\perp \mu}^b A_{\perp \nu}^c \left( \partial_\perp^\mu A_\perp^{\nu a} - \partial_\perp^\nu A_\perp^{\mu a} \right) + n\cdot A^b\, A_{\perp \mu}^c  (\nbp A_\perp^{a\mu}) \Big]\,.
 \end{align}
The first term does not depend on $n\cdot A$, so we can just simply rewrite it in terms of the light cone scalars:
 \begin{align}
 -\frac{1}{2}\,g \,f^{abc} A_{\perp \mu}^b A_{\perp \nu}^c \left( \partial_\perp^\mu A_\perp^{\nu a} - \partial_\perp^\nu A_\perp^{\mu a} \right)  = -g\, f^{abc}\,  \alc^{*b} \alc^c (\partial^* \alc^a - \partial \alc^{*a}) \,.
 \end{align}
The second term is more complicated.  First we rewrite the transverse components in terms of light cone scalars:
\begin{align}
-\frac{1}{2}\,g \,f^{abc}  n\cdot A^b\, A_{\perp \mu}^c  (\nbp A_\perp^{a\mu}) &=- \frac{1}{2}\,g\, f^{abc}  \,n\cdot A^b (\alc^{*a} \,\nbp \alc^c + \alc^{a}\, \nbp \alc^{*c})\,.
\end{align}
\newpage
\noindent Next, using \eref{eq:ncdotA} for the equation of motion of $n\cdot A^b$ in the adjoint representation, and putting it all together gives 
\begin{align}
\label{eq:Lg1} 
\mathcal{L}_{g,1} = & -g\, f^{abc}  \alc^{*b} \alc^c (\partial^* \alc^a - \partial \alc^{*a})  - g\, f^{abc} (\partial^* \alc^b + \partial \alc^{*b}) \left(  \frac{1}{\nbp} \Big[  \alc^{*a}\nbp \alc^c + \text{h.c.}   \Big] \right)  \nonumber \\  
&+g^2 f^{ebh}  f^{abc} \left(  \frac{1}{\nbp} \Big[  \alc^{*a}\nbp \alc^c + \alc^a \nbp \alc^{*c}   \Bigr] \right)  \left(  \frac{1}{\nbp} \Big[  \alc^{*e}\nbp \alc^h + \alc^e \nbp \alc^{*h}   \Big] \right)  \nonumber\\ 
&-2 \,g^2 f^{ebh} f^{abc} \left( \frac{1}{\nbp} \left( u^{\dag a}_2 u^c_2  \right) \right) \left(  \frac{1}{\nbp} \Big[  \alc^{*e}\nbp \alc^h + \alc^e \nbp \alc^{*h}   \Bigr] \right)
\end{align}

\subsubsection*{Deriving $\mathcal{L}_{g,2}$ in LCG}
The final piece is straightforward to derive.  Simply expanding the gauge fields in light cone coordinates, and relying on the asymmetry of gauge indices to convert from the transverse gauge bosons to the light cone scalars yields
\bea
\label{eq:Lg2}
\mathcal{L}_{g^2} &=& -\frac{1}{4}\,g^2\, f^{abc} f^{ade} A_{ \mu}^b A_{  \nu}^c A^{\mu d} A^{\nu e} =-\frac{1}{4}\, g^2\, f^{abc} f^{ade} A_{ \perp \mu}^b A_{ \perp \nu}^c A_\perp^{\mu d} A_\perp^{\nu e} \\ \nonumber
&=& -\frac{1}{2} \,g^2\, f^{abc} f^{ade} A_1^b A_2^c  A_1^d  A_2^e = -\frac{1}{2} \,g^2\, f^{abc} f^{ade}  \alc^{*b} \alc^c \alc^{*d} \alc^e\,.
 \eea

\subsubsection*{Summarizing $\mathcal{L}_g$ in LCG}
Putting together Eqs.~(\ref{eq:Lg0}), (\ref{eq:Lg1}), and (\ref{eq:Lg2}), the LCG Lagrangian is:
\bea
\nonumber
& \mathcal{L}_{g} =  -\alc^{*b} \Box \alc^b - g\, f^{abc}  \alc^{*b} \alc^c (\partial^* \alc^a - \partial \alc^{*a}) - \frac{1}{2}\,g^2 f^{abc} f^{ade}  \alc^{*b} \alc^c \alc^{*d} \alc^e  \\  \nonumber
 &\,\,\,\,\,\, - g\, f^{abc} (\partial^* \alc^b + \partial \alc^{*b})  \left(\frac{1}{\nbp} \Big[  \alc^{*a}\nbp \alc^c + \alc^a \nbp \alc^{*c}   \Bigr] \right)  \\ \nonumber 
  &\,\,\,\,  \, + \frac{3}{2}\, g^2\, f^{ebh}  f^{abc} \left(\frac{1}{\nbp} \Big[  \alc^{*a}\nbp \alc^c + \alc^a \nbp \alc^{*c}   \Bigr] \right)  \left(\frac{1}{\nbp} \Big[  \alc^{*e}\nbp \alc^h + \alc^e \nbp \alc^{*h}   \Bigr] \right)  \\ \nonumber
 &\,\,\,\, \, +  2 \,g^2\,  f^{abc} f^{ebh}    \left( \frac{1}{\nbp} \left( u^{\dag a}_2 u^c_2  \right) \right)   \left( \frac{1}{\nbp} \left( u^{\dag e}_2 u^h_2  \right) \right)  \\
  &\,\,\,\, \,   -4 \,g^2\,  f^{abc} f^{ebh}  \left( \frac{1}{\nbp} \left( u^{\dag a}_2 u^c_2  \right) \right)   \left(\frac{1}{\nbp} \Big[  \alc^{*e}\nbp \alc^h + \alc^e \nbp \alc^{*h}   \Bigr] \right) \, , 
   \label{eq:LGsummary}
\eea

\newpage
\subsubsection*{Gauge Boson -- Fermion Interactions in LCG}
Fixing the LCG condition, $\bar n\cdot A_n = 0$, in \eref{eq:LSCET} and explicitly writing out the gauge structure, the gaugino Lagrangian becomes:
\begin{align}
\mathcal{L}_u=\,& u_n^{\dagger}\left( i\, n \cdot \mathcal{D}_{n,s} + i \,\bar{\sigma} \cdot \mathcal{D}_{\perp,n} \frac{1}{i \,\bar{n}\cdot \mathcal{D}_n} i\, \sigma \cdot \mathcal{D}_{\perp,n} \right) \frac{\bar{n}\cdot \bar{\sigma}}{2}\,  u_n \\ \nonumber%
=\,&u^{\dag a}_2 \, (i \np \,  \delta^{ac} \, u_{2}^c) + u^{\dag a}_2 \Big(g \,  n\cdot A^e \,  (t^e)^{ac}  \, u_2^c\Big) - 2\,i \, u_{2}^{\dag a} \bigg[  \nabla^{ab}\frac{1}{\nbp} \nabla^{*bc}  u_2^c\bigg] \\ \nonumber
=\,& u_2^{\dag a} \, (i \,\nbp \, u_2^a) - i\,g\,  f^{eac} \, u_2^{\dag a} \,  (n\cdot A^e)\, u_2^c \\[3pt] \nonumber
&- 2\,i \, u_2^{\dag a} \, \left(  \frac{\partial \partial^*}{\nbp} \, u_2^a \right) +2\,i\,  g  f^{eab} \,  u_2^{\dag a} \, \bigg[   \alc^e \frac{\partial^*}{\nbp} u_2^b+ \frac{\partial}{\nbp} \,  (\alc^{*e} u_2^b)  \bigg] \\[3pt] \nonumber
&- 2\,i \, g^2 f^{eab}  f^{hbc} \, u_2^{\dag a}\, \alc^e \, \bigg[ \frac{1}{\nbp} \, (\alc^{*h} u_2^c) \bigg]\,,
\end{align}
where again we have taken the gaugino to be in the adjoint representation.  Reorganizing the above as
\begin{align}
\mathcal{L}_u =&\,\, u_2^{\dag a}\left( i \,\np  + \frac{2\, \partial \partial^*}{i\,\nbp} \right) u_2^a  +2\,i\,g f^{eab} u_2^{\dag a}\bigg[   \alc^e \frac{\partial^*}{\nbp} u_2^b+ \frac{\partial}{\nbp} \Big(\alc^{*e} u_2^b\Big)  \bigg] \\  &\,- 2\,i\,g^2 f^{eab} f^{hbc} u_2^{\dag a} \alc^e \bigg[ \frac{1}{\nbp}\Big(\alc^{*h} u_2^c\Big) \bigg] - i\,g\, f^{eac} u_2^{\dag a} (n\cdot A^e) u_2^c \,, \notag
\end{align}
we can plug in the equation of motion of $n\cdot A^e$ from \eref{eq:ncdotA}:
\bea
\label{eq:LuSummary}
&&\mathcal{L}_u = u_2^{\dag a}\left(i \,\np  + \frac{2\, \partial \partial^*}{i\,\nbp}  \right) u_2^a\\ \nonumber
&&\,\,\,\,\,\, +2\,i\,g\, f^{bac} \bigg(   u_2^{\dag a}\alc^b \frac{\partial^*}{\nbp}   u_2^c +u_2^a \,  \alc^{*b} \frac{\partial}{\nbp}  u_2^{\dag c} +  u_2^{\dag a}\, u_2^c \left(     \frac{\partial^*}{\nbp}  \alc^b +\frac{\partial}{\nbp}   \alc^{*b}   \right)  \bigg) \\ \nonumber
&&\,\,\,\,\,\, -4\,i\,g^2\, f^{bac}  f^{ebh} u_2^{\dag a}  u_2^c  \left(  \frac{1}{(\nbp)^2} (u^{\dag e}_2 u^h_2) \right)   - 2\,i\,g^2 \,f^{eab} f^{hbc} u_2^{\dag a} \alc^e \bigg( \frac{1}{\nbp}(\alc^{*h} u_2^c) \bigg)  \\ \nonumber
&&\,\,\,\,\,\, -2\, i\,g^2\, f^{bac} f^{ebh} u_2^{\dag a} u_2^c   \left( \frac{1}{(\nbp)^2} \Big(  \alc^{*e}\, \nbp \alc^h + \alc^e\, \nbp \alc^{*h} \Big) \right) \, ,
\eea
\noindent  where derivative operators act to the right.
Note that $\partial_\perp^2 = - 2 \,\partial \partial^* $, so we recover the expected kinetic gaugino term \eref{eq:Lfree}.

This completes the derivation of LCG SCET.  Ultimately, we will show that this Lagrangian can be formulated in superspace, thereby demonstrating its SUSY invariance.  But first, the next section provides the necessary background to understand SUSY in the collinear limit.

\section{Supersymmetry and SCET}
\label{sec:susy}
This section explores the formalism of supersymmetric SCET.
To begin, we show how the supercharges expand in terms of the SCET power counting parameter, from which we will see that half of the supercharges are sub-leading. Likewise we show that two out of the four $\mathcal{N} = 1$ superspace Grassmann coordinates have virtuality beyond the validity of the EFT, and therefore should not be present. Next, we construct the collinear chiral and vector multiplets of the EFT, \emph{i.e.}, we derive the chiral and vector representations of collinear SUSY. We do so analogous to the usual way, by building up states that form representations of the Super-Poincar\'e algebra on the light cone, see \emph{e.g.} \cite{Wess:1992cp, Terning:2006bq} for a review. Then we provide a discussion of the residual SUSY algebra cast in the language of RPI.   For completeness, we briefly discuss SUSY Wilson lines in App.~\ref{app:wilson}.

Having established this framework, Sec.~\ref{sec:CollSuperspace} will show that the consequence of these power counting rules corresponds to ``integrating out" half of superspace. In other words; integrating out fields at the component level is equivalent to selecting a shell of superspace, on which the EFT lives. Working in ``collinear superspace"  will yield the tools necessary to construct an algorithm for deriving a collinear superspace Lagrangian for general theories~\cite{Cohen:2016jzp}.  We will then apply this explicitly for two interacting models: non-Abelian gauge theory in Sec.~\ref{sec:ymsect} and to the Wess-Zumino model in Sec.~\ref{sec:WZModel}. 

\subsection{Power Counting the Supersymmetry Algebra}
${\cal N}=1$ SUSY is defined by the graded algebra
\begin{equation}
\left\{
Q_\alpha,Q^\dagger_{\dot \alpha}
\right\} = 2\,\sigma^\mu_{\alpha\dot\alpha} P_\mu \,.
\end{equation}
The corresponding algebra for the EFT can be derived by considering the collinear scaling for the generator of translations $P_\mu \sim (1, \lambda^2, \lambda)$:
\bea
\Big\{Q_\alpha, Q_{\dot{\alpha}}^\dagger \Big\} = 2\, \sigma^{\mu}_{\alpha \dot{\alpha}} P_\mu = 2\,i
\left[ \begin{array}{cc}
\np & \sqrt{2}\,\partial^*  \\ 
\sqrt{2}\, \partial & \nbp  \end{array} \right] _{\alpha \dot{\alpha}} \sim \,\, \left[ \begin{array}{cc}
\mathcal{O}(\lambda^2) &\mathcal{O}(\lambda)  \\ 
\mathcal{O}(\lambda) & \mathcal{O}(1) \end{array} \right] \,.
\eea
Truncating to leading power in the EFT, it is clear that only one supercharge survives.  For the concrete choice of $n^\mu$ and $\bar{n}^\mu$ we are using here, this prescription yields $Q_2 \sim \mathcal{O}(1)$, while the second supercharge is higher order in $\lambda$, see Table~\ref{table:superScaling}.
This scaling can also be inferred by constructing the multiplets directly, as discussed below in Sec.~\ref{sec:SUSYReps}.

\begin{table}[h!]
\renewcommand{\arraystretch}{1.7}
\setlength{\arrayrulewidth}{.3mm}
\centering
\setlength{\tabcolsep}{1.3em}
\begin{tabular}{|c  c  c  c |}
    \hline
  \multicolumn{4}{|c|}{Supercharges}  \\
   $Q^1 = Q_2$ &  $Q^{\dagger \dot{1}} = Q^{\dagger}_{\dot{2}}$ &  $Q^2 = - Q_1$  & $Q^{\dagger \dot{2}} = - Q^{\dagger}_{\dot{1}}$ \\[3pt] \hline 
  $1$ & $1$ & $\lambda$ & $\lambda$  \\ \hline
\end{tabular}
\\[5pt]
\setlength{\tabcolsep}{1.62em}
\begin{tabular}{|c  c  c  c |}
    \hline
  \multicolumn{4}{|c|}{Superspace Coordinates}  \\   
  $\theta^1 = \theta_2$ &  $\theta^{\dagger \dot{1}} = \theta^{\dagger}_{\dot{2}}$ &  $\theta^2 = - \theta_1$  & $\theta^{\dagger \dot{2}} = - \theta^{\dagger}_{\dot{1}}$ \\[3pt] \hline 
   $\lambda^{-1}$ & $\lambda^{-1}$ & $1$ & $1$  \\ \hline
\end{tabular}
\\[5pt]
\setlength{\tabcolsep}{1.6em}
\begin{tabular}{|c  c  c  c |}
    \hline
  \multicolumn{4}{|c|}{Super-transformation Parameters}  \\ 
 $\eta^1 = \eta_2$ &  $\eta^{\dagger \dot{1}} = \eta^{\dagger}_{\dot{2}}$ &  $\eta^2 = - \eta_1$  & $\eta^{\dagger \dot{2}} = - \eta^{\dagger}_{\dot{1}}$ \\[3pt] \hline
$1$ & $1$ & $\lambda$ & $\lambda$  \\ \hline
\end{tabular}
\caption{The non-trivial scaling of the supercharges $Q$, the superspace coordinates $\theta$, and the super-transformation parameters $\eta$.}
\label{table:superScaling}
\end{table}

Lifting the theory to superspace provides a manifestly supersymmetric formulation of the EFT.  To this end, it will be useful to understand how power counting applies to the differential form of the supercharges.  Recall that in addition to the usual space-time coordinates $x^\mu$, superspace is expressed using anti-commuting spinor-valued Grassmann coordinates $\theta^\alpha$ and $\theta^{\dagger \dot{\alpha}}$ and $y^\mu = x^\mu - i \theta \sigma^\mu \theta^\dagger$.  Then the supercharges may be written as differential operators in the usual way:\footnote{\footnotesize{See pages 449-453 of \cite{Binetruy:2006ad} for a review of conventions.}}
\begin{align}
&& Q_\alpha =  i \frac{\partial}{\partial \theta^\alpha} + \left( \sigma \cdot \partial \right)_{\alpha \dot{\alpha}} \theta^{\dag \dot{\alpha}} \,,
&& Q^\dag_{\dot{\alpha}} = - i \frac{\partial}{\partial \theta^{\dag \dot{\alpha}}} - \theta^\alpha \left( \sigma \cdot \partial \right)_{\alpha \dot{\alpha}}\,,
\end{align}
which in light-cone coordinates become
\begin{align}
&Q_2 =  \left(i \frac{\partial}{\partial \theta^2} + \theta^{\dagger \dot{2}} \nbp +\sqrt{2}\, \theta^{\dagger \dot{1}}  \partial \right) \,, \qquad Q_1 = \left( i \frac{\partial}{\partial \theta^1} + \theta^{\dagger \dot{1}}\np + \sqrt{2} \, \theta^{\dagger \dot{2}}  \partial^*\right) \,,
\end{align}
with analogous expressions for $Q_{\dot{\alpha}}^\dag$.  To maintain the consistency of the scalings for the supercharges and momentum operators, the superspace coordinates must have non-trivial $\lambda$ scaling as well, see Table \ref{table:superScaling}.

In SCET, the scaling of space-time coordinates $\left(n\cdot x, \bar{n} \cdot x, x_\perp \right) \sim \left(1,1/\lambda^2, 1/\lambda \right)$ is fixed by requiring that the exponent of the Fourier transform $p^\mu x_\mu \sim \mathcal{O}(1)$. For a general superfield  $\mathcal{F}$ we can write down the analogous super-Fourier transform \cite{duplij2000noncommutative}
\begin{equation}
\label{eq:superfourier}
\mathcal{F}(x,\theta, \theta^\dagger ) = \frac{1}{(2\,\pi)^2} \int \text{d}^4 p\, \text{d}^2 \theta\, \text{d}^2 \theta^\dagger \, \tilde{\mathcal{F}}\big(p,\eta,\eta^\dagger\big) \, e^{i \left( p^\mu x_\mu + \eta^\alpha \theta_{\alpha} + \eta^{\dagger}_{ \dot{\alpha}} \theta^{\dagger \dot{\alpha}}  \right)}\,,
\end{equation}
where the spinor $\eta$ is the SUSY transformation parameter:
\bea
\delta_S\, \mathcal{F} (x^\mu, \theta, \theta^\dagger) = - i\, x^\mu \, P_\mu \, \mathcal{F} - i \,\eta^\alpha \, Q_\alpha \,  \mathcal{F} - i\, \eta^\dagger_{\dot{\alpha}} \, Q^{\dagger \dot{\alpha}} \,  \mathcal{F}\,.
\eea
In analogy with the bosonic space-time coordinates, the non-trivial scaling of the superspace coordinates fixes the scaling of the SUSY transformation parameters, see Table~\ref{table:superScaling}. 
Additionally, the notion of a well-defined super-Fourier transform will allow us to write inverse derivative operators in superspace in terms of their eigenvalues analogous with the interpretation of $1 / \nbp$.

Finally, we can infer the scaling of the superspace derivatives by starting with their definition 
\begin{align}
&D_\alpha = \frac{\partial}{\partial \theta^\alpha} - i\, (\sigma \cdot \partial)_{\alpha \dot{\alpha}} \theta^{\dag \dot{\alpha}} \,,
& \bar{D}_{\dot{\alpha}} =  \frac{\partial}{\partial \theta^{\dag \dot{\alpha}}}- i \,\theta^\alpha (\sigma \cdot \partial)_{\alpha \dot{\alpha}}\,.
\end{align} 
The leading order superspace derivatives (in terms of $x^\mu$ variables) in the EFT are:
\begin{align}
&D_2 = \frac{\partial}{\partial \theta^2} -i \,\theta^{\dag \dot{2}} (\nbp) - i\, \sqrt{2}\, \partial \theta^{\dag \dot{1}} \sim \mathcal{O}(\lambda^0 )\,,\\
&\bar{D}_{\dot{2}} =  \frac{\partial}{\partial \theta^{\dag \dot{2}}} - i\,\theta^2 (\nbp) - i\, \sqrt{2} \,\theta^1 \partial^*\sim \mathcal{O}(\lambda^0 )\,,
\end{align}
which obey
\bea
\Big\{D_\alpha, \bar{D}_{\dot{\alpha}}\Big\}= -2 \,i \, \left(\sigma \cdot \partial  \right)_{\alpha \dot{\alpha}} = -2\, i \,
\left[ \begin{array}{cc}
\np & \sqrt{2}\,\partial^*  \\ 
\sqrt{2} \,\partial & \nbp  \end{array} \right] _{\alpha \dot{\alpha}} \sim \left[ \begin{array}{cc}
\mathcal{O}(\lambda^2) &\mathcal{O}(\lambda)  \\ 
\mathcal{O}(\lambda) & \mathcal{O}(1) \end{array} \right] \,.
\eea
This provides the full dictionary required for working with SCET in superspace.

\subsection{Super-RPI; the Super-Poincar\'e Group on the Light Cone}
The super-Poincar\'e group of the full theory is defined by extending the bosonic algebra Eqs.~(\ref{eq:poincare1})-(\ref{eq:poincare3}) with
\bea
\label{eq:superpoincare1}
\Big[ Q_\alpha, P^\mu \Big]  = &0&  = \Big[ Q_{\dot{\alpha}}^\dagger, P^\mu \Big]\,; \\ 
\label{eq:superpoincare2}
\Big[ M_{\mu \nu}, Q_\alpha \Big] = i \,\big(\sigma_{\mu \nu}\big)^{\beta}_\alpha Q_\beta  \,\,\,\,\, &\textrm{and}&  \,\,\,\,\, \Big[ M_{\mu \nu}, Q_{\dot{\alpha}}^\dagger \Big] = i \,\big(\bar{\sigma}_{\mu \nu}\big)^{\dot{\beta}}_{\dot{\alpha}} Q_{\dot{\beta}}^\dagger \,; \\
\label{eq:superpoincare3}
 \Big\{Q_\alpha, Q_{\dot{\alpha}}^\dagger \Big\} &=& 2\, \sigma^\mu_{\alpha \dot{\alpha}} P_\mu \,.
\eea
We can choose the momentum eigenvalue for a massless collinear field as $\lambda\rightarrow 0$ to be $p_\mu = (E,0,0,E)$; the supersymmetry algebra reduces to that of the leading power EFT:
\begin{align}\label{eq:frame1}
\left\{
Q_1,Q^\dagger_{\dot 1}
\right\} &= 0 \,,\\
\label{eq:frame2}
\left\{
Q_2,Q^\dagger_{\dot 2}
\right\} &= 4\,E \,.
\end{align}
Using \eref{eq:superpoincare2} and the SUSY algebra,
\begin{align}
\Big[
R_1^\nu,Q_\alpha
\Big] &=-i\,\bigg[
\sigma_\perp^\nu \frac{\bar n\cdot \bar \sigma}{2}
\bigg]_\alpha^\beta Q_\beta \,,\\
\Big[
R_2^\nu,Q_\alpha
\Big] &=-i\,\bigg[
\sigma_\perp^\nu \frac{n\cdot \bar \sigma}{2}
\bigg]_\alpha^\beta Q_\beta \,,\\
\Big[
R_3,Q_\alpha
\Big] &=i\,\bigg[
\frac{n\cdot \sigma}{2} \frac{\bar n\cdot \bar \sigma}{2}-1
\bigg]_\alpha^\beta Q_\beta \,.
\end{align}
Expanding these explicitly for $n_\mu = (1,0,0,1)$ and $\bar n_\mu = (1,0,0,-1)$, which is equivalent to the frame choice used for \Eqs{eq:frame1}{eq:frame2}, the leading power anti-commutator of the supersymmetry generators is given by
\begin{equation}
\left\{
Q_\alpha,Q^\dagger_{\dot \alpha}
\right\} = (n\cdot\sigma)_{\alpha\dot \alpha} \, \bar n\cdot P \,,
\end{equation}
which is just a rewriting of \Eqs{eq:frame1}{eq:frame2}.  The algebraic relations between the RPI and supersymmetry generators becomes
\begin{align}
&\Big[
R_1^\nu,Q_1
\Big] =-i\,\bar a^\nu Q_2 \,;
\qquad\qquad\qquad\qquad\qquad\quad\,
\Big[
R_1^\nu,Q_2
\Big] =0 \,;\\
&\Big[
R_2^\nu,Q_1
\Big] =0 \,;
\qquad\qquad\qquad\qquad\qquad\qquad\qquad
\Big[
R_2^\nu,Q_2
\Big] =-i\,a^\nu Q_1 \,;\\
&\Big[
R_3,Q_1
\Big] =-i \,Q_1 \,;
\qquad\qquad\qquad\qquad\qquad\qquad\,
\Big[
R_3,Q_2
\Big] =0\,,
\end{align}
where $a^\nu=(0,1,i,0)$ and $\bar a^\nu = (0,1,-i,0)$.  

\newpage

\subsection{Representations of the Light-Cone Super-Poincar\'e Algebra}
\label{sec:SUSYReps}
In this section we construct the representations of the super-Poincar\'e algebra on the light cone which are relevant for the SYM and Wess-Zumino models discussed below; the free collinear vector and chiral supermultiplets respectively.  These will be the asymptotic states built from the non-interacting Clifford vacuum $|\Omega \rangle$.  

\vspace{10pt}
\noindent \textbf{Lorentz Invariant Theory:}  To begin, we will review the procedure for constructing SUSY multiplets in a Lorentz invariant theory, for details see \emph{e.g.}~\cite{Wess:1992cp, Terning:2006bq}.  The irreducible representations of the Super-Poincar\'e algebra are characterized by the eigenvalues of the momentum squared $P^2$ operator and the Casimir operator $C^2 = 2\, m^4\, \mathcal{J}_k \mathcal{J}^k$, where $\mathcal{J}_k$ is related to the spin operator $S_k$ by 
\begin{align}
m\,\mathcal{J}_k = m \,S_k - \big(Q^\dagger \bar{\sigma} Q\big)\,.
\label{eq:RelateJandS}
\end{align}  
Given this definition, $\mathcal{J}_k$ satisfies the algebra for angular momentum $\bigl[ \mathcal{J}_k, \mathcal{J}_l \bigr] = i\, \epsilon_{k l m} \mathcal{J}_m$, such that $\mathcal{J}^2 =\mathcal{J}_k \mathcal{J}^k$ is an invariant operator with eigenvalues $j(j+1)$.  Additionally, $\bigl[ \mathcal{J}_k, Q_\alpha \bigr] =  \bigl[ \mathcal{J}_k, Q^\dagger_{\dot{\alpha}} \bigr] = 0 $, implying that $\mathcal{J}_k$ and $P^2$ are compatible observables.  Thus, the irreducible representations of the Super-Poincar\'e algebra are characterized by the eigenvalues of the Casimir operators $m^2$ and $j(j+1)$.  Additionally, the states in a specific representation have an additional quantum number $j_3$, the eigenvalue of $\mathcal{J}_3$ which takes values $j_3 = - j, -j+1, \textrm{. . . } j-1, j$ as usual.  

In general, these states are not eigenstates of ordinary spin $S^2$ and $S_3$.  However, our goal is to find the corresponding spin eigenstates in order to understand the building blocks of a supermultiplet. From the algebra
\bea
\Bigl[ M^{\mu \nu}, Q^{\dagger}_{\dot{\alpha}} \Bigr] = i\, Q^{\dagger}_{\dot{\beta}} \big(\bar{\sigma}^{\mu \nu}\big)^{\dot{\beta}}_{\dot{\alpha}}\,,
\eea
with $S_3 = M^{12}$, and $\bar{\sigma}^{\mu \nu} \equiv \frac{1}{4} \left( \bar{\sigma}^{\mu} \sigma^\nu - \bar{\sigma}^\nu \sigma^\mu   \right)$. We find that
\bea
\Bigl[ S_3, Q^{\dagger}_{\dot{1}} \Bigr] = \frac{1}{2} Q^{\dagger}_{\dot{1}}\,; \qquad\qquad \Bigl[ S_3, Q^{\dagger}_{\dot{2}} \Bigr] = -\frac{1}{2} Q^{\dagger}_{\dot{2}} \,.
\label{eq:SpinAlg}
\eea

We will work out the simple example of a massive particle.  Boosting to its rest frame $p^\mu = (m,0,0,0)$, the SUSY algebra becomes
\bea
\Big\{Q_\alpha, Q_{\dot{\alpha}}^\dagger \Big \} = 2 \,\sigma^{\mu}_{\alpha \dot{\alpha}} P_\mu =  \left[ \begin{array}{cc}
2\,m & 0  \\ 
0 & 2\,m  \end{array} \right]_{\alpha \dot{\alpha}} \,.
\eea
This defines the Clifford Algebra with $Q^\dagger$ and $Q$ as creation and annihilation operators respectively.  A state $|m,j,j_3\rangle$ which is annihilated by $Q_\alpha$ with $\alpha  = 1,2$ is the Clifford Vacuum:
\bea
Q_\alpha |\Omega \rangle = 0\,.
\eea
This provides the vacuum state of the spinor representation.  Using \eref{eq:RelateJandS}, we see that the Clifford Vacuum is an also an eigenstate of the operators $S^2$ and $S_3$, that is it can be characterized by its spin:
\bea
|\Omega \rangle = |m,s,s_3 \rangle\,.
\eea
To build up the rest of the multiplet, consider the action of the creation operators on the Clifford Vacuum; using \eref{eq:SpinAlg},
\bea
\label{eq:raisingloweringspin}
&& S_3\,  Q^{\dagger}_{\dot{1}} | \Omega \rangle = (s_3 + 1/2)\,Q^{\dagger}_{\dot{1}} | \Omega \rangle\,;  \\  \nonumber
&& S_3  \,Q^{\dagger}_{\dot{2}} | \Omega \rangle = (s_3 - 1/2)\,Q^{\dagger}_{\dot{2}} | \Omega \rangle\,,
\eea
so that $Q^{\dagger}_{\dot{1}}$ raises and $Q^{\dagger}_{\dot{2}}$ lowers the value of $s_3$ by $1/2$.  Additionally,
\bea
S_3 \,Q^{\dagger}_{\dot{1}}  Q^{\dagger}_{\dot{2}} | \Omega \rangle = s_3 \,Q^{\dagger}_{\dot{1}} \, Q^{\dagger }_{\dot{2}} | \Omega \rangle\,.
\eea
Finally, applying $ Q^{\dagger}_{\dot{\alpha}}\, Q^{\dagger}_{\dot{1}} \, Q^{\dagger}_{\dot{2}} | \Omega \rangle = 0$, truncates the construction of the multiplet.

Hence, for each pair of values $(m,j)$ of the Casimir operators, we obtain an irreducible representation of the Super-Poincar\'e algebra. The states corresponding to specific values of $m$ and $j$ contain $2j+1$ subspaces corresponding to possible values of $j_3$. This implies that for fixed $j_3$ each subspace contains four eigenstates of spin $S_3$ namely $s_3 = j_3, j_3 + 1/2, j_3 - 1/2$ and again $j_3$. These four states correspond to $1|\Omega\rangle$, $Q^{\dagger}_{\dot{1}}|\Omega\rangle$, $Q^{\dagger}_{\dot{2}}|\Omega\rangle$, and $Q^{\dagger}_{\dot{1}}Q^{\dagger}_ {\dot{2}}|\Omega\rangle$, thereby forming an irreducible $4(2j+1)$ dimensional massive representation of the Clifford algebra.

\vspace{10pt}
\noindent \textbf{Light-Cone Theory:} 
Now we can move on to apply the same logic to the model as defined on the light-cone.  The main result will be to demonstrate the self-consistency of the scalings of the supercharges and the collinear fields.  This can be made explicit by constructing the representations.

Consider a massive chiral multiplet.  The Clifford vacuum of the free theory $|\Omega \rangle$ is identified with the scalar state.  Boosting the full theory massive multiplet along the collinear direction will allow us to identify states in the collinear sector of the EFT.  In the rest frame of a massive particle $p^\mu = (m,0,0,0)$, we construct the states of the chiral multiplet:
\begin{align}
\nonumber
 | \Omega \rangle &= |m,0,0\rangle \\ \nonumber
 Q^{\dagger}_{\dot{1}} | \Omega \rangle &= |m,1/2,1/2\rangle \\ \nonumber
 Q^{\dagger}_{\dot{2}} | \Omega \rangle &= |m,1/2,-1/2\rangle \\ 
 Q^{\dagger}_{\dot{1}} \,Q^{\dagger}_{\dot{2}} | \Omega \rangle &= |m,0,0 \rangle\,.  
\end{align}
The state $Q^{\dagger}_{\dot{1}} | \Omega \rangle$ corresponds to a left handed Weyl fermion with spin-up along the $\hat{z}$ direction, which we identify as the spacial direction for the light-light vector $n^\mu$. Similarly, $Q^{\dagger}_{\dot{2}} | \Omega \rangle$ corresponds to a fermion in a spin-down state along the $\hat{z}$ direction.\footnote{\footnotesize{We can write the eigenstates of spin along the $\hat{z}/n^\mu$ direction as $\xi_{1/2} = (1,0)^\text{T}$ and $\xi_{-1/2} = (0,1)^\text{T}$.}}  The remaining states correspond to the two bosonic degrees of freedom of the multiplet.
Now perform a boost along $\hat{z}$. According to \eref{eq:uCollinearLimit}, half the spin states will be projected out upon boosting; a spin-up left handed Weyl fermion when boosted along the $\hat{z}$ direction is suppressed, while a spin-down left handed fermion is not.  This identification allows us to construct collinear and anti-collinear states by acting with the SUSY charges on the Clifford vacuum of the full theory and then boosting along the collinear direction. 

In particular, the collinear and anti-collinear projection operators act on the fermionic states of the multiplet as follows
\begin{align}
& P_n \,Q^{\dagger}_ {\dot{1}} | \Omega \rangle = 0 \, , \qquad\qquad\qquad
 P_n \,Q^{\dagger}_{\dot{2}} | \Omega \rangle = Q^{\dagger}_{\dot{2}} | \Omega \rangle  \,, \\
& P_{\bar{n}}\, Q^{\dagger}_{\dot{2}} | \Omega \rangle =0 \, ,\qquad\qquad\qquad
 P_{\bar{n}}\, Q^{\dagger}_{\dot{1}} | \Omega \rangle = Q^{\dagger}_{\dot{1}} | \Omega \rangle  \,.\notag
\end{align}
Then in the collinear sector $P_n$ projects out $Q^\dagger_{\dot{1}} |\Omega \rangle$ so that the remaining state
\bea
  Q^\dagger_{\dot{2}}  |\Omega \rangle  = |m,1/2,-1/2 \rangle \quad\longrightarrow\quad  \sqrt{2\,E} \, \xi_{-1/2}  
 =  u_n
\eea
is a left handed two-component massless spinor with spin-down along the collinear direction.   Applying the rules for the supercharges given in Table~\ref{table:superScaling}, we can infer that $u_n \sim \mathcal{O}(\lambda)$.  Furthermore, comparing to Table~\ref{table:scaling}, we see that $u_n \sim \mathcal{O}(\lambda)$ and can therefore be identified with the collinear fermion of the EFT multiplet. For the anti-collinear sector, $P_{\bar{n}}$ projects out $Q^\dagger_{\dot{2}} |\Omega \rangle $ so that the remaining collinear state
\bea
Q^\dagger_{\dot{1}} |\Omega \rangle = |m,1/2,1/2 \rangle  \quad\longrightarrow\quad
\sqrt{2\,E} \, \xi_{1/2}  =
u_{\bar{n}} \,.
\eea
Again, the power counting of the supercharge and of the field both yield $u_{\bar{n}}\sim \mathcal{O}(\lambda^2)$.  Therefore we have constructed the collinear and anti-collinear states of the chiral superfield and have demonstrated a consistent picture for the power counting of the SUSY collinear multiplet.

Finally, we can consider the vector multiplet. The Clifford vacua is the spin one half Weyl fermion state. In a massless theory we construct the on-shell physical gauge degrees of freedom as follows 
\begin{align}
& Q^{\dagger}_{\dot{2}} \big|\Omega\big\rangle  = Q^{\dagger}_{\dot{2}}  \big|h = -1/2\big\rangle = \big|h = -1\big\rangle \sim \lambda \, ,
\end{align}
where for massless states the appropriate quantum number is helicity $h$.  Completing the multiplet requires a similar construction with the CPT conjugate. These can be identified as the two transverse gauge degrees of freedom $\alc$ and $\alc^*$.  Again comparing the scalings given in Table~\ref{table:scaling} and Table~\ref{table:superScaling}, we find a consistent picture for the EFT vector multiplet. 

\subsection{Non-interacting SUSY SCET Lagrangian}
\label{sec:FreeSUSYLag}
We have now demonstrated that there is a consistent way to construct both a SUSY algebra and non-interacting scalar and vector multiplets in the collinear limit. Hence, for concreteness we provide the explicit Lagrangian for a free collinear scalar and a free collinear fermion along with the SUSY transformations that leave this Lagrangian invariant.  To leading power, the collinear theory of a chiral multiplet, a complex scalar and a left handed Weyl fermion, is given by
\bea\label{eq:FreeLag}
{\cal L}_{u_n} + {\cal L}_{\phi_n} &=&  u_n^\dagger \left(i\,n\cdot \partial - \frac{\partial_\perp^2}{i\,\bar n \cdot \partial}  \right) \frac{\bar n\cdot \bar \sigma}{2}\,u_n 
-  \phi_n^* \left( \nbp\, \np + \partial^2_\perp \right)\phi_n \\ \nonumber
&=& u^\dagger_{n,\dot{2}} \left(i\,n\cdot \partial - \frac{\partial_\perp^2}{i\,\bar n \cdot \partial}  \right) u_{n,2} 
-  \phi_n^* \left( \nbp \,\np + \partial^2_\perp \right)\phi_n\,.
\eea
We see that the EFT has two fermionic and two bosonic degrees of freedom, which can be seen as the components of an on-shell chiral supermultiplet.  Then \eref{eq:FreeLag} is invariant under the SUSY transformations 
\begin{align}
\delta\phi_n &= \eta^2\, u_{n,2} \,, \qquad   \delta u_{n,2} = -\eta^{\dagger}_{\dot{2}} (i \,\bar{n} \cdot \partial)\, \phi_n,\nonumber \\
\delta\phi_n^*&=  \eta^{\dagger \dot{2}} \,u^\dagger_{n, \dot{2}} \,, \qquad   \delta  u^\dagger_{n, \dot{2}} = \eta_2\, \phi_n^* \Big(i\,\bar n \cdot \overleftarrow{\partial}\Big)  \,,
\label{eq:SUSYtransformationsChiSupermultiplet}
\end{align} 
where we have included implicit projection operators. This is straightforward to check: 
\bea
\delta \mathcal{L} &=& i\,u^\dagger_{n} \frac{\Box}{\nbp} \frac{\bar n\cdot \bar \sigma}{2} \,\delta u_{n}  - \delta \phi_n^* \Box  \phi_n + \textrm{h.c.} \nonumber \\ 
&=& -i\,u^\dagger_{n} \frac{\Box}{\nbp} \frac{\bar n\cdot \bar \sigma}{2} \left( \frac{n\cdot \sigma}{2}\eta^\dagger(i\,\bar n \cdot \partial) \phi_n \right)  -  \eta^\dagger\, u_n^\dagger \Box  \phi_n \, + \textrm{h.c.} \\ 
&=& -i\,u^\dagger_{n} \frac{\Box}{\nbp}  \eta^\dagger(i\,\bar n \cdot \partial)\, \phi_n   -  \eta^\dagger \,u_n^\dagger \Box  \phi_n \, + \textrm{h.c.} =   \eta^\dagger\, u^\dagger_{n} \Box \phi_n   -  \eta^\dagger \,u_n^\dagger \Box  \phi_n  + \textrm{h.c.}  = 0 \,, \nonumber
\eea
where we have used the projection operators given in \eref{eq:uProjections} and have integrated by parts, dropping total derivatives, to demonstrate that $\delta \mathcal{L}$ vanishes. Note that since the SUSY transformation parameter scales as $\eta^2 \sim \mathcal{O}(\lambda)$ the SUSY transformations are higher order with respect to the fields themselves, that is $\delta  u_n$ and $\delta   \phi$ scale as $ \sim \mathcal{O}(\lambda^2)$ while the fields scale as $ \sim \mathcal{O}(\lambda)$.

Note that $\phi$ can be either interpreted as either a complex scalar field restricted to the light-cone, or it could also describe a gauge boson in LCG with $\phi \rightarrow \mathcal{A}^a$  and $u_n \rightarrow u_n^a$ (where $a$ is a gauge index and note that the partial derivative does not get promoted to a covariant derivative).  While these are interchangeable for a non-interacting theory, we will show in what follows that the models with non-trivial couplings are quite different.  Specifically, a consistent treatment of $\mathcal{N}=1$ SYM SCET will be provided in Sec.~\ref{sec:ymsect}, while the Wess-Zumino model will prove to be problematic as described in Sec.~\ref{sec:WZModel}.

\newpage

\section{Collinear Superspace}
\label{sec:CollSuperspace}
In the previous section, we found that in the SCET expansion half the supercharges have sub-leading $\lambda$-scaling. Additionally, two out of the four $\mathcal{N} = 1$ superspace Grassmann coordinates were shown to have high virtuality, implying that they should not be present in the EFT. In the present section, we demonstrate how to derive a consistent leading power EFT by  ``integrating out" half of superspace. In other words; integrating out anti-collinear fields at the component level is equivalent to selecting a ``collinear slice" of superspace, on which the theory lives.  Beyond packaging the Lagrangian in a simpler from, writing the theory in this so-called collinear superspace makes SUSY manifest.\footnote{\footnotesize There is an interesting construction called ``$\mathcal{N} = 1/2$ superspace" which is defined on non-commutative backgrounds~\cite{Seiberg:2003yz}.  In these theories $Q^{\dagger}_{\dot{\alpha}}$ is not present in contrast with collinear superspace where $Q_1$ and $Q_{\dot{1}}^\dagger$ are removed.} 

The procedure for deriving a collinear superspace Lagrangian can be summarized with the following general algorithm~\cite{Cohen:2016jzp}:
\begin{itemize}
\item Find projection operators that separate the superfield into collinear/anti-collinear superfields, \emph{e.g.} see \eref{eq:ChiralSuperfieldProjection}.
\item Starting with the superspace action for the full theory, integrate out the entire anti-collinear superfield. This will yield a constraint equation, \emph{e.g.} see \eref{eq:collinearFreeChiralConstraint}.
\item Show that the constraint equation reproduces the expected equations of motion at the component level. This works both as a sanity check and will prove useful later on, \emph{e.g.} see \eref{eq:checkcomponent}.
\item Based on the constraint equation, guess an ansatz for the equation of motion for the anti-collinear superfield in terms of collinear degrees-of-freedom. Check that ansatz satisfies any additional constraints (like chirality or reality), \emph{e.g.} see \eref{eq:freeChiralSuperfieldAnsatz}.
\item Plug the ansatz into the full theory action to yield the superspace action of the effective theory, \emph{e.g.} see \eref{eq:freeChiralCollinearL}.
\end{itemize}

In the what follows, we will follow this algorithm to derive the collinear superspace Lagrangian for a few examples.  First, we will apply the algorithm to the simple case of the non-interacting chiral superfield, which we will then apply to Abelian gauge theory. We will discuss the case of soft-collinear SYM in Sec.~\ref{sec:ymsect}, and verify that the collinear superspace Lagrangian is equivalent to the component LCG SCET Lagrangian derived in Sec.~\ref{sec:LCSCET}.  The superspace derivation of the interacting Wess-Zumino model will be discussed in Sec.~\ref{sec:WZModel}.

\subsection{Non-Interacting Chiral Multiplet}
\label{sec:freeWZcollinearsuperspace}
We begin our study of collinear superspace with the formulation a non-interacting chiral supermultiplet discussed above in Sec.~\ref{sec:FreeSUSYLag}.  This is an instructive starting point not only for its simplicity, but also because (as discussed in Sec.~\ref{sec:LCSCET} and as we verify in detail below) the vector multiplet LCG degrees of freedom reorganize into a chiral multiplet. Hence, the machinery presented here is applicable to both gauge and matter theories. 

The first step to formulate the Lagrangian in superspace is to identify the anti-collinear chiral superfield to be integrated out. Recall that the superspace action for a massless non-interacting chiral superfield is $S = \int \text{d}^4 x \, \text{d}^4 \theta \,\,  \Phi^\dagger \left( x, \theta^\dagger \right) \Phi \left(x, \theta \right)$, where $\Phi$ is a chiral superfield, subject to the constraint $\bar{D}_{\dot{\alpha}} \Phi = 0 = D^\alpha \Phi^\dagger$. We can apply projection operators to the fermionic component of the full theory chiral multiplet to separate out collinear and anti-collinear modes. Therefore in a supersymmetric theory we expect the entire superfield to obey the following decomposition; 
\bea
\label{eq:ChiralSuperfieldProjection}
\Phi  = \Phi_n + \Phi_{\bar{n}} \, ,
\eea
where $\Phi_n$ is defined as the on-shell superfield built from $\phi_n$ and $u_n = P_n u$, where $P_n$ is the projection operator given in \eref{eq:uProjections}, and $\Phi_{\bar{n}}$ is defined with $P_n \rightarrow P_{\bar{n}}$; see \eref{eq:collinearFreeChiral} below for a detailed expression. The action becomes
\begin{equation}
\mathcal{L} = \int \text{d}^4 \theta \Big(\Phi_n + \Phi_{\bar{n}} \Big)^\dagger \Big(\Phi_n + \Phi_{\bar{n}} \Big)= \int \text{d}^4 \theta \left( \Phi_n^\dagger \Phi_n + \Phi_n^\dagger \Phi_{\bar{n}} + \Phi_{\bar{n}}^\dagger \Phi_n + \Phi_{\bar{n}}^\dagger \Phi_{\bar{n}} \right)\,. 
\end{equation}
We can now supersymmetrically integrate out\footnote{\footnotesize{Recall we can exchange integration and differentiation over superspace: $\text{d}^2  \theta \leftrightarrow -\frac{1}{4} D D$, or simply $\text{d} \theta^1  \leftrightarrow D_1$.}} the anti-collinear superfield $\Phi_{\bar{n}}$. This yields a constraint equation
\begin{equation}
\label{eq:freeChiralconst}
\frac{\delta S \left( x, \theta, \theta^\dagger \right)}{\delta \Phi_{\bar{n}}^\dagger \left( x, \theta^\dagger \right)} =0 \,\,\,\,\, \Longrightarrow \,\,\,\,\, -\frac{1}{4} D D \Phi_{\bar{n}}-\frac{1}{4} D D \Phi_n = 0 \, .
\end{equation}
This completes the first two steps outlined in our general algorithm. Now, it must be the case that \eref{eq:freeChiralconst} encodes the anti-collinear fermionic component equation of motion \eref{eq:anticollEOM}. Indeed this is trivial to see; simply act on \eref{eq:freeChiralconst} with $\bar{D}_{\dot{\alpha}}$ to pick out the fermionic component: 
\begin{equation}
\label{eq:collinearFreeChiralConstraint}
\bar{D}_{\dot{\alpha}} DD \Phi_{\bar{n}} = -\bar{D}_{ \dot{\alpha}}  DD \Phi_{n} \,\,\,\,\,\, \Longrightarrow \,\,\,\,\,\,  4\, i \left( \bar{\sigma} \cdot \partial \right)^{\dot{\alpha} \alpha} D_\alpha \Phi_{\bar{n}} = -  4 \,i \left( \bar{\sigma} \cdot \partial \right)^{\dot{\alpha} \alpha} D_\alpha \Phi_{n}\, ,
\end{equation}
where we have used $D_{\alpha} \Phi |_{\theta = 0 = \theta^\dagger} = u_\alpha$.  Note that the upper index $\dot{\alpha} =1$ corresponds to the superspace coordinates that is present in the EFT.  We can then take the $\bar{D}^{\dot{1}}$ projection: 
\begin{equation}
\label{eq:checkcomponent}
 \left( \bar{\sigma} \cdot \partial \right)^{\dot{\alpha} 1} u_{\bar{n},1} = -  \left( \bar{\sigma} \cdot \partial \right)^{\dot{\alpha} 2} u_{n,2} \,\,\,\,\,\, \Longrightarrow \,\,\,\,\,\, u_{\bar{n},1} = \frac{\sqrt{2}\, \partial}{\nbp} u_{n,2}\,,
\end{equation}
where we have used $u_{\bar{n},2} = 0 = u_{n,1}$, and recall $\partial$ is now a LC derivative.

Additionally, \eref{eq:checkcomponent} motivates an ansatz for the superfield solution to the free equation of motion:
\bea
\label{eq:freeChiralSuperfieldAnsatz}
\Phi_{\bar{n}} = - \Phi_n - \frac{1 }{\left(\nbp \right) D_1}\bar{D}^{\dot{1}} D D \Phi_n \, ,
\eea
which satisfies the constraint equation \eref{eq:freeChiralconst} and the chirality condition (this is trivial to show; see App.~\ref{app:notation} for a summary of the relevant super derivative identities). 

This is the first time we have come across the unusual inverse superspace derivative.  In analogy with the non-local terms $1/\nbp$ present in the collinear fermion Lagrangian, we can define $1/D_\alpha$ operators by taking the Super-Fourier transform -- see \eref{eq:superfourier}. Additionally in SCET, integration by parts is well defined for the inverse derivative operator $1/\nbp$, because it can be cast in momentum space. By analogy we can extend this argument and use integration by parts on $1/D_\alpha$ operators. Recall also that $\int \text{d} \theta^\alpha D_\alpha (\textrm{. . . })$ is a total derivative in real space -- we will drop surface terms when using integration by parts under the assumption that they vanish sufficiently fast at infinity.  

At this stage, we have completed the first four steps of the general algorithm.  To finally derive the superspace action for the EFT of a collinear chiral multiplet, simply plug the ansatz \eref{eq:freeChiralSuperfieldAnsatz} back into the Lagrangian:
\bea 
\label{eq:freeChiralCollinearL}
\nonumber
\int \text{d}^4 \theta \left(\Phi_n^\dagger \Phi_n + \Phi_n^\dagger \Phi_{\bar{n}}  +  \Phi_{\bar{n}}^\dagger \Phi_n +  \Phi_{\bar{n}}^\dagger  \Phi_{\bar{n}} \right) &=& 
\int \text{d}^4 \theta  \left( D^1 \bar{D} \bar{D} \Phi^{\dagger}_n\right)  \frac{1}{(\nbp)^2 \bar{D}_{\dot{1}} D_1} \left( \bar{D}^{\dot{1}} DD \Phi_n \right) \\ \nonumber
 &=& \frac{1}{4} \int \text{d} \theta_1\, \text{d}\theta_2\, \text{d} \theta^{\dagger}_{\dot{1}} \,\text{d} \theta^{\dagger}_{\dot{2}}  \frac{i\, \Phi_{n}^\dagger \Box \Phi_n}{\left(\nbp\right) \bar{D}_{\dot{1}} D_1} \\ 
 &=&  \frac{1}{2}\int  \text{d}\theta^2 \,\text{d} \theta^{\dagger \dot{2}}\,  \Phi_{n}^\dagger \frac{i\, \Box}{\nbp} \Phi_n \, ,
\eea
where we have used various superspace derivative identities (see App.~\ref{app:notation}) and integrated by parts. 
Note that the result \eref{eq:freeChiralCollinearL}, depends only on the two superspace coordinates, $\theta^2$ and $\theta^{\dagger \dot{2}}$, which scale as $\mathcal{O}(\lambda^0)$. That is integrating out the anti-collinear fermion translates to integrating out the two high virtuality coordinates; $\theta^1\sim 1/\lambda$ and $\theta^{\dagger {1}} \sim 1/\lambda$. The EFT only depends on half the supercharges, which can be interpreted as the result of having integrated out half of superspace.  We refer to this SCET subsurface as collinear superspace~\cite{Cohen:2016jzp}.

The collinear superfield satisfies the chirality condition in the EFT:
\bea
D_2 \Phi_n^\dagger = 0 = \bar{D}_{\dot{2}} \Phi_n \, ,
\eea
which is solved by the following: 
\bea
\begin{array}{l}
\Phi_n = e^{-i\, \theta^2 \theta^{\dagger \dot{2}}  \nbp}\, \big( \phi(y) + \sqrt{2}\, \theta^2 u_2(y) \big) = \phi(x) + \sqrt{2}\, \theta^2 u_2(x) - i\, \theta^2 \theta^{\dagger \dot{2}} \, \nbp \,  \phi(x),  \\ [4pt]
\Phi_n^\dagger = e^{i\, \theta^2 \theta^{\dagger \dot{2}}  \nbp} \, \big( \phi^*(y)+ \sqrt{2}\, \theta^{\dagger \dot{2}} u_2^\dag(y) \big) = \phi^*(x) + \sqrt{2}  \, \theta^{\dagger \dot{2}} u_2^\dag(x) + i\,\theta^2 \theta^{ \dagger \dot{2}} \, \nbp \, \phi^*(x).
\end{array}
\label{eq:collinearFreeChiral}
\eea

Note that the auxiliary $F$-term, the $ \theta^1 \theta^2$ term in the full theory expansion, is not present. Counting degrees of freedom, we have a complex scalar and a single component of a Weyl spinor (which we can think of as an anti-commuting complex scalar); we have two fermionic and two bosonic degrees of freedom, consistent with an on-shell chiral multiplet.

To summarize the collinear superspace action for a non-interacting collinear chiral multiplet, which reduces to the expected component EFT Lagrangian \eref{eq:Lfree} for the free theory, is
\bea
\label{eq:collinearFreeChiralL}
 \mathcal{L} = \frac{1}{2}\int  \text{d}\theta^2 \text{d} \theta^{\dagger \dot{2}}\,  \Phi_{n}^\dagger \frac{i\,\Box}{\nbp} \Phi_n \,\,\,\,\,\, \Longrightarrow \,\,\,\,\,\, \mathcal{L} = u_{2}^\dag \left( i\,n \cdot \partial - \frac{\partial_{\perp}^2}{i \,\nbp} \right) u_{2} -  \phi^* \Box \phi\,.
\eea
Since only $\big\{Q_{2}, Q^\dagger_{\dot{2}} \big\} =  2\, i\,  \nbp$ is non-vanishing at leading order in the power counting parameter, \eref{eq:collinearFreeChiralL} is clearly supersymmetric under the subset of supersymmetry transformations that are linearly realized, namely \eref{eq:SUSYtransformationsChiSupermultiplet}.

 \subsection{Abelian Gauge Theory}
We now consider the vector multiplet of Abelian gauge theory. Since the degrees of freedom for an on-shell vector multiplet in LCG can be organized into a chiral multiplet, we will build on the intuition developed in Sec.~\ref{sec:freeWZcollinearsuperspace}.

The $\mathcal{N} = 1$ vector field obeys the reality condition $V = V^\dagger$, which in Wess-Zumino gauge can be written as the following:
\bea
V(y) &=& e^{-\frac{i}{2} \theta \sigma^\nu \theta^\dag \partial_\nu} \left(   - \theta \sigma^\mu \theta^\dagger\, A_\mu(x) + i \,\theta \theta \theta^\dag\, u^{\dag}(x) - i \,\theta^\dag \theta^\dag \theta \,u (x) \right) \nonumber \\  
&=& - \theta \sigma^\mu \theta^\dag \,A_\mu(y) + i \theta \theta \theta^\dag\, u^\dag(y) - i\, \theta^\dag \theta^\dag \theta\,u (y) - \frac{i}{2}\, \theta \theta \theta^\dag \theta^\dag \,\partial^\mu A_\mu (y) \, .
\eea
As with the chiral multiplet derivation, we omit auxiliary fields since the EFT is formulated on-shell.  SUSY dictates that the projection operators should act on the full vector multiplet, such that the superfield will obey the following decomposition:\footnote{\footnotesize{This is consistent since all anti-collinear gauge fields are Lorentz contracted in the vector superfield.}}
\bea
\label{eq:SuperfieldProjections}
V = V^\dagger = P_n\,V+P_{\bar{n}}\,V =  V_n + V_{\bar{n}}\,.
\eea 
The projection operators eliminate half of the fermionic spin states from the EFT; they enforce $u_{n,1} = 0 =u_{\bar{n},2}$. Therefore, the collinear and anti-collinear vector superfields (in terms of $x^\mu$ coordinates) are;
\bea
\label{eq:VectorSuperfields}
\nonumber
V_n  &=& - \,\, \theta^1 \theta^{\dag \dot{1}}\, n \cdot A_n - \sqrt{2}\, \left(\theta^1\theta^{\dag \dot{2}}  \alc_n^* + \theta^2 \theta^{\dag \dot{1}} \, \alc_n\right) + 2\,i\, \theta^1 \theta^2  \theta^{\dag \dot{2}}\, u^\dag_{n, \dot{2}} - 2\,i\, \theta^{\dag \dot{1}} \theta^{\dag \dot{2}}\, \theta^2\, u_{n,2}\,,    \\ 
V_{\bar{n}}   &=& -\,\, \theta^1 \theta^{\dag\dot{1}}\, n \cdot A_{\bar{n}} 
- \, \theta^2 \theta^{\dag\dot{2}}\, \bar{n} \cdot A_{\bar{n}} - \sqrt{2}\,\left( \theta^1 \theta^{\dag\dot{2}} \, \alc^*_{\bar{n}}  + \theta^2 \theta^{\dag\dot{1}}\,  \alc_{\bar{n}}\right) \nonumber \\ 
&& \quad + \,\, 2\,i\, \theta^1\theta^2 \theta^{\dag\dot{1}}\,u^\dag_{\bar{n}, \dot{1}}  - 2\,i\, \theta^{\dag\dot{1}}\theta^{\dag\dot{2}} \theta^1\, u_{\bar{n},1}   \,, 
\eea 
where we have explicitly expanded out the spinor indices, and written the gauge field $A^\mu (y)$ using the LCG notation introduced in Sec.~\ref{sec:LCSCET}.  Additionally, we have gauge fixed $\bar{n}\cdot A_n = 0$, but have not yet integrated out the non-propagating modes.

The action for the Abelian theory is 
\bea
\label{eq:fulltheoryabelian}
S = \int \text{d}^4 x\, \text{d}^2 \theta  \,\mathcal{W}^\alpha \mathcal{W}_\alpha + \int \text{d}^4 x\, \text{d}^2 \theta^\dag \,\bar{\mathcal{W}}_{\dot{\alpha}} \bar{\mathcal{W}}^{\dot{\alpha}} \, ,
\eea
where the chiral superfield $\mathcal{W}$ in Wess-Zumino gauge is given by 
\bea
\mathcal{W}_\alpha =-\frac{i}{4} \bar{D} \bar{D}D_\alpha \Big(V_n + V_{\bar{n}} \Big) \,.
\eea
To integrate out the anti-collinear vector superfield simply take the variation of the superspace action (for brevity, we take $z \equiv (x^\mu, \theta^\alpha, \theta^{\dagger \dot{\alpha}})$):
\bea
\nonumber
\frac{\delta S (z)  }{\delta V_{\bar{n}} (z' ) } &=& -\frac{1}{2} \int \text{d}^4 x\, \text{d}^2 \theta \frac{\delta \mathcal{W}^\alpha (x,\theta)}{\delta V_{\bar{n}} (z' )} \mathcal{W}_\alpha (x, \theta) - \frac{1}{2} \int \text{d}^4 x\, \text{d}^2 \theta^\dag \,\frac{\delta  \bar{\mathcal{W}}_{\dot{\alpha}} (x, \theta^\dagger)}{\delta V_{\bar{n}} (z')} \bar{\mathcal{W}}^{\dot{\alpha}} (x,\theta^\dagger) \\ \nonumber
&=& \frac{1}{32} \int \text{d}^4 x \,\text{d}^2 \theta\, \bar{D} \bar{D} D^\alpha \bar{D} \bar{D} D_\alpha (V_n + V_{\bar{n}}) + \frac{1}{32} \int \text{d}^4 x\, \text{d}^2 \theta^\dag\,  D D \bar{D}_{\dot{\alpha}} DD \bar{D}^{\dot{\alpha}}(V_n + V_{\bar{n}}) \\  
&=& -\frac{1}{8} \int \text{d}^4 x \,\text{d}^2 \theta\, \text{d}^2 \theta^\dag \left(  D^\alpha \bar{D} \bar{D} D_\alpha + \bar{D}_{\dot{\alpha}} D D \bar{D}^{\dot{\alpha}}  \right)(V_n + V_{\bar{n}})   = 0 \,.
\eea
Therefore, the constraint that yields the equation of motion for $V_{\bar{n}}$ is given by
\bea
\label{eq:VectorSuperfieldConstraint}
&&D^\alpha \bar{D} \bar{D} D_\alpha \left(V_n + V_{\bar{n}} \right)=  \left(-\!16\, \Box + 4\,i\, D^\alpha \left(\sigma \cdot \partial \right)_{\alpha \dot{\alpha}}\, \bar{D}^{\dot{\alpha}} \right) \left(V_n + V_{\bar{n}}\right)  = 0  \, .
\eea

Next, we can verify that the component equation of motion (for both $u_{\bar{n}}$ and $n\cdot A_n$) can be extracted from \eref{eq:VectorSuperfieldConstraint}. To find the equation of motion for the anti-collinear gaugino, simply project with $\bar{D}^{\dot{1}}$: 
\bea
\bar{D}^{\dot{1}} D^\alpha (\sigma \cdot \partial)_{\alpha \dot{\alpha}} \bar{D}^{\dot{\alpha}}V_{\bar{n}} = - \bar{D}^{\dot{1}} D^\alpha (\sigma \cdot \partial)_{\alpha \dot{\alpha}} \bar{D}^{\dot{\alpha}}V_n \,\,\,\,\, \Longrightarrow \,\,\,\,\, u_{\bar{n}, 1} = \frac{\sqrt{2}\, \partial^*}{\nbp} u_{n,2},
\label{eq:EOMgaugino}
\eea
which reproduces the expected equation of motion for the anti-collinear gaugino. The other choice $\bar{D}_{\dot{\alpha}} DD \bar{D}^{\dot{\alpha}} (V_n + V_{\bar{n}})$ yields the conjugate equation of motion for the anti-gaugino.  This shows how the superspace derivatives pick out the various components of the vector superfield  \eref{eq:VectorSuperfields}. 
 
Next, we check that \eref{eq:VectorSuperfieldConstraint} additionally reproduces the equation of motion for the unphysical gauge polarization $n\cdot A_n$.  It will be convenient to work with the chiral superfield $\mathcal{W}_\alpha$.  It is straightforward to rewrite the constraint equation as $D^\alpha \mathcal{W}_{n \alpha} + D^\alpha \mathcal{W}_{\bar{n} \alpha}=  0$. Recall that before going to LCG,  $D^\alpha \mathcal{W}_\alpha  =  \big(\sigma^{\mu \nu}\big)^\alpha_\alpha F_{\mu \nu} + 2 i \theta^\alpha (\sigma \cdot \partial)_{\alpha \dot{\alpha}} u^{\dagger \dot{\alpha}} + \dots\,$.  We can isolate $n\cdot A_n$ by taking the $\alpha, \dot{\alpha} = 1$ components of $\big(\sigma^{ \mu \nu}\big)^\alpha_\alpha F_{\mu \nu}$, then upon simplification this yields the expected result \eref{eq:eomfree}.

Working our way through the procedure outlined above, next we write down an ansatz for the equation of motion for the anti-collinear vector superfield: 
\bea
\label{eq:VectorAnsatz}
\,\, V_{\bar{n}}= - V_n -  \frac{1}{\nbp \,D_2 \,\bar{D}_1 \, D_1   }  \bar{D}_{\dot{2}} \,DD \,\Big(\bar{D}_{\dot{2}}\, D_1\,  V_n\Big)  - \frac{1}{\nbp  \,  \bar{D}_{\dot{2}}  \,D_1 \, \bar{D}_{\dot{1}}}  D_2\, \bar{D} \bar{D}\, \Big(D_2\, \bar{D}_{\dot{1}}\, V_n\Big) .\,\,\,\,\,\,\,\,\,\,\,\,\, 
\eea
Here both terms are required to ensure the reality condition $V_{\bar{n}} = V_{\bar{n}}^\dagger$. Using the relations such as $DD D_{\dot{\alpha}} DD = 0 = \bar{D}\bar{D} D_\alpha \bar{D}\bar{D}$, it can be shown that \eref{eq:VectorAnsatz} satisfies the constraint equation \eref{eq:VectorSuperfieldConstraint}. In addition, \eref{eq:VectorAnsatz} encodes various equations of motion.  For example, the gaugino constraint can be derived by acting with $\bar{D}_1 D_2 D_1$ on \eref{eq:VectorSuperfieldConstraint} to select $u_{\bar{n},1}$.  

In the LCG EFT the degrees of freedom $u_{2}$ and $\alc$ form a chiral superfield.  This is expressed in superspace by taking the appropriate projections on the vector superfield \eref{eq:VectorSuperfields}; for instance
\bea
\!\!\!\!\!\!\Phi\big(x^\mu\big) \equiv \bar{D}_{\dot{2}}\, D_1\, V_n \Big|_{\theta^1 = 0 = \theta^{\dagger \dot{1}}} = \sqrt{2} \,\alc^*(x^\mu)\! +2\,i\, \theta^2 u_{2}^\dag(x^\mu) -  i\, \sqrt{2}\, \theta^2 \theta^{\dagger \dot{2}} \,\nbp \alc^*(x^\mu) \, , 
\label{eq:ChiralFromVector}
\eea
which obeys the EFT chirality constraint: 
\bea
\bar{D}_{\dot{2}} \Phi = -    i\, \sqrt{2}\,  \frac{\partial}{\partial \theta^{\dagger \dot{2}}}\theta^2 \theta^{\dagger \dot{2}} \,\nbp \alc^{*} - i\, \sqrt{2} \,\theta^2\, \nbp    \alc^{*} =0\,.
\eea
For the anti-chiral multiplet, simply take the conjugate of \eref{eq:ChiralFromVector}. Therefore, the ansatz \eref{eq:VectorAnsatz} can be expressed in terms of the chiral and anti-chiral superfields,\footnote{\footnotesize{Note that in order to make contact with the chiral superfields we wrote down in the previous section we need to absorb the factor of $i$ into the gaugino, and do a superspace rotation to express the field in terms of $x^\mu = y^\mu - i \theta \sigma \cdot \partial \theta^\dag$.}} and we will see below that this simplifies calculations and allows us to leverage results from Sec.~\ref{sec:freeWZcollinearsuperspace}. This remarkable organization of the gauge degrees of freedom into a chiral multiplet is a well-known feature of working in LCG~\cite{Mandelstam:1982cb}. A different gauge choice would require a vector multiplet with the usual constraint  $V = V^\dagger$. 

The full theory action from \eref{eq:fulltheoryabelian}, can be rewritten as
\begin{align}
S = - 2 \int \text{d}^4 x\, \text{d}^2 \, \theta\, \text{d}^2 \, \theta^\dag\, \bar{D}_{\dot{1}}  D^\alpha \, (V_n + V_{\bar{n}}) \,  \bar{D}_{\dot{2}}  D_\alpha \, (V_n + V_{\bar{n}}) + \textrm{h.c.}  \, , 
\end{align}
where we have expressed \eref{eq:fulltheoryabelian} in terms of the vector superfields, integrated by parts, and exchanged superspace derivatives for integration of superspace coordinates.  Now we can plug in the ansatz superfield \eref{eq:VectorAnsatz}:
\begin{align}
\nonumber
\mathcal{L} \, & \supset    \int \text{d}^4 \, \theta  \left( \frac{1}{\nbp \, \bar{D}_{\dot{1}} } D^1 \bar{D} \bar{D} \, \Phi_n^\dag \right) \left(\frac{1}{\nbp \,D_1}  \bar{D}^{\dot{1}}DD \,\Phi_n   \right)   \\ 
&= \frac{1}{4} \int \text{d}\theta^2 \, \text{d} \theta^{\dagger \dot{2}} \,   \text{d} \theta^{\dagger \dot{1}}\, \text{d} \theta^1  \, \frac{1}{ \,D_1   \bar{D}_{\dot{1}}} \, \Phi_{n}^\dagger \, \frac{ \bar{D}\bar{D} D_2 \bar{D}_{\dot{2}} DD }{\left( \nbp \right)^2}\Phi_{n}=  \frac{1}{4}  \int  \text{d}\theta^2\, \text{d} \theta^{\dagger \dot{2}}  \,\Phi_{n}^\dagger \frac{i \,\Box}{\nbp}\Phi_n \, .
\label{eq:VectorMultCollinearSuperspace}
\end{align}
Once again we see that integrating out the anti-collinear superfield is equivalent to integrating out the two highly virtual superspace coordinates $\theta^1\sim 1/\lambda$ and $\theta^{\dagger {1}} \sim 1/\lambda$.  Note that in this derivation, it was important to recall that $\{D_2, \bar{D}_{\dot{2}} \} = -2 i \nbp \sim \mathcal{O}(\lambda^0)$ while $\{ D_1, \bar{D}_{\dot{1}} \} \sim \mathcal{O}(\lambda) $ and $\{ D_1, \bar{D}_{\dot{2}} \} \sim \mathcal{O}(\lambda^2) $; to leading order $D_1$ and $\bar{D}_{\dot{\alpha}}$ anti-commute, which led to many simplifications.

Expanding out component fields and integrating over superspace, we arrive at the component Lagrangian for the free non-Abelian theory in LCG; 
\bea
\mathcal{L}=   \frac{1}{4} \int  \text{d}\theta^2\, \text{d} \theta^{\dagger \dot{2}}  \,\Phi_{n}^\dagger \,\frac{i \,\Box}{\nbp}\Phi_n  \,\,\,\, \Longrightarrow \,\,\,\, u_{2}^\dag \left( i\,n \cdot \partial - \frac{\partial_{\perp}^2}{i\, \nbp} \right) u_{2} - \alc^* \Box \alc \, ,
\label{eq:superspaceToComponentsKineticTerm}
\eea
which is invariant under the SUSY transformations to leading order in SCET power counting parameter for the single component of an adjoint Weyl fermion and a complex scalar,\footnote{\footnotesize{The supersymmetries that remain linearly realized on the light cone are referred to as ``kinematical" supersymmetry~\cite{Belyaev:2009rj}. The remaining  ``dynamical" supersymmetries, associated with the superspace coordinates that have been integrated out, along with some of the Lorentz symmetries, are no longer manifest and are realized non-linearly. In other words, we may think of $Q_2$ as generating the kinematical supersymmetry, while the power suppressed $Q_1$ charges as generating the dynamical supersymmetry.}} \eref{eq:SUSYtransformationsChiSupermultiplet} with $\phi \rightarrow \alc$.

\subsection{A Candidate Supercurrent}

Before moving to collinear SYM in the next section, we will explore the possible existence of a supercurrent for the free vector model.  With a conserved supercurrent, we can then couple our collinear SUSY theory to gravity in a gauge invariant way.  This could be the first step in developing a soft-collinear theory of supersymmetric gravity.  We will present a current that is both conserved on-shell and reproduces the supercharges of the collinear SUSY theory.  However, we do not know how to derive this candidate supercurrent from a Noether procedure.  This could be a sign of an inconsistency inherent in this proposal, or it might point to the need for an improvement term to make it clear how the current descends from deeper symmetry principles.  Nevertheless, we believe this object is sufficiently compelling to warrant presenting it here.  

A candidate supercurrent for the free LCG vector multiplet (or a free chiral multiplet) is given by
\begin{align}
-iJ^\mu &= n^\mu \, u_n \,\bar n\cdot \partial  \, \alc^*+ u_n\, \partial_\perp^\mu  \alc^*-\Big[
\partial_{\perp}^\mu,u_n
\Big] \alc^*-\bar n^\mu \Big[
n\cdot \partial,u_n
\Big]  \alc^*  \,. 
\end{align}
This supercurrent is conserved: $\partial\cdot J = 0$, when the on-shell condition $p^2=0$ is imposed.  It has the corresponding conserved charge
\begin{equation}
Q = \int \text{d}(n\cdot x) \, \text{d}^2x_\perp\, \frac{\bar n \cdot J}{2}=i\int \text{d}(n\cdot x) \, \text{d}^2x_\perp\, u_n\,\big( \bar n\cdot \partial  \alc^*\big)\,,
\end{equation}
where $\bar n^\mu/2$ picks out the light-cone time and the domain of the integral is light-cone space.
Using the light-cone canonical (anti-)commutation relations 
\begin{align}
\Big[
\alc(n\cdot x, \, x_\perp), \, \bar n\cdot \partial\alc^*(n\cdot y,y_\perp)
\Big] &= i  \, \delta( n\cdot x- n\cdot y)\, \delta^{(2)}( x_\perp- y_\perp)\,,\\
\Big\{
u_n(n\cdot x, \, x_\perp), \, u_n^\dagger(n\cdot y,y_\perp)
\Big\}&=\frac{n\cdot \sigma}{2}  \, \delta( n\cdot x- n\cdot y)  \, \delta^{(2)}  ( x_\perp- y_\perp)\,,
\end{align}
it can be shown that the charge $Q$ and its conjugate $Q^\dagger$ generate the SUSY transformations of the fields $u,u^\dagger,\alc, \alc^*$.  For example, consider the SUSY transformation with parameter $\eta$ of the LCG collinear gauge field:
\begin{align}
\Big[\eta \,  Q,\alc\Big]=i\,\eta \int \text{d}(n\cdot x) \, \text{d}^2x_\perp\, u_n \, \Big[\bar n\cdot \partial \, \alc^*(x), \,\alc(y)\Big]=\eta \, u_n
\,,
\end{align}
as expected from \eref{eq:SUSYtransformationsChiSupermultiplet} with $\phi \rightarrow \alc$.

\section{Soft-Collinear SUSY Yang-Mills}
\label{sec:ymsect}
As we have now emphasized many times, SUSY SCET lives in collinear superspace.  In this section, we will apply the formalism developed so far to the case of $\mathcal{N}=1$ SYM.  It is interesting to note that our derivation reproduces previous results derived in so-called light cone superspace \cite{Gates:1980az}; an on-shell superspace that has been widely studied for instance in the context of deriving the UV finiteness of 4-D $\mathcal{N} = 4$ SYM~\cite{Mandelstam:1982cb}, and more recently in the context of supergravity~\cite{Kallosh:2009db, Kallosh:2008mq}.  While these models are well studied, to our knowledge our work is the first time an algorithm has been proposed for deriving such an action in a generic way.

\subsection{SUSY Yang-Mills in Collinear Superspace}
We now summarize the steps for deriving the collinear superspace Lagrangian for a non-Abelian gauge theory.  Since this derivation parallels the examples given in the previous section, we will keep the details to a minimum.

The superspace action of the full theory is;
\bea
S = \int \text{d}^4 x \, \text{d}^2  \, \theta \,  \textrm{Tr} \bigl[ \mathcal{W}^\alpha \, \mathcal{W}_\alpha \bigr] + \int \text{d}^4 x \, \text{d}^2 \, \theta^\dagger \,\textrm{Tr} \bigl[  \bar{\mathcal{W}}_{\dot{\alpha}} \,\bar{\mathcal{W}}^{\dot{\alpha}} \bigr] \, ,
\eea
where $\mathcal{W}_\alpha = - \frac{i}{4}\bar{D} \bar{D} e^{-V} D_\alpha e^{V}$ is a matrix valued chiral superfield.
We again decompose the full theory non-Abelian vector superfield into collinear and anti-collinear pieces $V = V_n + V_{\bar{n}}$: 
\begin{align}
&\mathcal{W}_\alpha = - \frac{i}{4}  \, \bar{D}  \bar{D} \, e^{- (V_n + V_{\bar{n}})} \, D_\alpha  \, e^{(V_n + V_{\bar{n}})} \,, 
& \qquad \bar{\mathcal{W}}_{\dot{\alpha}} = - \frac{i}{4} \, D  D \, e^{- (V_n + V_{\bar{n}})} \, \bar{D}_{\dot{\alpha}} \, e^{(V_n + V_{\bar{n}})}
 \, .
\end{align}
Expanding the exponents and enforcing Wess-Zumino gauge yields
\begin{align}
\nonumber
\label{eq:nonabelianW}
 \mathcal{W}_\alpha &= -\frac{i}{4} \, \bar{D} \bar{D} \, D_\alpha (V_n + V_{\bar{n}}) + \frac{i}{4} \, \bar{D} \bar{D} \,\Big[(V_n + V_{\bar{n}}), \, D_\alpha \, (V_n + V_{\bar{n}}) \Big] \,  \\ 
\Longrightarrow \quad  \mathcal{W}_{\alpha}^a &= -\frac{i}{4} \, \bar{D}  \bar{D} \, D_\alpha (V_n + V_{\bar{n}})^a + \frac{g}{4} \, f^{abc} \,\bar{D}  \bar{D} \, (V_n + V_{\bar{n}})^b \, D_\alpha \, (V_n + V_{\bar{n}})^c \, ,
\end{align}
where the second line is expressed in the adjoint representation with $\mathcal{W}_\alpha = 2 g_a t^a \mathcal{W}_\alpha^a$.

Next, we integrate out the anti-collinear superfield $V_{\bar{n}}$.  The variation of the action is
\bea
\label{eq:nonabeliandeltaS}
\frac{\delta  S (z)}{\delta  V_{\bar{n}}^e(z')} = 2  \int \text{d}^4 x \,  \text{d}^2 \, \theta \,  \frac{\delta \mathcal{W}^{a \alpha}}{\delta  V_{\bar{n}}^e(z')} \mathcal{W}^a_\alpha + 2 \int \text{d}^4 x \, \text{d}^2 \, \theta^\dagger \, \frac{\delta \bar{\mathcal{W}}^a_{\dot{\alpha}}}{\delta  V_{\bar{n}}^e(z')}  \bar{\mathcal{W}}^{a \dot{\alpha}}  =0 \, , 
\eea
with
\bea
\frac{\delta \mathcal{W}^{a \alpha}}{\delta  V_{\bar{n}}^e(z')}  = -\frac{i}{4} \, \bar{D}\bar{D} \, D^\alpha \, \delta^{ae} + \frac{g}{4} \, f^{abc} \, \bar{D}\bar{D} \, \delta^{eb} D^\alpha \, V^c + \frac{g}{4} \, f^{abc} \,\bar{D}\bar{D} \, V^b \, D^\alpha \,  \delta^{ec} \, .
 \label{eq:nonabeliandeltaW}
\eea

The constraint equation is then found by combining Eqs.~(\ref{eq:nonabelianW}), (\ref{eq:nonabeliandeltaS}), and (\ref{eq:nonabeliandeltaW}):
\begin{align}
0 \, = \, &- D^\alpha   \bar{D}\bar{D} \,  D_\alpha  \, V^e + ig \, f^{ace} \, \left(  D^\alpha \bar{D}\bar{D}\, V^c + \bigl[ D^\alpha, \,V^c \bigr] \, \bar{D}\bar{D}  \right) \, D_\alpha \, V^a  \notag \\ 
&-  g^2 f^{ace} f^{adh} \bigl[D^\alpha, V^c \bigl] \bar{D}\bar{D} V^d D_\alpha V^h  \,.
\label{eq:nonabelianconstraint}
\end{align}

After gauge fixing to LCG and integrating out the anti-collinear gaugino and unphysical gauge modes, the remaining two gaugino and two gauge degrees of freedom ($u_2$ and $\alc$) can be organized into a collinear chiral superfield $\Phi^a$:
\bea
 \Phi^a_n(x)  =  \sqrt{2} \,\alc^{*a}(x) \! +2\,i\, \theta^2 u_{2}^{\dag a}(x)  -  i\, \sqrt{2}\, \theta^2 \theta^{\dagger \dot{2}} \,\nbp \alc^{*a}(x) \, . 
\eea
Following the general algorithm of Sec.~\ref{sec:CollSuperspace} we now use the constraint equation \eref{eq:nonabelianconstraint} to construct the superspace Lagrangian of the EFT in LCG:\footnote{\footnotesize{Note that similar results are written down in the literature, for instance in \cite{Belitsky:2004yg} where the $\mathcal{N} = 4$ Lagrangian is reduced to $\mathcal{N}=1$.  Our approach is novel in the sense that we show that this is the equivalent $\mathcal{N}$ = 1 Lagrangian achieved by integrating out superspace coordinates.}}
\bea
\label{eq:VectorSuperspaceLagrangian}
\nonumber
 \mathcal{L} &=& \frac{i}{4} \int  \text{d} \theta^2 \, \text{d} \theta^{\dagger \dot{2}} \, \biggl[ \Phi_n^{\dagger a} \frac{\Box}{\nbp} \Phi_n^a +  g\,\sqrt{2}\, f^{abc} \Big(  \Phi^a_n  \Phi_n^{\dagger b} \frac{\partial}{\nbp} \Phi_n^c -  \Phi_n^{\dagger a}  \Phi_n^b \frac{\partial^*}{\nbp} \Phi_n^{\dagger c}   \Big) \\ 
&&\qquad\qquad\qquad\,\,\,\,\,\,\,\,+ 2\, g^2\, f^{abc}f^{ade} \frac{1}{\nbp} \Big( \Phi_n^b\, \bar{D}_{ \dot{2}}  \Phi_n^{\dagger c}  \Big) \frac{1}{\nbp} \Big(  \Phi_n^{\dagger d} \,D_2  \Phi_n^e \Big) \bigg] \, .
\eea
\noindent We refer the reader to App.~\ref{app:intWZcollinearsuperspace} for an example derivation of interaction terms in collinear superspace.  This theory as formulated in collinear superspace is manifestly supersymmetric under half of the supersymmetries of the original theory, \emph{i.e.}, the non-Abelian generalization the LCG SUSY transformations given in \eref{eq:SUSYtransformationsChiSupermultiplet} with $\phi \rightarrow \alc$, which are linearly realized in the EFT. Therefore, \eref{eq:VectorSuperspaceLagrangian} is invariant under transformations generated by the charges $Q_2$ and $Q_{\dot{2}}^\dagger$.

\subsection{SUSY Yang-Mills in Components}
Now that we have \eref{eq:VectorSuperspaceLagrangian} -- the collinear superspace Lagrangian for a vector superfield in LCG -- we will show that it reproduces the expected component LCG Lagrangian Eqs.~(\ref{eq:LGsummary}) and (\ref{eq:LuSummary}).
The kinetic term of \eref{eq:VectorSuperspaceLagrangian}, encodes the usual kinetic EFT Lagrangian for the single component of a Weyl fermion and a complex scalar as was shown in \eref{eq:superspaceToComponentsKineticTerm}.  In the rest of this section, we will demonstrate that the $\mathcal{O}(g)$ terms reproduce the component Lagrangian, call these terms $\mathcal{L}_{\mathcal{O}(g)}$.  For the sake of brevity, we will not show the same exercise for the $\mathcal{O}(g^2)$ terms, although we have checked that these agree as well.

To begin, we need identities such as 
\begin{align}
\Phi_n^a \Phi_n^{\dagger b}   \frac{\partial}{\nbp} \Phi_n^c &=2 \Big(\alc^{*a}\alc^b -i\sqrt{2} \theta^{\dagger \dot{2}} \alc^{*a} u^b_2 + i \theta^2 \theta^{\dagger \dot{2}} (\alc^{*a} \nbp \alc^b) + i\sqrt{2} \theta^2 u^{\dag a}_2 \alc^b +2 \theta^2 \theta^{\dagger \dot{2}} u^{\dag a}_2 u^b_2 \nonumber \\ 
&\, - i \theta^2 \theta^{\dagger \dot{2}}  ( \alc^b \nbp \alc^{*a})\Big) \Big( \sqrt{2} \,  \frac{\partial}{\nbp}  \alc^{*c} \! -2i \theta^2  \frac{\partial}{\nbp} u_{2}^{\dag c}  +  i \sqrt{2} \theta^2 \theta^{\dagger \dot{2}} \partial \alc^{*c}  \Big).
\end{align}  
Expanding in components and integrating over superspace yields
\begin{align}
\label{eq:collL2}
\frac{1}{g}\mathcal{L}_{\mathcal{O}(g)} &= f^{abc}\Big( \alc^{*a} \alc^b \partial \alc^{*c} +   \alc^a \alc^{*b} \partial^* \alc^c \Big) \\ \nonumber
&+  f^{abc} \Big( \frac{\partial}{\nbp} \alc^{*c} + \frac{\partial^*}{\nbp} \alc^c \Big)  \Big( \alc^b(\nbp \alc^{*a})   + \alc^{*b} (\nbp \alc^a) \Big) \nonumber \\ 
&- 2\,i f^{abc} \left(\!\alc^{*a} u^b_2 \frac{\partial}{\nbp} u^{\dag c}_2 -  u^{\dag a}_2 u^b_2 \frac{\partial}{\nbp} \alc^{*c} \!  \right) + 2\,i f^{abc} \left(\!\alc^a u^{\dag b}_2 \frac{\partial^*}{\nbp} u^c_2 - u^a_2 u^{\dag b}_2 \frac{\partial^*}{\nbp} \alc^c  \! \right). \nonumber
\end{align}
By permutation of asymmetric indices the first term becomes $- f^{abc}  \alc^{*b} \alc^c (\partial^* \alc^a - \partial \alc^{*a})$, while the second term reduces to
$f^{abc} /\nbp (\partial \alc^{*c} +  \partial^*\alc^c)  \Big[ \alc^b(\nbp \alc^{*a})   + \alc^{*b} (\nbp \alc^a) \Big] = - f^{abc} /\nbp  (\partial \alc^{*b} + \partial^* \alc^b)  \Big[ \alc^c(\nbp \alc^{*a})   + \alc^{*c} (\nbp \alc^a) \Big] $. With these substitutions, we find exact agreement with the relevant terms in \eref{eq:LGsummary}. Now consider the gauge-gaugino interaction given by the final line of \eqref{eq:VectorSuperspaceLagrangian}. After some standard manipulations (integration by parts, permutation of gauge indices, and anti-commutation of the collinear fermion fields), we find that these terms reproduce the second line of \eref{eq:LuSummary}. 

This verifies that the full LCG super Yang Mills theory, formulated in collinear superspace \eref{eq:VectorSuperspaceLagrangian}, reproduces the expected component Lagrangian for LCG SCET with an adjoint fermion, \eref{eq:LGsummary} and \eref{eq:LuSummary}. The SUSY invariance can be checked by showing that the component Lagrangian is unchanged by the transformations given in \eref{eq:SUSYtransformationsChiSupermultiplet} with $\phi \rightarrow \alc^a$ and $u_2 \rightarrow u^a_2$.

\newpage

 \section{The Collinear Wess-Zumino Model}
\label{sec:WZModel}
Thus far we have focused our study of the interplay between SUSY and SCET in the context of the collinear limit for the theory of a single vector multiplet.   As a next step towards understanding the broader applicability of this formalism, it is interesting to study the simplest interacting theory involving a single chiral superfield, the Wess-Zumino model.

The main result presented in this section will be an argument that the Lagrangian for the propagating degrees of freedom that make up the (single flavor) Wess-Zumino model, as derived following Sec.~\ref{sec:twocomp}, does not account for the full infrared divergence structure of the full theory.  There are a number of ways to understand why this must have been the case, and elucidating these will provide us with the opportunity to explore the kinds of cross checks that one would want to perform when deriving a new SCET.  We will also provide an argument that RPI acting on this Lagrangian must be realized non-linearly, again in contrast with the theory for gauge interactions.  Finally, we will argue that many of these features could change in the presence of multiple flavors.

The massless Wess-Zumino model can be simply expressed in superspace as
\bea
\mathcal{L}=\int \text{d}^4 \theta\, \Phi^\dagger \Phi + \int \text{d}^2 \theta\, \frac{y}{3!}\Phi^3  + \textrm{h.c.}\,,
\eea
where $\Phi(y) = \phi + \sqrt{2}\theta \psi + \theta^2 F$ is a chiral superfield consisting of a complex scalar $\phi$, a left-handed Weyl fermion $\psi$, and an auxiliary field $F$.  Expanded in components and integrating out the auxiliary field yields
\begin{align}
\mathcal{L} =-\phi^* \Box \phi  + i \,\psi^\dag \bar{\sigma}^\mu\, \partial_\mu \psi  - \frac{|y|^2}{4} |\phi|^4 - \left(\frac{y}{2} \,\phi\, \psi\, \psi +\frac{y^*}{2}\, \phi^\dag \,\psi^\dag \,\psi^\dag \right)\,.
\label{eq:WZComponentLagrangian}
\end{align}
The rest of this section is devoted to exploring the IR of this Lagrangian.

\subsection{Infrared Divergences}
\label{sec:IRstructureYukawa}
As with the Yang-Mills case, we begin by exploring the IR singularity structure of a theory with a Yukawa coupling between a massless scalar and a massless Weyl fermion. This model has collinear singularities, motivating our attempt to write down an effective theory that describes the collinear limit.  However, we will see that the full theory has IR divergences which are not captured by the Lagrangian for the propagating degrees of freedom of the single flavor collinear Wess-Zumino EFT. 

The Yukawa coupling in \eref{eq:WZComponentLagrangian} leads to the following interaction vertex:
\begin{equation}
\raisebox{-0.15\height}{\includegraphics[width=2.5cm]{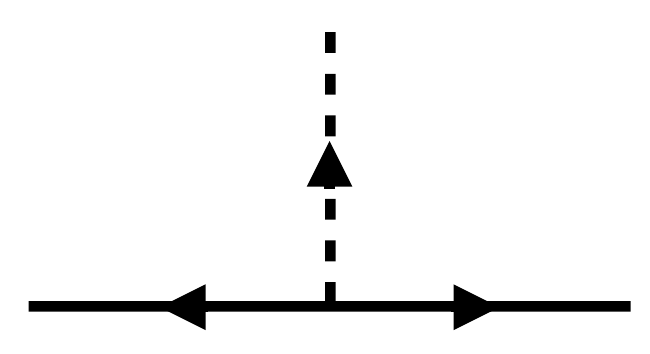}}= -i\,y\,.
\end{equation}
To explore the IR, we analyze various splitting amplitudes in the soft or collinear limits.  

\vspace{10pt}
\noindent \textbf{Scalar emission:} For the emission of a scalar from a fermion current, the amplitude is
\begin{align}
\raisebox{-0.25\height}{\includegraphics[width=3.7cm]{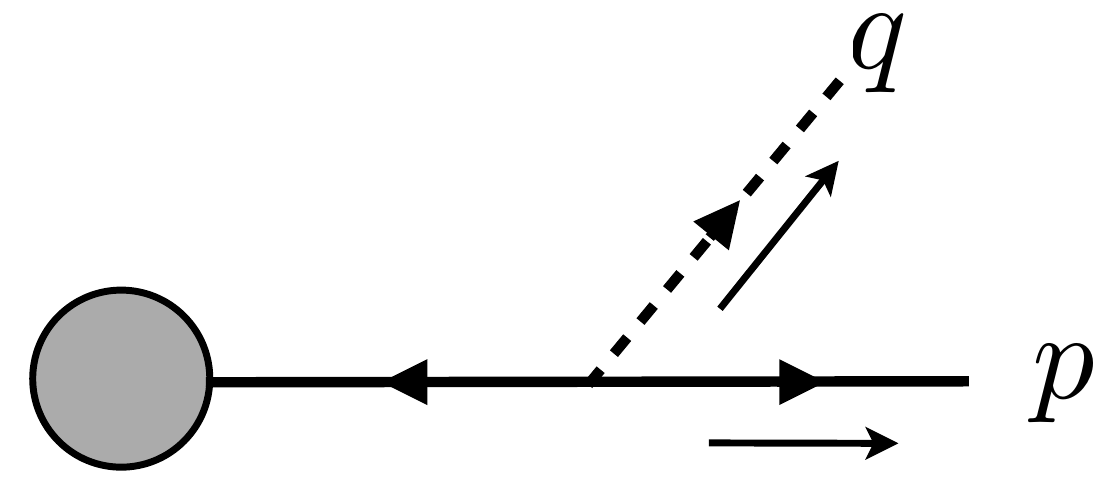}} &= (-i\,y) x^{\dagger}_{\dot{\alpha}} (p,s)\frac{i(p+q) \cdot \bar{\sigma}^{\dot{\alpha} \beta}}{2\,p\cdot q}  {\cal M}_{\beta}(p+q)  \notag \\
&= y\, x^{\dagger}_{\dot{\alpha}} (p,s) \frac{q \cdot \bar{\sigma}^{\dot{\alpha} \beta}}{2\,p\cdot q}  {\cal M}_{\beta}(p+q)  \,, 
\label{eq:ferm_amp}
\end{align}
where ${\cal M}_{\beta}(p+q)$ is the fermion current and in the second line we have used the Weyl equation of motion $ x^\dagger (p) p \cdot \bar{\sigma} =0$. 
It is clear at the diagrammatic level that this interaction flips the helicity. 

We can explore the collinear and soft limits to see if there are any IR divergences.  This is most straightforward to do using \eref{eq:ferm_amp}.  We can write $\mathcal{M}_{\psi \rightarrow \psi\phi} = x_{\dot{\alpha}}^\dag \mathcal{M}^{\dot{\alpha}}_{\psi \rightarrow \psi\phi} = x_{\dot{2}}^\dag \mathcal{M}^{\dot{2}}_{\psi \rightarrow \psi\phi} $, where the last equality is due to the fact that only $x^\dag_{\dot{2}}$ survives in the collinear limit.  Then the scalings are
\begin{align}
\mathcal{M}^{\dot{2}}_{\psi \rightarrow \psi\phi} \sim \frac{1}{2 \,p\cdot q} q\cdot\bar{\sigma}^{\dot{2}\beta}  {\cal M}_{\beta}(p+q) \quad\xrightarrow[\text{collinear}]{\makebox[0.8cm]{}} \quad
\frac{1}{\lambda}\,,
\end{align}
where we have taken the collinear limit for both $p$ and $q$ and truncated to the leading divergence.  Similarly, taking $p$ collinear and $q$ soft yields
\begin{align}
\mathcal{M}^{\dot{2}}_{\psi \rightarrow \psi\phi} \sim \frac{1}{2\, p\cdot q} q\cdot\bar{\sigma}^{\dot{2}\beta} {\cal M}_{\beta}(p+q) \quad\xrightarrow[\text{soft}]{\makebox[0.8cm]{}} \quad
1\,,
\end{align}
We see that there are collinear but not soft divergences.

\vspace{10pt}
\noindent \textbf{Fermion splitting:}  For the case of a scalar current splitting to two fermions, we find the amplitude
\begin{align}
\raisebox{-0.45\height}{\includegraphics[width=3.2cm]{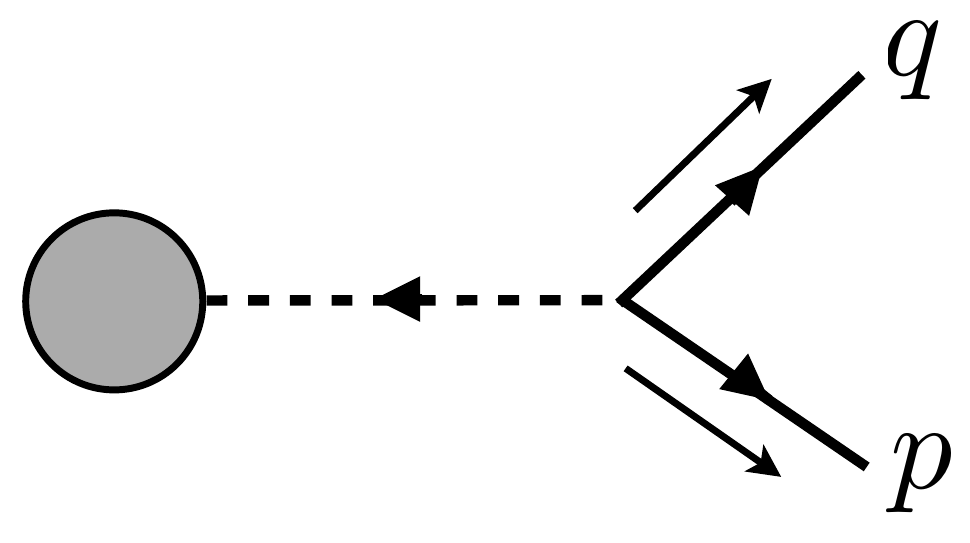}} &=y\frac{1}{2\,p\cdot q} x^\dag_{\dot{\alpha}} (p,s) x^{\dag{\dot{\alpha}}}(q,s){\cal M}(p+q)\,.
\end{align}
Assuming that the momentum flowing through the current $p+q\sim 1$, there is no soft divergence.  With this assumption, the soft limit can only occur if one of the two external fermions goes soft.  However, this violates fermion number conservation and so the amplitude cannot be singular in the soft region of phase space.  The amplitude can be soft divergent if both fermions go soft simultaneously, but then the momentum flowing through the current is small.  

Next, we take the limit in which the fermions and scalar all become collinear.  The scaling of the spinor product $x^{\dagger}_{\dot{\alpha}} (p,s) x^{\dag\dot{\alpha}}(q,s)$ can be derived by noting that
\begin{align}
\sum_\text{spins}\tr\big[x^\dagger(p) x^\dag(q) x(q) x(p)\big] &= - (p\cdot \bar{\sigma})(q\cdot \bar{\sigma}) =(\bar n \cdot p)(n\cdot q) \,, \nonumber
\end{align}
which implies
\begin{equation}
x^\dag(p) x^\dag(q) \sim \sqrt{
(\bar n \cdot p)(n\cdot q)} \sim \lambda \,.
\end{equation}
Hence, in the this limit
\begin{equation}
\label{eq:1to2collinear}
y \frac{1}{2\,p\cdot q} x^\dag  (p,s) x^{\dagger }(q,s){\cal M}(p+q)\sim \frac{1}{\lambda}\,,
\end{equation}
demonstrating that the splitting of a scalar to collinear fermions is singular.

We see that in both examples of $1\to2$ splittings, the IR singularities contained in \eref{eq:1to2collinear} require the presence of more than one fermion spin state. Therefore, we do not expect these singularities to manifest in the Lagrangian of the EFT.  In particular, it should not be possible to write down any tri-linear interaction terms, without the inclusion of an explicit external source as will be discussed in Sec.~\ref{sec:collWZL}. This motivates us to study the $1\to3$ collinear structure of the full theory IR singularities to verify that it agrees with the non-trivial EFT couplings which appear in the SCET Lagrangian. 

\vspace{10pt}
\noindent \textbf{Double scalar emission:}  Next, we focus our attention on the IR divergences that result from the $1\rightarrow 3$ process of a collinear fermion emitting a pair of collinear scalars:  
\begin{align} \label{eq:123full}
\raisebox{-0.24\height}{\includegraphics[width=4cm]{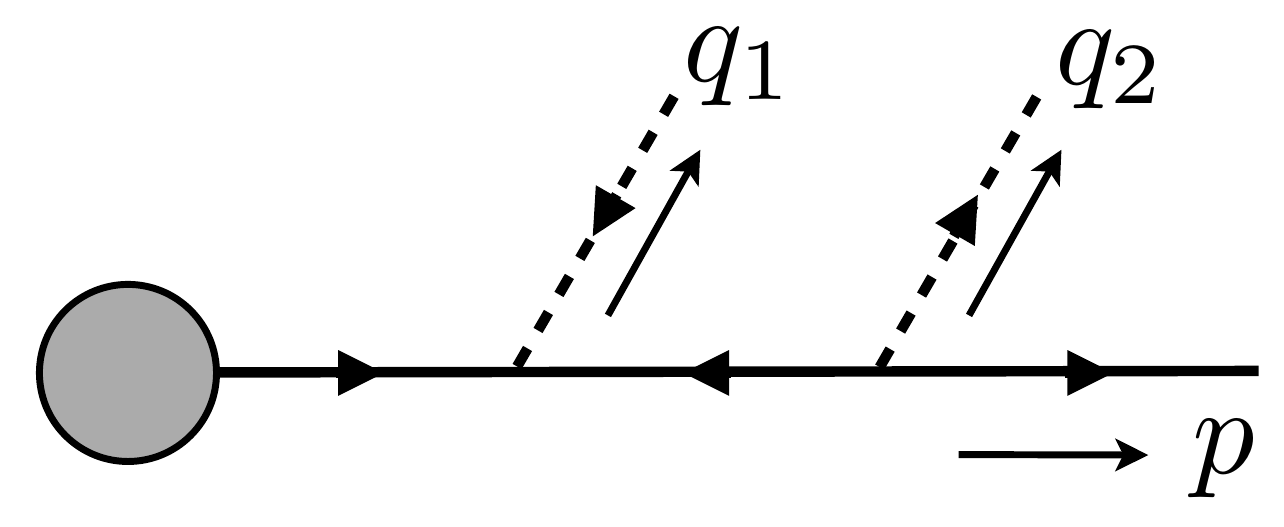}}
=\frac{|y|^2}{4}x^\dagger(p)\frac{(p+q_2)\cdot \bar \sigma}{2\,p\cdot q_2}\frac{(p+q_1+q_2)\cdot \sigma}{(p+q_1+q_2)^2} \,{\cal M}(p+q)\,.
\end{align}
Without loss of generality, we  assume $p$ and $q_2$ have no net $\perp$ momentum:
$(p+q_2)_\perp=0$.

\subsection{Collinear Lagrangian for the Wess-Zumino Model}
\label{sec:collWZL}

We are interested in the properties of the $1\rightarrow 3$ splitting function
\begin{align}
\mathcal{S}_{1\rightarrow 3} = \frac{(p+q_2)\cdot \bar \sigma}{2\,p\cdot q_2}\frac{(p+q_1+q_2)\cdot \sigma}{(p+q_1+q_2)^2} \,,
\end{align}
in the limit when the particles become collinear.  Expanding in light-cone coordinates, we have
\begin{align}
\mathcal{S}_{1\rightarrow 3}=&\frac{\left[
\frac{n\cdot \bar \sigma}{2}\bar n\cdot(p+q_2)+\frac{\bar n\cdot \bar \sigma}{2} n\cdot(p+q_2)
\right]}{\bar n\cdot(p+q_2)\,n\cdot (p+q_2)}\times\notag\\
&\frac{\left[
\frac{n\cdot \sigma}{2}\bar n\cdot(p+q_1+q_2)+\frac{\bar n\cdot \sigma}{2}n\cdot(p+q_1+q_2)+q_{1\perp}\cdot \sigma_\perp
\right]}{(p+q_1+q_2)^2}\,.
\end{align}
In the collinear limit, the constraint on the spinor $x^\dagger(p)$ is
\begin{equation}
x^\dagger(p) \frac{n\cdot \bar \sigma}{2} = 0\,,
\end{equation}
which eliminates terms in $\mathcal{S}_{1\rightarrow 3}$.  Additionally, the $\sigma_\perp$ matrix flips helicity, and so projects onto components of the matrix element ${\cal M}(p+q)$ that are zero in this collinear limit.  Using these constraints, the splitting function dramatically simplifies:
\begin{equation}
\mathcal{S}_{1\rightarrow 3}=\frac{\bar n\cdot \bar \sigma}{2}\frac{n\cdot \sigma}{2}\frac{1}{\bar n\cdot(p+q_2)}\frac{
\bar n\cdot(p+q_1+q_2)
}{(p+q_1+q_2)^2}\,,
\end{equation}
where we also have used $(\bar n\cdot \bar \sigma)(\bar n\cdot \sigma) = 0$.  Inserting this expression into the matrix element, we then find, in the collinear limit,
\begin{equation}\label{eq:S1to3FullTheory}
\frac{|y|^2}{4}\,x^\dagger(p)\,\mathcal{S}_{1\rightarrow 3}\,{\cal M}(p+q) = \frac{|y|^2}{4}x^\dagger(p)\left[
1+\frac{\bar n\cdot q_1}{\bar n\cdot(p+q_2)}
\right]\frac{1}{(p+q_1+q_2)^2}{\cal M}(p+q)\,.
\end{equation}
The term in the square brackets scales like $\lambda^0$, and the propogator factor scales like $\lambda^{-2}$.  We see that there is a collinear divergence associated with the $1\rightarrow 3$ splitting function that must be reproduced by a valid Wess-Zumino EFT.  Next, we will analyze the candidate interacting SCET to reproduce this divergence. 

\begin{figure}[!]
	\begin{center}
		\includegraphics[width=12cm]{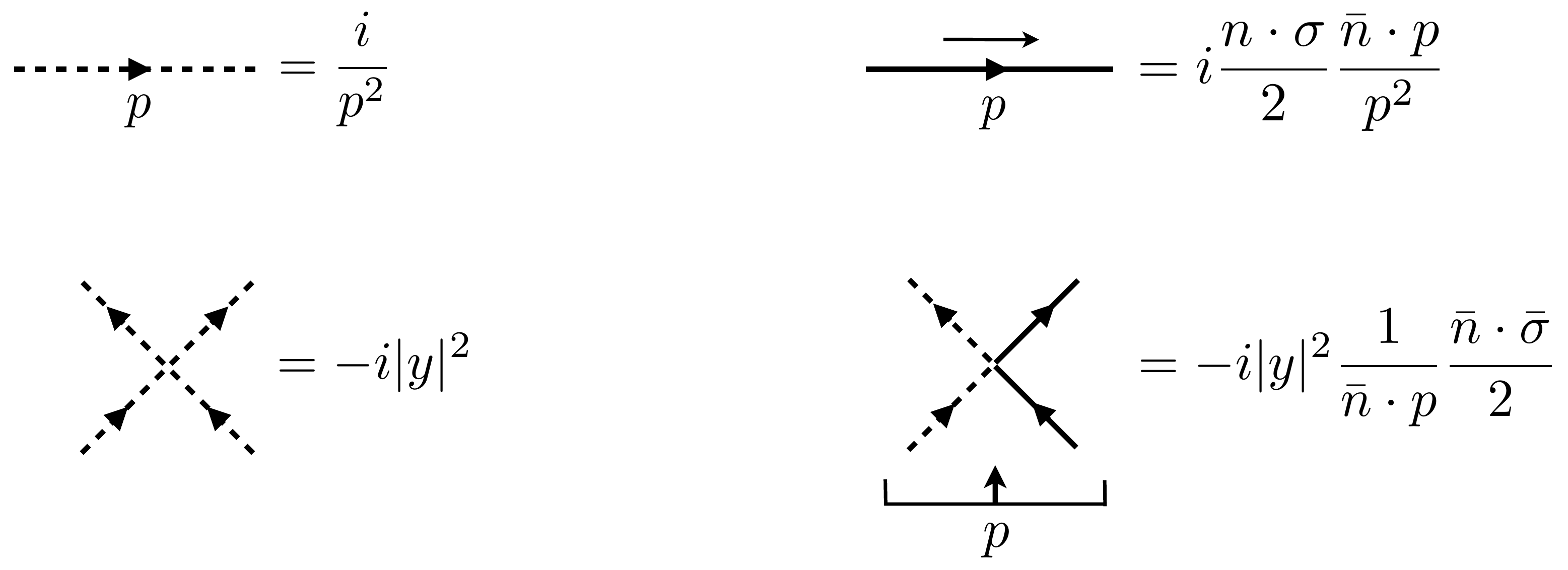}
	\end{center}
 	\caption{
Feynman rules from the supersymmetric collinear Lagrangian of \eqref{eq:colllag}.  The momentum $p$ is incoming. 
	}
 	\label{fig:scet0_feyn}
 \end{figure}

We now construct a collinear EFT for the interacting Wess-Zumino model. Starting with \eref{eq:WZComponentLagrangian}, we can run this model through the SCET procedure.  Separating out collinear and anti-collinear fermionic modes via the two-component projection operators $u = P_{n} u + P_{\bar{n}} u= u_n + u_{\bar{n}}$, and then integrating out the anti-collinear fermion by solving for the classical equation of motion for the anti-collinear field
\bea
\label{eq:IntWZcompEOM}
u_{\bar{n},\alpha} = \left(\sigma \cdot \partial_\perp\right)_{\alpha \dot{\alpha}} \frac{1}{\bar{n} \cdot \partial} \left(\frac{\nbsb}{2}\right)^{\dot{\alpha}\beta} u_{n,\beta} + y^* \left(\frac{\nbs}{2}\right)_{\alpha \dot{\beta}} \frac{1}{i\, \bar{n}\cdot \partial} \Bigl[ \phi_n^*   u_n^{\dagger \dot{\beta}} \Bigr] \,,
\eea
we arrive at the following collinear Lagrangian: 
\begin{align}
\!\!{\cal L}_n\!= -\phi_n^* \Box \phi_n + u_n^\dagger \left(
i\,n\cdot \partial - \frac{\partial_\perp^2}{i\,\bar n \cdot \partial}
\right)
\frac{\bar n\cdot \bar \sigma}{2}u_n  -\frac{|y|^2}{4}|\phi_n|^4 -|y|^2\phi_n^* u_n^\dagger \frac{1}{i\,\bar n\cdot \partial}\frac{\bar n\cdot \bar \sigma}{2}u_n\phi_n\,.
\label{eq:colllag}
\end{align}
The soft Lagrangian is trivially identical to the full-theory Lagrangian at leading power.  Since we will use them below, we provide the Feynman rules for this Lagrangian in \Fig{fig:scet0_feyn}.

One interesting feature of this Lagrangian is the absence of a cubic coupling.  Recall that in the full theory Lagrangian, this cubic coupling has the form
\begin{align}
\label{eq:WZinteractionComponents}
-\mathcal{L}\supset \frac{y}{2} \phi \, u^\alpha u_\alpha   =  \frac{y}{2} \phi_n \left( u_{n}^{\alpha} +  u_{\bar{n}}^{\alpha}\right) \left( u_{n \alpha} +  u_{\bar{n} \alpha}\right) = y\, \phi_n\,   u_{\bar{n}}^\alpha u_{n \alpha} =  y \,\phi_n \, u_{\bar{n}1} \,u_{n2} \,.
\end{align}
 The absence of an analogous coupling in \eref{eq:colllag} can be understood from the anti-commuting nature of the spinors. Fourier transforming to momentum space makes it is clear that this term vanishes due to the anti-commuting nature of the spinors:
\bea
x_{n,\dot{2}}^\dagger x^{\dagger}_{n, \dot{2}} \,   \frac{\left( \bar{\sigma}\cdot p_\perp \right)^{21}}{\bar{n}\cdot p} \phi^*  = 0
\eea
Physically the issue is clear  since the full theory interaction involves a helicity flip.  However, the fermionic degree of freedom needed for these processes is precisely the one integrated out in the collinear limit.  This manifests in the relationship between helicity and the EFT projection operators.  The EFT Lagrangian alone does not contain the appropriate degrees of freedom to reproduce the IR of the full theory.

\subsubsection*{Infrared Structure: $\mathbf{1\rightarrow 2}$}
It is possible to account for the divergences from $1\rightarrow 2$ splittings that would result from a cubic coupling by working with a background source for the anti-collinear fermions.  An external source couples to the infinite tower of EFT Lagrangians defined for any possible choice of $n$ direction.  This allows us to model the helicity flipping interaction in the collinear limit, which is not present within any given EFT sector on its own for the reasons given previously.

What follows is a  derivation of the interacting theory in the presence of a background source $J$.  Begin with the fermion Lagrangian
\bea
\mathcal{L}(J) = \mathcal{L} + u_{\bar{n}} \,J + u_{\bar{n}}^\dagger \,J^\dagger\,,
\eea
where $\mathcal{L}$ is the Lagrangian given in \eref{eq:WZComponentLagrangian}, and $J$ is a non-Hermitian operator with leading order scaling dimension that models a fermionic source term.  Solving for the equation of motion for $u_{\bar{n}}$ and plugging it back into the Lagrangian yields  
\bea
\label{eq:genericfermionL}
\mathcal{L}_u &=& \mathcal{L}_n   - y\,J^\dagger\frac{i}{\nbp} \frac{\bar{n}\cdot \bar \sigma}{2}  \phi_n\, u_{n} + y^*\phi_n^*\, u_{n}^\dagger\frac{i}{\bar n\cdot \overleftarrow\partial} \frac{\bar{n}\cdot  \bar \sigma}{2} \,J \\ \nonumber
&- & J^\dagger \frac{\bar{n}\cdot \bar\sigma}{2}   \frac{\sigma\cdot \partial_\perp}{\bar{n}\cdot \partial} u_n^\dagger  - J \frac{\sigma \cdot \partial_\perp}{\bar{n}\cdot \partial} \frac{\bar{n}\cdot \bar{\sigma}}{2} u_n  - J^\dagger  \frac{i}{\bar{n}\cdot \partial} \frac{\bar{n}\cdot \bar \sigma}{2} J \,.
\eea
Here, $\mathcal{L}_n$ is the collinear Lagrangian for the propagating mode $u_n$, \eref{eq:colllag}.  This enables our ability to write down the matrix element for $1\to 2$ splitting in this collinear theory with sources:
\begin{align} \label{eq:123full}
\raisebox{-0.32\height}{\includegraphics[width=4cm]{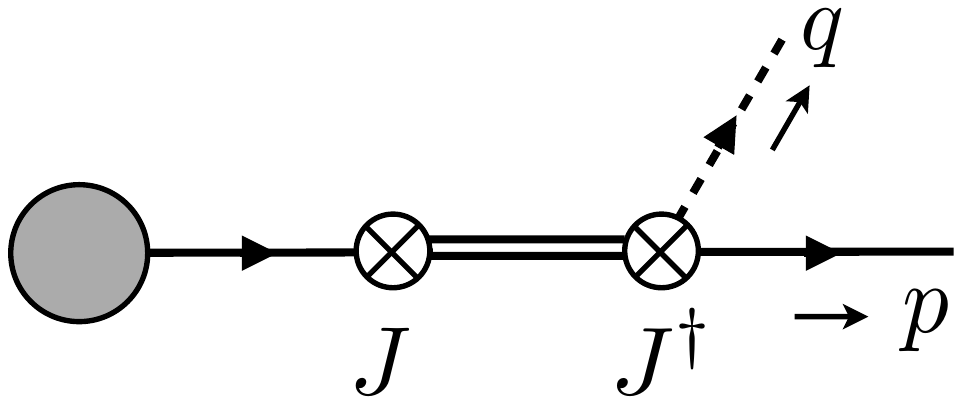}}
= &\left[ -iy x_n^\dagger\frac{1}{\bar n\cdot (p+q)}\frac{\bar n\cdot \bar \sigma}{2} \right] \,\left( -i\bar n\cdot (p+q)\frac{n\cdot \sigma}{2} \right) \\
& \times \left[ i  \frac{\sigma_\perp\cdot(p+q)_\perp}{\bar n\cdot(p+q)}  \frac{\bar n\cdot \bar \sigma}{2} \right]  \left(i \frac{n\cdot \sigma}{2}\frac{\bar n\cdot(p+q)}{2\,p\cdot q}  \right) {\cal M}\nonumber \\
= &\, y\,x_n^\dagger \frac{\bar \sigma_\perp \cdot(p+q)_\perp}{2\,p\cdot q}{\cal M}\,.\nonumber
\end{align}
In this expression, vertices from the Lagrangian \eref{eq:genericfermionL} are written within the square brackets, and propagators are within parentheses, where the double line connecting $J$ to $J^\dag$ represents the current-current 2-point function.  For example, the first factor in square brackets on the right of \eref{eq:123full} is the vertex from the term in the Lagrangian that couples the current $J^\dagger$ to the fields $\phi_n$ and $u_n$.  Note that the seemingly confusing $\sigma$-matrix structure in the above is valid: the two-point function coming from $J^\dagger J$ carries spin structure which is implicit in \eref{eq:123full}. This conspires to reproduce the correct structure for instance using the identity $ \bar{\sigma}_{\dot{\alpha}\alpha}^\mu = \epsilon_{\alpha \beta} \epsilon_{\dot{\alpha} \dot{\beta}} \sigma^{\mu \beta \dot{\beta}}$.

This result can be directly compared to the collinear limit of \eref{eq:ferm_amp}.  We can choose the frame where $p_\perp=0$; that is the external fermion lies exactly along the $n$ direction.  In \eref{eq:ferm_amp}, making this choice leaves only the product $q_\perp\cdot \bar \sigma_\perp$ at leading power in the collinear limit, demonstrating that the two approaches agree.

\subsubsection*{Infrared Structure: $\mathbf{1\rightarrow 3}$}
\begin{figure}[t!]
	\begin{center}
		\includegraphics[width=10cm]{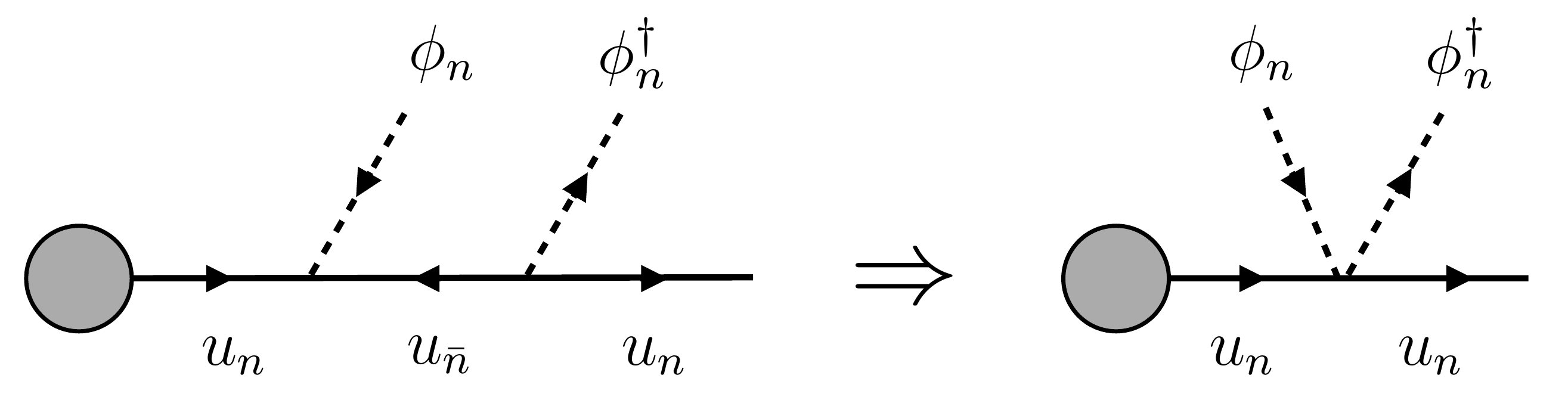}
	\end{center}
 	\caption{
Integrating out the anti-collinear field yields a fermion-scalar four point interaction. 
	}
 	\label{fig:IntegrateOut}
 \end{figure}

As discussed above, there are $1\rightarrow 3$ splittings that should match onto a consistent SCET.  In principle, one can write down a term involving two fermions as long as one is represented by a spinor and the other by its conjugate.  Then the appropriate $\sigma$-matrix structure is required in order to contract the relevant indices.  A term of this type is in fact generated when integrating out the anti-collinear fermion $u_{\bar{n}}$.  However, as illustrated in Fig.~\ref{fig:IntegrateOut}, the full theory Yukawa interaction generates the four point scalar-fermion interaction in \eref{eq:colllag}, without generating any three-point vertices, as it must have.  Hence, the only collinear singularities present in the Lagrangian of the EFT are those that map onto double scalar emission in the full theory.   This motivates analyzing the interaction derived for the EFT, so that we can compare it with the full Wess-Zumino theory collinear structure discussed in Sec.~\ref{sec:IRstructureYukawa}. 

\newpage

In the collinear effective theory described by the Lagrangian in \Eq{eq:colllag}, the corresponding amplitude is
\begin{align} \label{eq:123eft}
\raisebox{-0.2\height}{\includegraphics[width=3cm]{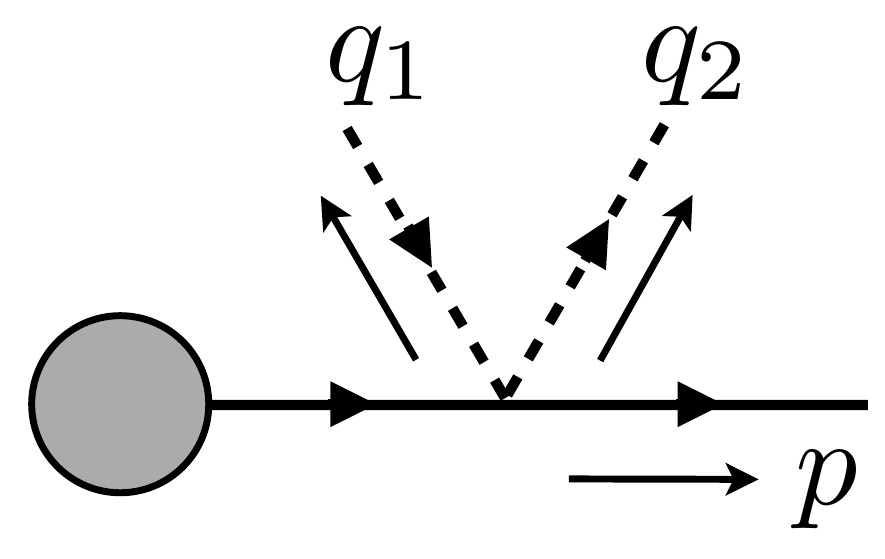}}
&=   x_n^\dagger(p)(-i|y|^2)\frac{1}{\bar n\cdot(p+q_2)}\frac{\bar n\cdot \bar \sigma}{2}i\frac{n\cdot \sigma}{2}\frac{\bar n\cdot (p+q_1+q_2)}{(p+q_1+q_2)^2}{\cal M}(p+q)\\
&=|y|^2x_n^\dagger(p)\left[
1+\frac{\bar n\cdot q_1}{\bar n\cdot(p+q_2)}
\right]
\frac{1}{(p+q_1+q_2)^2}{\cal M}(p+q)
\,,\nonumber
\end{align}
from which we can read off $\mathcal{S}_{1\rightarrow 3}^\text{EFT}$.  This agrees with \eref{eq:S1to3FullTheory}, the explicit calculation in the full theory.

\subsection*{Reparameterization Invariance}
The key step in deriving the fermion terms in the EFT Lagrangian is to construct projection operators that can be used to decompose fields into collinear and anti-collinear components.  For example, we decompose the Weyl spinor into $u_n$ and $u_{\bar n}$ by projecting with $n\cdot \sigma$ or $\bar n\cdot \sigma$, see \eref{eq:uProjections}.  Because the vector $n^\mu$ breaks Lorentz symmetry, each component $u_n$ and $u_{\bar n}$ does not respect full Lorentz covariance.  The RPI transformations are the Lorentz generators that are broken when fixing $n^\mu$, which in turn feeds into the projections.  One way to determine how the spinor components $u_n$ and $u_{\bar n}$ transform under RPI can be inferred by demanding that the full spinor $u$ is RPI invariant.  This was the procedure used in Sec.~\ref{sec:RPItrans} where we determined the RPI transformations for gauge theories; we apply the same arguments here to the Wess-Zumino model.

From \eref{eq:IntWZcompEOM}, the full spinor $u$ can be written in terms of the collinear spinor $u_n$ as
\begin{equation}\label{eq:fullWZspinor}
u_\alpha=u_{n,\alpha}  + u_{\bar{n},\alpha}=u_{n,\alpha}+\left(\sigma \cdot \partial_\perp\right)_{\alpha \dot{\alpha}} \frac{1}{\bar{n} \cdot \partial} \left(\frac{\nbsb}{2}\right)^{\dot{\alpha}\beta} u_{n,\beta} + y^*  \frac{1}{i \bar{n}\cdot \partial} \Bigl[ \phi_n^* \left(\frac{\nbs}{2}\right)_{\alpha \dot{\beta}} u_n^{\dagger \dot{\beta}} \Bigr] \,.
\end{equation}
To determine the RPI transformations of $u_n$, we demand that \eref{eq:fullWZspinor} is invariant.  Note that RPI-III does not have any effect on  \eref{eq:fullWZspinor} because there is no net $n$ or $\bar n$ number in this expression.  Therefore, $u_n$ transforms trivially under RPI-III as it did in the models with gauge boson interactions, see Table~\ref{table:RPI}.  RPI-I and RPI-II will be non-trivial, and as we will now argue that there does not exist a consistent linear realization for these transformations.

For concreteness, we can consider RPI-I.  A linear transformation implies that 
\begin{align}
u_{n} \rightarrow \mathcal{T}\,u_{n}\,; \qquad \qquad \qquad u_{n}^\dagger \rightarrow \overline{\mathcal{T}}\,u_{n}^\dag\,,
\end{align}
for some matrices $\mathcal{T}$ and $\overline{\mathcal{T}}$.  The index structure of the spinors implies that $\mathcal{T} \neq \overline{\mathcal{T}}$.  Note that $\phi_n^* \rightarrow \phi_n^*$ since it is a scalar.  It is possible that the transformations of the two terms that are proportional to $u_{n,\alpha}$ work in concert to yield an invariant expression (in fact this is precisely how RPI is preserved by the non-interacting theory).  This leaves the term proportional to $u_n^\dag$.  Noting that RPI-I does not change the definition of $\bar{n}$, see Table~\ref{table:RPI}, the only possible impact this transformation can have is to change $u_n^\dag$.  But then there is nothing left to cancel against it.  The implication is that presence of the non-trivial interaction breaks a linearly-realized RPI-I.  A similar argument holds for RPI-II, although for this case the $\bar{n}$ vector does shift, so the argument is not as simple.

The fact that RPI cannot be linearly realized in the Wess-Zumino model can be traced back to the projections.  These operators pick out the different spin states, and since the free Wess-Zumino Lagrangian respects chiral symmetry, RPI can be linearly realized in that case.  However, in the presence of any non-zero Yukawa interaction, chiral symmetry is broken and so RPI (if it exists at all) must be non-linearly realized in the interacting theory.\footnote{Cf.~theories with spontaneously symmetry breaking.  The pion field $\pi^a$ transforms non-linearly under the broken generators, but the exponentiated field $U=\exp[i\pi^a T^a]$ transforms linearly under the broken generators $T^a$.}

\subsection{The Wess-Zumino Model in Collinear Superspace}
\label{app:intWZcollinearsuperspace}
Next, we will derive the collinear Wess-Zumino Lagrangian directly in superspace.  This illustrates another example of deriving a collinear superspace Lagrangian via the general algorithm outlined in Sec.~\ref{sec:CollSuperspace}.  This will further verify the absence of a three-point function in the propagating Lagrangian for the EFT fields.  Furthermore, it is interesting to compare this result to the similar interactions that arise in $\mathcal{N} \geq 1$ SUSY Yang-Mills theories, see \emph{e.g.} \eref{eq:VectorSuperspaceLagrangian}. The intuition we gain in the simple case at hand will be useful for future work.

As in previous sections we begin by noting that supersymmetry requires $\Phi = \Phi_n+ \Phi_{\bar{n}}$. Therefore we expand the full theory Lagrangian as
\begin{equation}
\mathcal{L} = \int \text{d}^4 \theta \left( \Phi_n^\dagger \Phi_n + \Phi_n^\dagger \Phi_{\bar{n}} + \Phi_{\bar{n}}^\dagger \Phi_n + \Phi_{\bar{n}}^\dagger \Phi_{\bar{n}} \right) +  \int \text{d}^2\theta \,W\!\left( \Phi_n, \Phi_{\bar{n}}  \right) + \textrm{h.c.}\,.
\end{equation}
Taking the variation in superspace with respect to $\Phi^\dag_{\bar{n}}$ yields the constraint equation;
\begin{equation}
-\frac{1}{4} D D\, \Phi_{\bar{n}}-\frac{1}{4} D D\, \Phi_n + \frac{\delta W^* (\Phi^{\dagger}_n , \Phi^{\dagger}_{\bar{n}})}{\delta \Phi^{\dagger}_{\bar{n}}} = 0\,,
\end{equation}
with a similar expression for the conjugate equation.
For the superpotential of the Wess-Zumino Model,\footnote{Note that $\Phi_n^3$ and $\Phi_{\bar{n}}^3$ vanish due to the definition of these projected superfields.}
\bea
W = \frac{y}{3!} \,\Phi^3 \,\,\,\,\,\,\, \Rightarrow  \,\,\,\,\,\,\,   W = \frac{y}{2} \left( \Phi_n^2 \Phi_{\bar{n}} + \Phi_{\bar{n}}^2 \Phi_n \right),
\eea
taking $\delta \mathcal{L} / \delta \Phi_{\bar{n}}^\dagger$ yields  
\bea
\label{eq:IntWZsuperspaceEOM}
-\frac{1}{4}DD\,\Phi_{\bar{n}}-\frac{1}{4}DD\,\Phi_{n} + y^* \left( \Phi^{\dagger 2}_n +  2\,\Phi^{\dagger}_{\bar{n}} \,\Phi^{\dagger}_n \right) = 0\,.
\eea
 To check that this reproduces the expected equation of motion for the fermion component \eref{eq:IntWZcompEOM} we act on the above with $\bar{D}^{\dot{1}}$. Making use of some identities and the chirality condition, this reproduces the expected equation of motion, \eref{eq:IntWZcompEOM}.  Note that the $\phi_{\bar{n}}$ term has a higher scaling and should be dropped. 
 
In terms of superfields we postulate the solution
\bea
\Phi_{\bar{n}} = - \Phi_n + \frac{\bar{D}^{ \dot{1}} DD}{4\, i\, (\nbp) D_1} \Phi_n  - y^* \,\Phi_n^\dag \frac{\bar{D}^{\dot{1}}}{i\, (\nbp) D_1} \Phi_{n}^\dagger,
\eea
 which obeys the chirality condition and solves the \eqref{eq:IntWZsuperspaceEOM}.  Once this is plugged back into the Lagrangian, an interaction term is inherited from the Yukawa coupling: 
\bea
\label{eq:superspaceWZL}
\mathcal{L}_n &=& \int \text{d}\theta^2\, \text{d} \theta^{\dag \dot{2}}\, i \,\Phi_n^\dagger \frac{\Box}{\nbp} \Phi_n +  |y|^2 \int \text{d}\theta^2\, \text{d} \theta^{\dag \dot{2}}\,  \frac{1}{\nbp} \left(\Phi_n\, D_2\, \Phi_n \right) \frac{1}{\nbp} \left(\Phi_n^\dagger \,\bar{D}_{\dot{2}}\, \Phi_n^\dagger \right) \\  \nonumber
\label{eq:componentWZL}
&=& \,\,  i \,u^{\dag}_{2} \frac{\Box}{\nbp} u_2 - \phi^{*} \Box  \phi+ |y|^2 \phi^*_n\, u^{\dag}_{2}\, \frac{1}{i \,\nbp}\, u_{2}\,\phi_n- \frac{|y|^2}{4} |\phi_n|^4\,.
\eea
This reproduces the desired interaction that we found above in \eref{eq:colllag}.

\newpage

\section{Outlook}
\label{sec:ConclusionsAndFuture}
Now that we have laid the groundwork for how to think about SUSY SCET, there are many interesting directions to pursue.   One could attempt to leverage some of the extraordinary results derived for SUSY models, for instance the exact NSVZ $\beta$-function~\cite{Novikov:1983uc}, Seiberg duality~\cite{Seiberg:1994pq}, Seiberg-Witten theory~\cite{Seiberg:1994aj,Seiberg:1994rs}, and the finiteness of $\mathcal{N} = 4$ SUSY Yang-Mills (SYM)~\cite{Mandelstam:1982cb}, in order to learn more about the fundamental properties of SCET.  Additionally, understanding how the construction presented here fits within the larger context of models that manifest SUSY in non-trivial ways, \emph{e.g.}~\cite{Dall'Agata:2015lek, Ferrara:2015tyn, Kahn:2015mla,Komargodski:2009rz, Festuccia:2011ws, Kallosh:2016hcm, Ferrara:2016een, Dall'Agata:2016yof}, should lead to a deeper understanding of collinear superspace and its possible extensions.

The biggest formal open question is to understand how to take the collinear limit of theories with extended SUSY, and in particular $\mathcal{N} = 4$ SYM.  Naively, we expect that applying this procedure to $\mathcal{N} = 4$ SYM would leave half the supercharges unbroken, as in the $\mathcal{N} = 1$ theories explored above.  However, it is clear that there will need to be a non-trivial connection with the $SU(4)_R$ symmetry, which will perhaps manifest by making a judicious choice of $R$-dependent projection operators.  

We additionally anticipate interesting features due to the fact that $\mathcal{N} = 4$ SYM includes Yukawa couplings.  It is possible that accounting for the complete infrared structure of this model will require including local operators as in the Wess-Zumino model.  However, we do have some preliminary evidence that multi-flavor Yukawa theories might include the $1\rightarrow 2$ collinear divergence structure directly in the Lagrangian for the propagating fields.  In particular, take the following Yukawa theory with multiple flavors, 
\begin{align}
\mathcal{L} =  i \,u_i^\dagger (\bar{\sigma}\cdot \partial) u_i + i \,v_k^\dagger (\sigma \cdot \partial) v_k - y_{lik} \,\phi_l\Big( v_k^\dagger u_i + u_i^\dagger v_k\Big) + \textrm{h.c.}
\end{align}
where $u_i$ $(v_k)$ are left (right) handed Weyl spinors.  The equation of motion for $u_i$ and $v_k$ are
\begin{align} 
& u_{\bar{n},i} = \frac{1}{\nbp} \left(\sigma \cdot \partial_\perp \right) \frac{\nbsb}{2} u_{n,i} + \frac{1}{i\, \nbp} \Big[ (y_{lik}\, \phi_l + y^*_{lik}\, \phi^*_l) \frac{\nbs}{2}v_{n,k}   \Big] \,; \notag\\
& v_{\bar{n},i} = \frac{1}{\nbp} \left(\bar{\sigma} \cdot \partial_\perp \right) \frac{\nbs}{2} v_{n,i} + \frac{1}{i\, \nbp} \Big[(y_{lik} \,\phi_l + y^*_{lik} \,\phi^*_l) \frac{\nbsb}{2}u_{n,k} \Big] \, ,
\end{align}
which lead to tri-linear interactions such as
\begin{align}
 \mathcal{L}^{(0)}_n \,\, \supset \,\, y_{ijk}\,  \phi_k v_{n,i}^{\dagger} \frac{\left(\sigma \cdot \partial_\perp \right) }{\nbp} \frac{\nbsb}{2} u_{n,j} \, .
\end{align}
This provides a candidate EFT interaction that could reproduce the collinear IR divergences of Yukawa theory.  These tri-linear couplings could not be written down for the single flavor Wess-Zumino model above, due to the fact that a collinear fermion anti-commutates with itself.  This provides a compelling hint that whatever form the collinear $\mathcal{N} = 4$ SYM model takes, it will involve the $SU(4)_R$ flavor structure in a non-trivial way.  Additionally, the interplay between gauge symmetry and RPI that we found for the $\mathcal{N}=1$ SYM theory should manifest in a similar way for the model with extended SUSY.  Furthermore, there is the interesting open question of constructing the local operators of an EFT for Higgsstrahlung off a top quark.

Once the $\mathcal{N} = 4$ SYM SCET model has been discovered, there will be a many interesting directions to explore.  There is so much structure in the full theory that can be applied to the EFT.  Understanding the connection with dual-conformal invariance, working out the collinear limit of the many ways of formulating this model -- the dualities among amplitudes, Wilson loops, and correlation functions --  and just leveraging the tremendous wealth of data on the amplitudes to get a deeper picture of SCET itself, are just some of the possible applications of such a formalism.  The work presented here has just begun to scratch the surface, and we are optimistic that SUSY SCET will lead to a deeper understanding of collinear EFTs and beyond.

\acknowledgments

We are especially grateful to our anonymous reviewer along with Andrew Cohen, Aneesh Manohar, Martin Schmaltz, Iain Stewart, and Jesse Thaler for insights that led to significant improvements for our treatment of the Wess-Zumino model. We thank Marat Freytsis, Ian Low, Duff Neill, Arvind Rajaraman, and David Shih for discussions.  TC and GE are supported by an LHC Theory Initiative Postdoctoral Fellowship, under the National Science Foundation grant PHY-0969510. GE was supported by the U.S. Department of Energy, under grant Contract Numbers DE-SC00012567. AL is supported an LHC Theory Initiative Postdoctoral Fellowship, under the National Science Foundation grant PHY-1419008.  This work was in part initiated at the Aspen Center for Physics, which is supported by National Science Foundation grant PHY--1066293.

\newpage
\addcontentsline{toc}{section}{$\,\,\,\,\,\,\,$ Appendices}


\appendix

\section*{Appendices}
What follows are three appendices.  Appendix~\ref{app:wilson} discusses Wilson lines in SUSY theories.  Appendix~\ref{sec:CovariantGaugeSingularities} derives the collinear splitting factor for SCET in a covariant gauge.  Finally, Appendix~\ref{app:notation} contains notation and conventions used throughout this paper.

\section{Wilson Lines and SUSY}\label{app:wilson}
Emission of multiple collinear gluons can be expressed as a Wilson line; for instance see Eq.~(2.33) of \cite{StewartNotes}. The QCD Lagrangian can therefore be written as;
\bea
\mathcal{L}_{u_n} = u_n^\dagger \left( i\, n\cdot \mathcal{D} + i \,\bs \cdot  \mathcal{D}_\perp W_n \frac{1}{i\, \bar{n} \cdot \partial} W_n^\dagger i\,\sigma \cdot  \mathcal{D}_\perp \right) \frac{\nbsb}{2} u_n \, ,
\eea
where
\bea
W_n = \sum_{\textrm{perms}}e^{-g \frac{1}{\nbp} \bar{n} \cdot A_n^a t^a} \, .
\eea
In other words, the field $n\cdot A_n$ can be summed into a collinear Wilson line. This has the advantage of making operators explicitly gauge invariant.  Note that in LCG, where $n\cdot A_n = 0$, the Wilson line is given by the identity operator. 

Given the goals of this paper, it is interesting to extend this formalism to SUSY theories; the collinear lines above now describe emission of collinear gauge fields from off-shell gauginos which have been integrated out.  The Wilson line transforms non-trivially under SUSY. For $\mathcal{N} = 1$ SYM, this can be deduced from the gauge and gaugino transformations:
\bea
W_n  \rightarrow && W_n \sum_{\textrm{perms}} e^{\frac{g}{\sqrt{2}} \frac{1}{\nbp} \Big(\eta^\dagger \nbsb u_n + u_n^\dagger \nbsb \eta\Big)^a t^a}  \\ \nonumber
&&= W_n + \frac{g}{\sqrt{2}} \frac{1}{\nbp} \Big(\eta^\dagger \nbsb u_n + u_n^\dagger \nbsb \eta\Big)^a t^a W_n \, ,
\eea
so that
\bea
\delta(W_n) &=& \frac{g}{\sqrt{2}} \frac{1}{\nbp} \Big(\eta^\dagger \nbsb u_n + u_n^\dagger \nbsb \eta\Big)^a t^a W_n\,; \nonumber\\ 
\delta(i  \mathcal{D}_\perp^\mu)&=&- \frac{g}{\sqrt{2}}\Big( \eta^\dagger \bar{\sigma}_\perp^\mu u_n  + u_n^\dagger \bar{\sigma}_\perp^\mu \eta\Big)^a t^a \,.
\eea
For additional discussions regarding SUSY Wilson lines see \cite{Zarembo:2002an} and \cite{Pestun:2008ora}.

\section{Collinear Factor from SCET in Covariant Gauge}
\label{sec:CovariantGaugeSingularities}
In this appendix, we compute the collinear factor arising from emission of a collinear gauge boson in Abelian SCET in the standard covariant gauge.  The result is a verification of the calculation for the full theory and for LCG SCET, see \eref{eq:FullThyCollinearFactor} and \eref{eq:LCGcollinearFactor} respectively, and acts to emphasize the gauge independence of this factor in the exactly collinear limit. The SCET Lagrangian gives rise to fermion-gauge interactions~\cite{StewartNotes}:
\begin{align}
\mathcal{L} \supset e\, u^{\dagger}_n  n\cdot A  \left(\frac{\nbsb}{2}\right)  u_n- i \,u^\dagger_n \Big( (\bar{\sigma} \cdot \mathcal{D}_{\perp,n})  \frac{1}{\bar{n} \cdot \mathcal{D}_n} (\sigma \cdot \mathcal{D}_{\perp,n})  \left(\frac{\nbsb}{2}\right) u_n\Big) \,,
\end{align}
where we have expressed this in two-component notation.  The Feynman rule for the vertex is given by:
\begin{align}
\raisebox{-0.25\height}{\includegraphics[width=2.7cm]{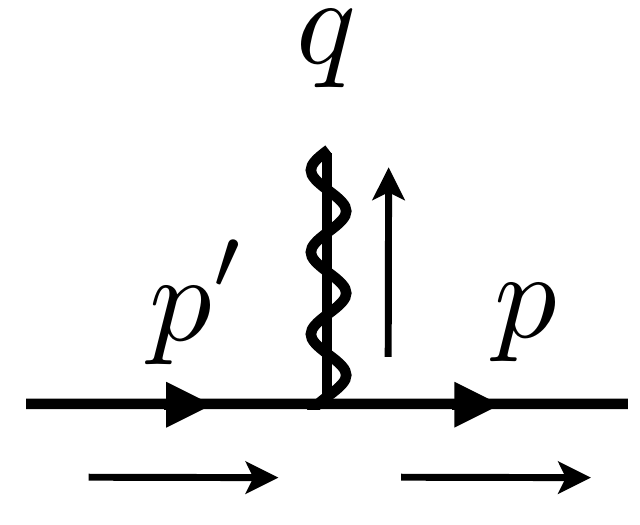}}
  &= -i\, e\, \left[ n^\mu - \frac{(\bar{\sigma} \cdot p_\perp)(\sigma \cdot p'_\perp)}{(\bar{n}\cdot p')(\bar{n}\cdot p)} \bar{n}_\mu +\bar{\sigma}_\perp^\mu \frac{\sigma_\perp \cdot p'_\perp}{\bar{n}\cdot p'}  + \frac{\bar{\sigma}_\perp \cdot p_\perp}{\bar{n}\cdot p} \sigma_\perp^\mu  \right] \frac{\nbsb}{2} \, .
\end{align}
Here the third and forth terms are due to the transverse gauge boson polarizations and agree with the corresponding terms shown in Fig.~\ref{fig:LCGFeynRules}.  However, the first and second terms are different, since they are due to the unphysical polarizations and are gauge dependent.

Now that we have the Feynman rules, we can compute the collinear factor.   As above, we will take the frame where the fermion is exactly collinear $p^\mu = \frac{n^\mu}{2} \bar{n}\cdot p$:
\begin{align}
\nonumber
\raisebox{-0.2\height}{\includegraphics[width=3.8cm]{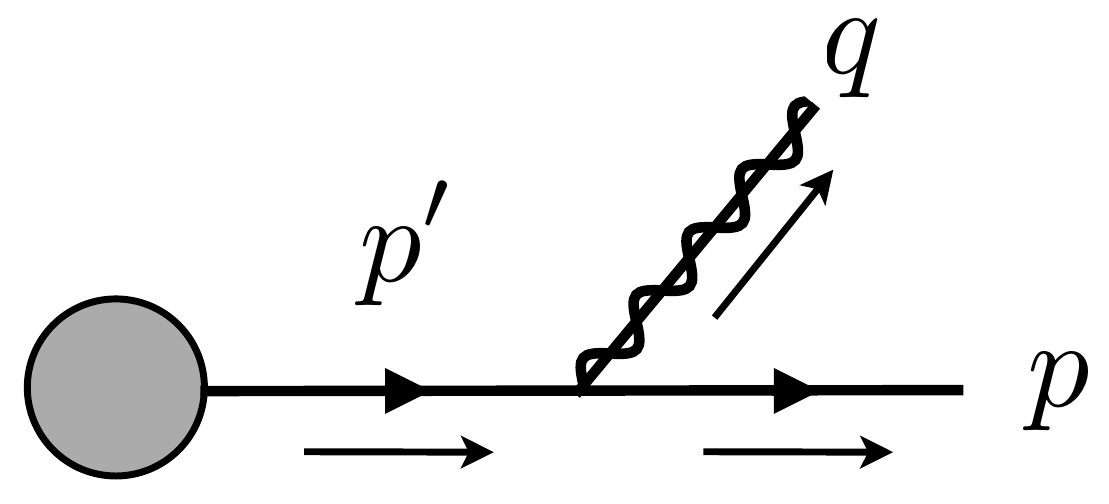}}
  &= -i\, e\, x^\dagger_n (p) \left[  \frac{n \cdot \epsilon^*   \bar{n} \cdot (p+q)}{(\bar{n}\cdot p)(n\cdot q)}+  \frac{(\bar{\sigma}_\perp \cdot \epsilon^* )(\sigma_\perp \cdot q_\perp)}{(\bar{n}\cdot p)(n\cdot q)}    \right]  i \mathcal{M}(p+q)\\
& = -\, e\, x^\dagger_n (p) \left[ \frac{(n\cdot \epsilon^*)}{(n \cdot q) } -    \frac{  (\bar{\sigma}_\perp \cdot q_\perp) (\sigma_\perp \cdot \epsilon^*)}{(\bar{n} \cdot p)(n \cdot q)}   \right]  \mathcal{M}(p+q),\,
\end{align}
where we have used conservation of momentum $p' = p + q$, have absorbed a factor of the projection operator $(\nbsb/2)(\ns/2) $ into the definition of $\mathcal{M}$, and have applied the sigma matrix identity $ \bar{\sigma}^\mu \sigma^\nu = -  \bar{\sigma}^\nu \sigma^\mu+ 2 g^{\mu \nu}$.  This expression matches the amplitude computed in \eref{eq:FullThyCollinearFactor} and \eref{eq:LCGcollinearFactor} as expected.

\section{Notation and Identities}\label{app:notation}
\noindent  We work in Minkowski space with metric signature $g^{\mu \nu} = \textrm{diag}\left(+1,-1,-1,-1\right)$. Throughout this paper we follow spinor conventions of \cite{Dreiner:2008tw} and \cite{Binetruy:2006ad}. For a useful review of the conventions relevant for SUSY see pages 449-453 of \cite{Binetruy:2006ad}.
We work in the Weyl Basis where the $\gamma$-matrices take the following form;
\bea
\gamma^\mu = \left[\begin{array}{cc}
0 & \sigma^\mu \\
\bar{\sigma}^\mu & 0 \\
\end{array}\right] \, .
\eea
Here $\sigma^\mu = \left(1,\sigma^i\right)$, $\bar{\sigma}^\mu = \left(1,-\sigma^i\right)$, and the Pauli matrices are:
\bea
\begin{array}{cccc}
\sigma^0 =   \left[ \begin{array}{cc}
1 & 0 \\
0 & 1
 \end{array} \right] \, ,
& \quad  \quad
\sigma^1 =  \left[ \begin{array}{cc}
0 & 1 \\
1 & 0
 \end{array} \right] \, ,
& \quad  \quad
\sigma^2 =  \left[ \begin{array}{cc}
0 & -i \\
i & 0
 \end{array} \right] \, ,
& \quad \quad
\sigma^3 =  \left[ \begin{array}{cc}
1 & 0 \\
0 & -1
 \end{array} \right] \, .
 \end{array}
\eea
The upper/lower spinor index convention is
\bea
\begin{array}{cc}
\left( \sigma^\mu \right)_{\alpha \dot{\alpha}}\, , & \quad \textrm{and} \quad \,  \left( \bar{\sigma}^\mu \right)^{\dot{\alpha} \alpha} \, .
 \end{array}\nonumber
\eea
Spinor indices are raised and lowered using the anti-symmetric $\epsilon$-matrix:
\bea
\epsilon_{\alpha \beta} = - \epsilon^{\alpha \beta} = 
\left[ \begin{array}{cc}
0 & -1 \\
1 & 0
 \end{array} \right] \, ,
\eea
and we contract pairs of spinors and anti-spinors follows the usual way:
\bea
\begin{array}{cc}
\xi^\alpha \psi_\alpha\, , & \quad \textrm{and} \quad \, \xi^\dag_{\dot{\alpha}}\psi^{\dag\dot{\alpha}}  \end{array}. \nonumber
\eea
\subsection*{Sigma Matrix Identities}
\noindent 
Throughout this work we make use of various sigma-matrix identities.  Those that were used most often are
\begin{align}
 \bar{\sigma}_{\dot{\alpha}\alpha}^\mu =& \,\epsilon_{\alpha \beta} \epsilon_{\dot{\alpha} \dot{\beta}} \sigma^{\mu \beta \dot{\beta}}  \, ,\\
 \sigma^\mu \bar{\sigma}^\nu + \sigma^\nu \bar{\sigma}^\mu =& \,2 g^{\mu \nu} \, ,\\
 \bar{\sigma}^\mu \sigma^\nu + \bar{\sigma}^\nu \sigma^\mu =& \,2 g^{\mu \nu} \, \\
\bs^\mu \sigma^\nu \bs^\rho =& \,g^{\mu \nu} \bs^\rho - g^{\mu \rho}\bs^\nu + g^{\nu \rho} \bs^\mu - i \epsilon^{\mu \nu \rho \kappa} \bs_\kappa \, ,\\
 \bs^\rho \sigma^\lambda \bs^\delta \sigma^\mu \bs^\nu =&\, g^{\rho \lambda} (g^{\delta \mu} \bs^\nu - g^{\delta \nu}\bs^\mu + g^{\mu \nu} \bs^\delta - i \epsilon^{\delta \mu \nu \omega} \bs_\omega) \\ \nonumber
&- g^{\rho \delta}( g^{\lambda \mu} \bs^\nu - g^{\lambda \nu} \bs^\mu + g^{\mu \nu} \bs^\lambda - i \epsilon^{\lambda \mu \nu \omega}\bs_\omega ) \\ \nonumber
 &+ g^{\lambda \delta}( g^{\rho \mu}\bs^\nu - g^{\rho \nu} \bs^\mu + g^{\mu \nu} \bs^\rho - i \epsilon^{\rho \mu \nu \omega} \bs_\omega ) \\ \nonumber
 &- i \epsilon^{\rho \lambda \delta \xi} g_{\xi \kappa} ( g^{\kappa \mu} \bs^\nu - g^{\kappa \nu} \bs^\mu + g^{\mu \nu}\bs^\kappa - i \epsilon^{\kappa \mu \nu \omega}\bs_\omega) \, .  \nonumber
\end{align}
For additional identities see \cite{Dreiner:2008tw} and \cite{Binetruy:2006ad}. 

\subsection*{Shorthand Conventions for Collinear Fields}
To keep the notation from being too cumbersome, we frequently drop the subscript ``$n$" on the collinear fields.  The potentially most confusing case is the fermion.  We always take ``$u$" to represent a full theory field; we only drop the collinear subscript when dealing with components:
 \bea
\begin{array}{cc}
u_{n,2} \equiv u_2 \, , & \quad \textrm{and} \quad \, u_{n,\dot{2}}^* \equiv u_2^*  \, . \end{array} \nonumber
\eea
Note we also drop the ``dot" on the subscript of the conjugate field.

\subsection*{Notation and Conventions in LCG}
\noindent It is useful to keep in mind the spinor structure of objects in LCG. For instance,
\begin{align}
& \sigma^\mu \partial_\mu = 
\left[ \begin{array}{cc}
\np & \sqrt{2}\partial^*  \\
\sqrt{2} \partial & \nbp  \end{array} \right]_{\alpha \dot{\alpha}},  \,
& \bar{\sigma}^\mu \partial_\mu = 
\left[ \begin{array}{cc}
\nbp & -\sqrt{2}\partial^*  \\
-\sqrt{2} \partial & \np  \end{array} \right]^{\dot{\alpha} \alpha}.
\end{align}
Note the slight abuse of notation; $\sqrt{2} \partial = \partial_1 - i \partial_2$ and $\sqrt{2} \partial^* = \partial_1 + i \partial_2$ refer to transverse degrees of freedom, while the derivative contracted with a four vector, \emph{e.g.} $n\cdot \partial = n^\mu \partial_\mu$, is relevant for the full Lorentz four vector. Similar expressions hold for other contractions such as $\sigma^\mu A_\mu$. These expressions are independent of the choice of $n^\mu$, and $\bar{n}^\mu$ direction. 

Note that throughout we include the Lorentz contraction in the definitions of $\partial_\perp^2$, as this is convenient when working with LCG scalars;
\bea
\partial_\perp^2 \equiv \partial_\perp^\mu \partial_{\perp \mu}  = - \partial_1^2 - \partial_2^2  = - 2 \partial \partial^* \, ,
\eea
where we have converted to LCG derivatives. This is in contrast to some places in the literature which relate $\partial_\perp^2$ to the explicit component expression with the opposite sign. In terms of this notation 
\bea
\Box = \partial^\mu \partial_\mu = \nbp \np + \partial_\perp^2 = \nbp \np - 2 \partial \partial^* \,.
\eea

\subsection*{Superspace Derivative Identities}
\noindent For manipulations involving superspace derivatives
\bea
&& D_\alpha = \frac{\partial}{\partial \theta^\alpha} - i (\sigma \cdot \partial)_{\alpha \dot{\alpha}} \bar{\theta}^{\dot{\alpha}}   \quad  \,\, \textrm{and} \quad \,\,  \bar{D}_{\dot{\alpha}} =  \frac{\partial}{\partial \bar{\theta}^{\dot{\alpha}}}+ i \theta^\alpha (\sigma \cdot \partial)_{\alpha \dot{\alpha}}\,, 
\eea
we make use of the following identities: 
\bea
\label{eq:superspacederivativeiden}
&&\Box \leftrightarrow -\frac{1}{16} \bar{D} \bar{D} D D \, , \\ 
&& D^\alpha \bar{D} \bar{D} D_\alpha V = \bar{D}_{\dot{\alpha}} D D \bar{D}^{\dot{\alpha}}V \, ,  \\ 
&& D D \bar{D}_{\dot{\alpha}} DD  = 0 = \bar{D} \bar{D} D_\alpha \bar{D} \bar{D} \, ,  \\ 
&& D_\alpha D_\beta D_\gamma = 0 = \bar{D}_{\dot{\alpha}} \bar{D}_{\dot{\beta}} \bar{D}_{\dot{\gamma}} \, ,  \\ 
&& [ DD, \bar{D}_{\dot{\alpha}}] = -4 i D^\alpha(\sigma^\mu)_{ \alpha\dot{\alpha}} \partial_\mu  \, ,  \\
&& [\bar{D}\bar{D}, D_\alpha] = 4 i (\sigma^\mu)_{ \alpha \dot{\alpha}} \partial_\mu \bar{D}^{\dot{\alpha}} \, , \\   \nonumber
&& D^\alpha \bar{D}\bar{D} D_\alpha = \bar{D}_{\dot{\alpha}} DD \bar{D}^{\dot{\alpha}}=  D D \bar{D}\bar{D}+ 4i D^\alpha (\sigma \cdot  \partial)_{\alpha \dot{\alpha}} \bar{D}^{\dot{\alpha}}  \\  
&&\quad \quad \quad \quad \,\, \,\,=  16 \Box+ 4i D^\alpha (\sigma \cdot \partial)_{\alpha \dot{\alpha}} \bar{D}^{\dot{\alpha}} \, . 
\eea
We use the shorthand $DD = D^\alpha D_\alpha$ and $\bar{D} \bar{D} = \bar{D}_{\dot{\alpha}} \bar{D}^{\dot{\alpha}}$ to imply spinor indices have been contracted. 
Since we often only work with the supercharges remaining in the EFT (or equivalently the superspace coordinates remaining in collinear superspace), it is often useful to explicitly write out the spinor indices of the above identities (as usual $D^2 = - D_1$ and $D^1 = D_2$), \emph{e.g.}
\eref{eq:superspacederivativeiden};
\bea
&& [ DD, \bar{D}_{\dot{1}}] = -4 i ( \np D^1 + \sqrt{2} \partial^* D^2) = -4i ( \np D_2 - \sqrt{2} \partial^* D_1) \, ,\\ \nonumber
&&[ DD, \bar{D}_{\dot{2}}] = -4i (\nbp D^2 + \sqrt{2} \partial D^1) = -4i (-\nbp D_1 + \sqrt{2} \partial D_2 ) \, ,
\eea 
which we made use of for the derivation of collinear superspace, \emph{e.g.}
\begin{align}
\bar{D} \bar{D} D ^ 1 \bar{D}^{ \dot{1}} DD  \, &\propto \, \bar{D} \bar{D} D^1 \Big[ (\nbp) D_1 - a D_2\Big] \left(D_2 \bar{D}_{\dot{1}} V_n\right) \notag\\
&\, \propto \, \bar{D} \bar{D} D D (\nbp)  \left(D_2 \bar{D}_{\dot{1}} V_n\right) \, \propto \, (\nbp)  \, \Box \,.
\label{eq:ManyDIdentity}
\end{align}

\subsection*{Yukawa Theory}
\noindent The Yukawa theory Lagrangian for a four component fermion is:
\bea
\mathcal{L}_Y = i \bar{\psi}_D \gamma^\mu \partial_\mu \psi_D - y \phi \bar{\psi}_D\psi_D + \textrm{h.c.}\,.
\eea
In terms of left and right handed Weyl spinors, the Dirac spinor is decomposed in the usual way
\bea
 \psi_D = {u \choose{v}} \,.
\eea
The Lagrangian becomes $\mathcal{L}_Y = i v^\dagger (\sigma \cdot \partial) v + i u^\dagger (\bar{\sigma}\cdot \partial) u - y \phi( v^\dagger u + u^\dagger v) + \textrm{h.c.}$

The collinear and anti-collinear fields are separated out using projection operators for left and right handed Weyl spinors:
\begin{align}
&u = \left(P_{n,L} + P_{\bar{n},L} \right) u = u_n + u_{\bar{n}}\,, &v = \left(P_{n,R} + P_{\bar{n},R} \right) v=  v_n + v_{\bar{n}}\,, 
\end{align}
where
\begin{align}
\begin{array}{ll}
P_{n,L} =  \frac{n \cdot \sigma}{2}   \frac{\bar{n} \cdot \bar{\sigma}}{2} \,; \quad\quad\quad\quad &P_{n,R} =   \frac{n \cdot \bar{\sigma}}{2}  \frac{\bar{n} \cdot \sigma}{2} \,; \\
P_{\bar{n},L} =  \frac{\bar{n} \cdot \sigma}{2}   \frac{n \cdot \bar{\sigma}}{2}\,; & P_{\bar{n},R} =   \frac{\bar{n} \cdot \bar{\sigma}}{2}  \frac{n \cdot \sigma}{2}\,,
\end{array}
\end{align}
and
\begin{align}
\begin{array}{ll}
(n\cdot \bar{\sigma}) u_n = 0 \,; \quad\quad\quad\quad &(n\cdot \sigma) v_n = 0 \,; \\
(\bar{n}\cdot \bar{\sigma}) u_{\bar{n}} = 0\,; & (\bar{n}\cdot \sigma) v_{\bar{n}} = 0\,.
\end{array}
\end{align}

\end{spacing}

\begin{spacing}{1.1}
\bibliography{SUSY_SCET}
\bibliographystyle{utphys}
\end{spacing}

\end{document}